\documentclass[aps,amsmath,amssymb,prx,twocolumn,longbibliography,floatfix]{revtex4-2}

\usepackage{mathrsfs} 
\usepackage{graphicx,color} 
\usepackage{mathtools}
\usepackage{bigints}
\usepackage{bm}
\usepackage[normalem]{ulem}
\usepackage{physics}
\usepackage{color}

\def\p{\partial}

\def\ve{\varepsilon}
\newcommand{\av}[1]{\langle{#1}\rangle}

\newcommand{\bi}[1]{\mbox{\boldmath$#1$}}
\newcommand{\pp}[2]{\frac{\partial {#1}}{\partial {#2}}}

\usepackage[colorlinks=true]{hyperref}
\hypersetup{
  colorlinks=true,
  linkcolor=blue,
  filecolor=magenta,
  citecolor=blue,
  urlcolor=blue,
}

\begin{document}
\title{Diffusionless relaxation of half-skyrmion liquid, hexatic, and crystalline states in a chiral molecular crystal}
\author{Kyohei Takae}
\email{takae@tottori-u.ac.jp}
\affiliation{Graduate School of Engineering, Tottori University, 4-101 Koyama-Cho Minami, Tottori, Tottori 680-8552, Japan}
\affiliation{Advanced Mechanical and Electronic System Research Center, Faculty of Engineering, Tottori University, 4-101 Koyama-cho-minami, Tottori, Tottori 680-8552, Japan}
\author{Kota Mitsumoto}
\affiliation{Graduate School of Arts and Sciences, The University of Tokyo, 3-8-1 Komaba, Meguro-ku, Tokyo 153-8902, Japan}
\date{\today}


\begin{abstract}
Particles in a crowded environment exhibit slow anomalous diffusion, and their efficient manipulation is important in controlling transport phenomena in complex materials.
Skyrmions and half-skyrmions, spatially localized quasiparticles observed in magnetic systems and liquid crystals, also exhibit diffusive motion. They exhibit normal diffusion in dilute conditions. However, the cooperative dynamics and diffusion of skyrmions and half-skyrmions in their condensed liquid, hexatic, and crystalline phases are elusive. Here we show in the half-skyrmion condensed phases that the fusion and fission of half-skyrmions, not their diffusion, are responsible for the primary structural relaxation. The fusion and fission occur due to the non-conserved nature of the quasiparticle number density. The diffusion, which contributes to the secondary structural relaxation, is suppressed by cages formed by surrounding half-skyrmions, whereas enhanced by Mermin-Wagner fluctuation characteristic to two-dimensional systems, leading to subdiffusive motion. Large displacement of half-skyrmions is locally excited by fusion-fission and bond-breaking between adjacent half-skyrmions via heterogeneous elastic fields. Furthermore, the motion of half-skyrmions couples with transverse and longitudinal sound wave excitation differently, where the transverse sound wave is more attenuated than the longitudinal one due to the coupling between transverse sound wave and half-skyrmion deformation. We also discuss the {relationship} between half-skyrmion diffusion in our system and skyrmion diffusion in magnetic systems. Our result provides a qualitative difference in dynamical properties between half-skyrmion gaseous and condensed phases, suggesting the efficient manipulation of high-density half-skyrmions and skyrmions.
\end{abstract}


\maketitle


\section{Introduction}
How a particle moves under the influence of surrounding media and external perturbation is essential for manipulating the particle and controlling the functions of particle-dispersed systems, such as biological cells~\cite{Franosch,Brangwynne}, polymer solutions~\cite{Amblard,Schmidt}, nanoporous materials~\cite{Karger}, glass-forming liquids and granular materials~\cite{Glasses}, and pedestrian movement in a crowd~\cite{Crowd}. An important property governing dynamical features is the mean-square-displacement (MSD), which exhibits $\av{|\Delta \bi r|^2}\sim t^\alpha$ where $\Delta \bi r$ is the displacement of each particle, the bracket represents statistical and time average, and $t$ is the time. $\alpha=1$ in simple liquids referred to as the normal diffusion to yield the diffusion constant $D=\av{|\Delta \bi r|^2}/2dt$ with $d$ be the spatial dimension~\cite{Hansen}. However, the above systems often exhibit $\alpha<1$, called subdiffusion, because of viscoelastic relaxation, disorder, and crowding~\cite{Bouchaud,Franosch}. In the subdiffusive systems, particle transport is drastically slowed down compared with the normal diffusion. Thus, understanding the mechanism of subdiffusion is crucial to controlling the particle motion and, accordingly, the transport properties in these complex media.

\begin{figure}[t!]
\centering
\includegraphics[width=8cm]{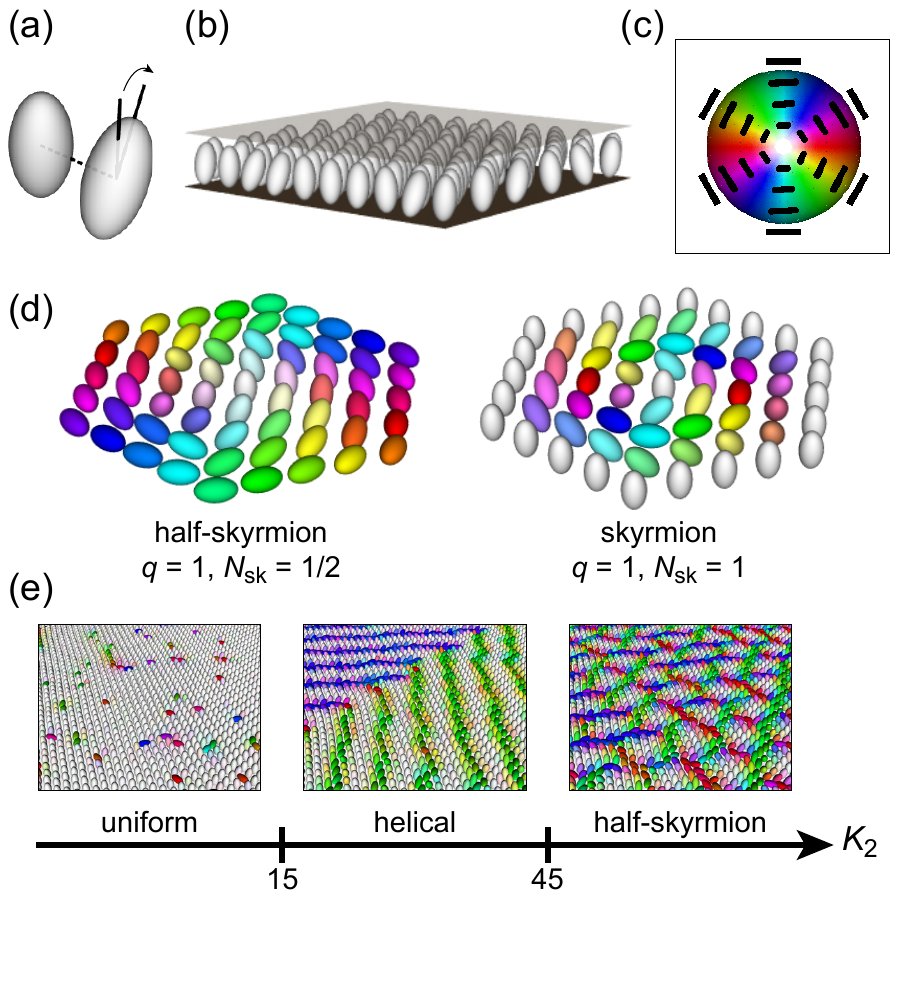}
\caption{
The numerical setup and color notation of our model.
(a) Adjacent molecules favor aligning with a twist.
(b) These molecules are confined in a monolayer.
(c) Color maps of molecular orientation adopted in this paper except for Fig.~\ref{fig:order}(a,c).
{(d) Schematic representation of half-skyrmions (left) and skyrmions (right). Molecular orientation becomes perpendicular (parallel) to the center in half-skyrmions (skyrmions). The vorticity $q$ and skyrmion number $N_{\rm sk}$ read $(q, N_{\rm sk})=(1, 1/2)$ for half-skyrmions, while $(1, 1)$ for skyrmions.}
(e) Typical molecular configuration at low temperature depending on the magnitude of the twist interaction with $\eta=2$, $q=0.3$, $\rho=0.3$, and $T=0.4$ {(see Sec.~IIA for their definition)}. The half-skyrmion phase is realized for $K_2\gtrsim 45$.
}
\label{fig:setup}
\end{figure}

Diffusion also manifests in quasiparticle systems.
Typical examples are skyrmions and half-skyrmions {(merons)}, spatially localized two-dimensional vortex-like objects formed by many particles. {Skyrmions are characterized by integer skyrmion number $N_{\rm sk}=1$, where $N_{\rm sk}$ in two dimension is defined by $N_{\rm sk}=\frac{1}{4\pi}\int dxdy~\bi n\cdot(\pp{\bi n}{x}\times\pp{\bi n}{y})$ with $\bi n$ be the orientation of spins or molecules. At the edge of the skyrmion, the orientation of spins or molecules becomes antiparallel to the center. On the other hand, half-skyrmions are characterized by half-integer skyrmion number $N_{\rm sk}=1/2$. The orientation at the edge becomes perpendicular to the center (see Fig.~\ref{fig:setup}(d)).} Skyrmions {and half-skyrmions} are observed in various condensed matter, including chiral and frustrated magnets~\cite{NagaosaTokura2013,Tokura2021review,Kawamura}, 
chiral liquid crystals~\cite{Fukuda2011,Fukuda2017}, Bose-Einstein condensates~\cite{OhmiMachida,Ho-BEC}, and dielectric materials~\cite{Ramesh-polar2019,Ramesh-polar2021}.
Among these systems, dynamics of skyrmions in chiral magnets are extensively studied, in which it has been known that magnetic skyrmions exhibit normal diffusive behavior in dilute conditions~\cite{Nagaosa2014,diffusion}. It has also been known that the existence of impurities under external drive, which results in competition between impurity pinning and driven motion, leads to subdiffusion in a particular direction while enhanced in another dimension~\cite{pinning}. These studies extract the conserved nature of skyrmions: skyrmions are treated as rigid particle-like objects because skyrmions are topologically protected objects {with fixed skyrmion number}. In skyrmion condensed phases, however, interacting skyrmions often deform and coalesce~\cite{Mochizuki,Buttner,Muller,Tokura2018NP,Litzius,Tokura2021}, indicating the non-conserved nature of skyrmions. That is, the shape and number density can vary over time. Although the manipulation of skyrmions in condensed phases suggests efficient high-density information carriers and storage~\cite {Fert,Klaui-review,Smalyukh}, the role of skyrmion deformation, aggregation, and crowding on skyrmion diffusion and relaxation remains elusive.
In other skyrmion and half-skyrmion systems, furthermore, the dynamical signature has yet to be measured quantitatively~\cite{Fukuda2017,blue,Musevic}.

In skyrmion condensed phases, skyrmions are arranged regularly to form two-dimensional lattice structures~\cite{Muhlbauer,Tokura2010}. Furthermore, phase transitions between liquid, hexatic, and crystalline phases have been observed recently~\cite{Huang}, though the existence of the hexatic phase is controversial~\cite{Zazvorka,Nishikawa,meisenheimer2023ordering}. The hexatic phase is peculiar to two-dimensional systems and is characterized by long-range bond-orientational ordering~\cite{KT1973,NelsonHalperin,Young,Krauth2015}. In isotropic molecular and colloidal systems, two-dimensional molecular dynamics have been extensively studied in liquid and hexatic phases. In these conserved systems where molecules and colloids are rigid particles without deformation and coalescence, long-wavelength cooperative translation is induced by strong thermal fluctuation, called Mermin-Wagner fluctuation~\cite{Shiba,Keim}. Although Mermin-Wagner fluctuation enhances molecular diffusion, it is not relevant to structural relaxation. The structural relaxation is primarily governed by local bond-breaking between adjacent molecules arising from displacement relative to Mermin-Wagner fluctuation. Revealing how this knowledge can be applied to non-conserved skyrmion and half-skyrmion systems is required for efficient manipulation of skyrmions and half-skyrmions. 

In our previous study, we studied liquid-hexatic-crystal transitions of half-skyrmions in two-dimensional chiral molecular systems~\cite{TakaeKawasaki}, {where each particle possesses a spheroidal shape. The spheroidal molecular shape indicates that} the steric interaction is anisotropic, resulting in the emergence of heterogeneous elastic fields when half-skyrmions are formed, which we call emergent elastic fields.
In this study, we elucidate the relationship between the structural relaxation and diffusion of half-skyrmions coupled with emergent elastic fields in this system. By applying the classical molecular dynamics model proposed in our previous study, we find that fusion and fission between adjacent half-skyrmions and deformation of each half-skyrmion are frequently excited by thermal fluctuation, leading to relaxation without diffusion. Cages formed by surrounding half-skyrmions suppress displacement of half-skyrmions, whereas translation-rotation coupling of particle motion and Mermin-Wagner fluctuation enhances their displacement, resulting in subdiffusion. Analysis of heterogeneous dynamics reveals strong correlations between fusion-fission, bond-breaking, and displacements of half-skyrmions via elastic interaction. We then propose half-skyrmion manipulation by sound wave perturbations, where transverse and longitudinal sound waves are coupled differently with half-skyrmion dynamics. The transverse sound wave exhibits strong attenuation due to the coupling between the sound wave propagation and half-skyrmion deformation. We also discuss the {relationship} between half-skyrmion diffusion in molecular systems and skyrmion diffusion in magnetic systems.
Our results provide a theoretical understanding of skyrmion dynamics in condensed phases.


\begin{figure}[t!]
\centering
\includegraphics[width=8cm]{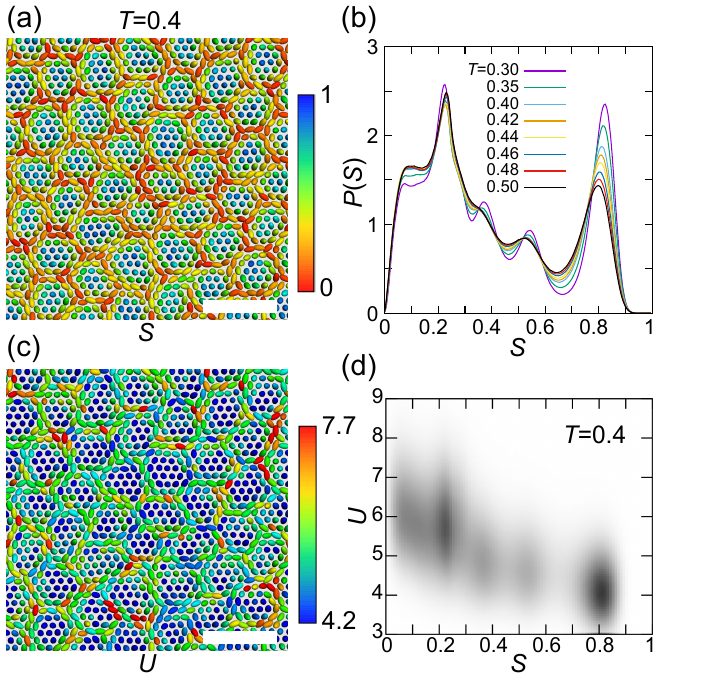}
\caption{
{Identification of half-skyrmions.}
(a) A snapshot of a half-skyrmion crystal at {temperature} $T=0.4$. Color represents the order parameter $S$ defined in Sec.IIB. $S$ becomes large inside half-skyrmion cores, whereas it becomes small for the particles at the edges of half-skyrmions. The contact points between three adjacent half-skyrmions represent $-1/2$ disclinations, where $S$ becomes the smallest.
(b) The probability distribution of the order parameter $S$ has distinct peaks. At the lowest temperature ($T=0.3$), there are four peaks in the distribution: $S\simeq 0.82$ (half-skyrmion cores), $S\simeq 0.55$ and $0.37$ (half-skyrmion edges), and $S\simeq 0.22$ ($-1/2$ disclinations). The first two peaks are separated at $S\simeq 0.65$. Thus, a particle can be regarded to belong to a half-skyrmion core when $S>0.65$.
(c) Potential energy $U$ of each particle represented by color in the same snapshot. $U$ becomes larger for the particles at the edges of half-skyrmions.
(d) Joint probability of the order parameter $S$ and the potential energy $U$ at $T=0.4$. The potential energy for $0.3<S<0.6$ (half-skyrmion edges) has a peak at $U\simeq 4.8$, whereas that for $S<0.3$ (particles at disclinations) has a peak at $U\simeq 5.7$. The energy difference is $\Delta U\simeq 2.25k_{\rm B}T$ at $T=0.4$.
The scale bar in (a,c) denotes $10\sigma$.
}
\label{fig:order}
\end{figure}

\begin{figure*}[t!]
\centering
\includegraphics[width=17cm]{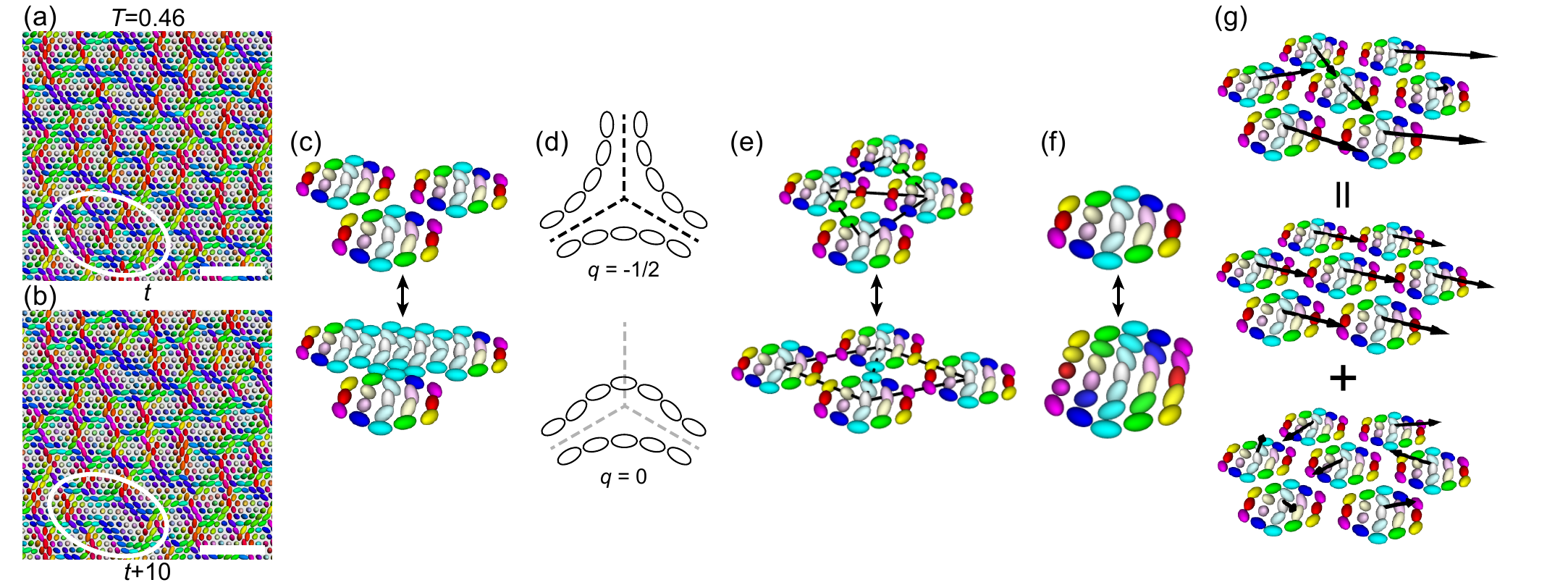}
\caption{
{Elementary processes governing structural relaxation of half-skyrmions.}
(a,b) Time evolution of a half-skyrmion liquid state at $T=0.46$. Color represents molecular orientation defined in Fig.~\ref{fig:setup}(c). The two half-skyrmions surrounded by the white open ellipse undergo fusion to form one large half-skyrmion. The scale bar denotes $10\sigma$.
(c-g) Schematic illustration of the elementary processes.
(c) Fusion and Fission of half-skyrmions, accompanying annihilation and creation of -1/2 disclinations. This process occurs in a diffusionless manner and becomes the dominant contribution to the structural relaxation as shown in Fig.~\ref{fig:relaxation}.
{(d) Schematic illustration of annihilation process of disclinations with vorticity $q=-1/2$.}
(e) Bond breaking and reconnection between half-skyrmions. This process is induced by cage-relative displacement of half-skyrmions without fusion-fission, as shown in (g).
(f) Deformation of individual half-skyrmion, where the size and the ellipticity of half-skyrmions vary.
(g) Displacement of half-skyrmions (top) can be decomposed into Mermin-Wagner fluctuation (middle) and cage-relative displacement (bottom). Arrows represent the displacement of half-skyrmions.
}
\label{fig:elementary}
\end{figure*}

\begin{figure*}[t!]
\centering
\includegraphics[width=15cm]{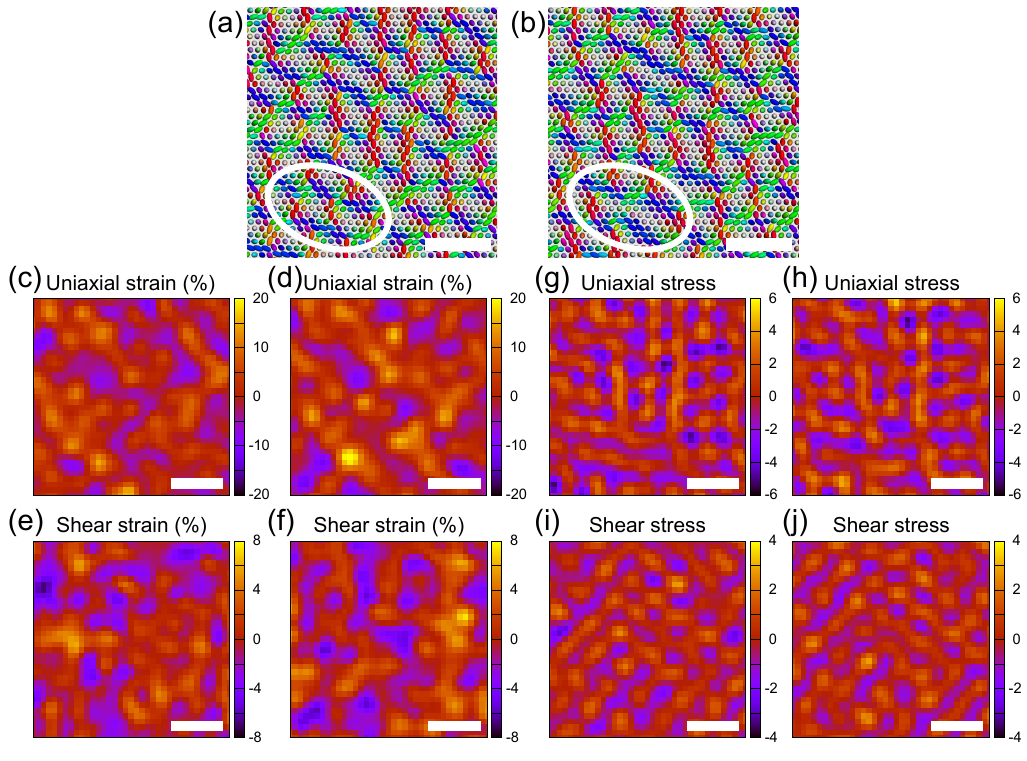}
\caption{
{Inherent structure before and after a fusion event.}
(a,b) Inherent (zero temperature) structure obtained from Fig.~\ref{fig:elementary}(a) and (b), respectively. The fused half-skyrmion still exists in (b), implying that the half-skyrmion fused structure is locally stable.
(c-j) Anisotropic strain and stress fields before (c,e,g,i) and after (d,f,h,j) the fusion event.
Large uniaxial strain emerges at the fused region (d), but the uniaxial stress does not grow in this region (h). This implies that the stress generated by the strain is relaxed by the translation and deformation of surrounding half-skyrmions, resulting in the reduction of the elastic energy associated with fusion events. Thus, the energy barrier of fusion-fission events is small and easily excited thermally. 
The scale bar denotes $10\sigma$.
}
\label{fig:fusionelastic}
\end{figure*}

\section{Methods}
\subsection{Molecular dynamics model}
We adopt a simple molecular model exhibiting phase transitions into helical and half-skyrmion phases formed by spheroidal molecules incorporating twist interaction between the adjacent molecules~\cite{TakaeKawasaki}. The potential energy of our system is given by
\begin{align}
U=&\sum_{i<j}4\epsilon(1+A_{ij}+B_{ij})\bigg(\frac{\sigma}{r_{ij}}\bigg)^{12} + U_{\rm wall}, \\
A_{ij}=&\eta[(\bi n_i\cdot \hat{\bi r}_{ij})^2+(\bi n_j\cdot \hat{\bi r}_{ij})^2],\\
B_{ij}=&\frac{K_2}{2}[(\bi n_i\cdot\bi n_j)(\bi n_i\times\bi n_j)\cdot\hat{\bi r}_{ij}-q_0]^2,
\end{align}
where $\epsilon$ and $\sigma$ denote the characteristic energy and length in our model, $r_{ij}=|\bi r_{ij}|$ and $\hat{\bi r}_{ij}=\bi r_{ij}/r_{ij}$ are the absolute value and unit vector of the intermolecular displacement, respectively, and $\bi n_i$ denotes the molecular orientation of the uniaxial molecules. $A_{ij}$ represents the symmetric steric repulsion to mimic spheroidal molecules in a condensed phase where $\eta$ represents the magnitude of the anisotropy~\cite{Takae2014origlass}.
$B_{ij}$ represents twisting interactions arising from molecular chirality that adjacent molecules favor to align with a twist as shown in Fig.~\ref{fig:setup}(a), where the favored twist angle and twist rigidity are given by $q_0$ and $K_2$, respectively. 

$U_{\rm wall}$ arises from the confinement due to the walls of the system. As shown in Fig.~\ref{fig:setup}(b), we assume a monolayer of spheroid-like molecules without surface anchoring as
\begin{equation}
U_{\rm wall}=\sum_i \epsilon[(\sigma/z_i)^{12}+(\sigma/(L_z-z_i))^{12}],
\end{equation}
where $z_i$ is the $z$-coordinate of $i$-th molecule and $L_z=2\sigma$ is the distance between the two walls.
By assuming densely packed systems, we realize monolayer (two-dimensional) hexagonal crystals in a broad temperature range up to the melting temperature $T_{\rm m}$, which is much greater than the temperature range studied in this paper.
{The half-skyrmions are realized for systems with sufficiently large $K_2$, whose canonical structure is presented in the left panel of Fig.~\ref{fig:setup}(d). The skyrmion structure shown in the right is not realized in this study. The disappearance of the skyrmions is consistent with the theoretical and numerical study for chiral liquid crystals~\cite{Duzgun2018}. It should be noted, however, that the skyrmion structure is not topologically prohibited, and indeed, it is observed experimentally~\cite{Smalyukh}.}

{
The equation of motion of each particle is given by~\cite{Allen}
\begin{align}
m_i\ddot{\bi r}_i&= - \frac{\p U}{\p \bi r_i},
\label{eq:eomt}
\\
I\ddot{\bi n}_i&=-\frac{\p U}{\p \bi n_i} + \lambda_i \bi n_i,
\end{align}
where $m$ is the particle mass, $I$ is the moment of inertia along the long axis, and $\lambda_i$ is the Lagrange multiplier arising from the constraint $|\bi n_i|^2=1$. By taking the inner product with $\bi n_i$ in the second equation, we obtain $\lambda_i=I\bi n_i\cdot\ddot{\bi n}_i+\bi n_i\cdot(\p U / \p \bi n_i)$. Thus, the equation of motion for the rotational motion reads
\begin{equation}
I(\tensor{1}-\bi n_i\bi n_i)\cdot\ddot{\bi n}_i=-(\tensor{1}-\bi n_i\bi n_i)\cdot\frac{\p U}{\p \bi n_i},
\label{eq:eomr}
\end{equation}
which is the projection onto the plane perpendicular to $\bi n_i$. By using $0=d^2 |\bi n|^2/dt^2=2\bi n\cdot\ddot{\bi n}+2|\dot{\bi n}|^2$, this equation can further be written as
\begin{equation}
I\ddot{\bi n}_i=-I|\dot{\bi n}_i|^2\bi n_i-(\tensor{1}-\bi n_i\bi n_i)\cdot\frac{\p U}{\p \bi n_i}.
\label{eq:eomr2}
\end{equation}
We numerically integrate Eqs.~(\ref{eq:eomt}) and (\ref{eq:eomr2}) under the $NVT$ ensemble, adopting the Nos\'e-Hoover thermostat and leap-frog algorithm.
}

In this study, we treat densely packed systems. The approximate volume fraction $\rho=\pi Np\sigma^3/6V$ is fixed to 0.3, where $N=64000$ is the number of the molecules, $p=(1+2\eta)^{1/12}$ is the corresponding aspect ratio, and $V$ is the system volume.
Temperature $T$ is noted in the unit of $\epsilon/k_{\rm B}$ where $k_{\rm B}$ is the Boltzmann constant. We adopt periodic boundary conditions in the $x$ and $y$ directions.
The phase diagram of this model was examined in our previous paper~\cite{TakaeKawasaki}. For $\eta=2$ and $q_0=0.3$, the phase boundary between the helical and the half-skyrmion phases at low temperature is located at $K_2\simeq 45$, as displayed in Fig.~1(e). We fix $K_2=50$, where half-skyrmion liquid, hexatic, and crystalline phases are realized for $T\gtrsim 0.46$, $T=0.44$, and $T\lesssim 0.42$, respectively~\cite{TakaeKawasaki}.
When preparing the system, the temperature is set to 4 to realize the liquid state. Then, the system is cooled to $T=0.01$ with the cooling rate $dT/dt=10^{-5}$ so that perfect crystalline states are realized below the melting temperature. Here, time is measured in the unit of $t_0=\sqrt{m\sigma/\epsilon}$.
The system is then heated to the target temperature and equilibrated for $10^6 t_0$. The data presented in this paper is measured after the equilibration is completed.

\subsection{Identification of half-skyrmions}
To identify half-skyrmions, we first calculate local order parameter $\tensor{Q}_i=(1/7)\sum_j(\bi n_j\bi n_j-\tensor{1}/3)$, where the summation is taken over $i$-th particle and its 6 nearest particles in the hexagonal lattice. The magnitude of the order parameter $S_i=(3/2){\rm tr}Q_i^2$ represents the local nematic order~\cite{deGennesProst}. Particles inside the core of half-skyrmions have $S_i>0.65$ for $S_i$ of each particle, as shown in Fig.~\ref{fig:order}(a,b). Among the particles with $S>0.65$, two particles are defined to belong to the same cluster if they are nearest neighbor particles. Then the position of a half-skyrmion core is defined by $\bi r_{\rm v}=(1/n_{\rm v})\sum_{j\in {\rm cluster}}\bi r_j$, where $n_{\rm v}$ is the number of particles in the same cluster. While this definition is convenient for determining the position of the half-skyrmion cores, the obtained cluster size does not correspond to the actual size of half-skyrmions because the particles at the edge of the half-skyrmions are not taken into account. To determine their size precisely, we regard a particle to belong to a half-skyrmion when the distance between the particle and any particle in the core of the half-skyrmion becomes smallest. In this way, every particle belongs to a half-skyrmion, with the size being the number of the particles in the half-skyrmion. The ellipticity of the $i$-th half-skyrmion is given by the ratio between the two eigenvalues of the matrix $\tensor{e}_i=\sum_{j\in i}(\bi r_j-\bi r_{{\rm v}i})(\bi r_j-\bi r_{{\rm v}i})$ (see Fig.~\ref{fig:texture} for the probability distribution of the size and the ellipticity).

\subsection{Time evolution and fusion-fission of half-skyrmions}
A typical time evolution of half-skyrmions is displayed in Fig.~\ref{fig:elementary}(a) and (b), where the color represents particle orientations as shown in Fig.~\ref{fig:setup}(c). Because the particles form a hexagonal crystal, they do not migrate translationally. Nevertheless, rotational motion is frequently excited because the translational displacement induced by the rotation is small, resulting in deformation and displacement of half-skyrmions.
A notable feature arising from half-skyrmion dynamics is the fusion and fission of half-skyrmions, as marked in Fig.~\ref{fig:elementary}(a) and (b), where two half-skyrmions merge into a large single half-skyrmion (fusion) and vice versa (fission). This process is schematically described in Fig.~\ref{fig:elementary}(c), accompanied by the annihilation and creation of two disclinations with {vorticity $q=-1/2$} at the vertex of three half-skyrmions. {The vorticity $q$ is a topological invariant defined as the molecular rotation angle along a closed loop: $q=\frac{1}{2\pi}\oint ~d\bi \ell\cdot\nabla\phi$, where $\phi=\tan^{-1}(n_y/n_x)$ denotes the molecular angle in $xy$-plane. $q=-1/2$ for the disclination in Fig.~\ref{fig:elementary}(d), whereas $q=1$ in half-skyrmions in Fig.~\ref{fig:setup}(d). When fusion occurs, the number of half-skyrmions ($q=1$) decreases by one, whereas two disclinations ($q=-1/2$) annihilate. Hence, the total {vorticity} is conserved.}
Half-skyrmion fusion and fission vary the number of half-skyrmions $N_{\rm v}$ and their positions $\bi r_{\rm v}$. Hereafter, half-skyrmions without fusion and fission are referred to as non-reacting half-skyrmions.
To identify the displacement of each half-skyrmion, we adopt the criterion proposed in ref.~\cite{Aoyama} to study the diffusion of $\mathbb{Z}_2$-vortices. By comparing the positions of the half-skyrmions between two different times with time interval $\Delta t$, a half-skyrmion is defined to be non-reacting when $|\bi r_{\rm v}(t+\Delta t)-\bi r_{\rm v}(t)|<2\sigma$. Otherwise, the half-skyrmion undergoes fusion or fission. By choosing small $\Delta t$ ($\lesssim 1$ in liquids and $\lesssim 10$ in crystals), non-reacting half-skyrmions with larger displacement do not appear. If both of the adjacent two half-skyrmions violate this criterion, they undergo fusion; else if an isolated half-skyrmion violates the criterion, fission occurs. Thus, the fusion and fission events are detected without ambiguity.
The velocity of a half-skyrmion $\bi v_{\rm v}$ is defined only for the non-reacting half-skyrmions as $\bi v_{{\rm v}i}(t)=(\bi r_{{\rm v}i}(t+\Delta t)-\bi r_{{\rm v}i}(t))/\Delta t$. Using $\bi v_{{\rm v}i}(t)$, their displacement and MSD read
\begin{align}
\Delta\bi r_{{\rm v}i}(\ell\Delta t)&=\sum_{m=0}^{\ell-1}\bi v_{{\rm v}i}(m\Delta t)\Delta t,\\
\av{|\Delta\bi r_{\rm v}|^2}(\ell\Delta t)&=\av{\sum_i |\Delta\bi r_{{\rm v}i}(\ell\Delta t)|^2 / N_{\rm nr}(\ell\Delta t)},
\end{align}
where the summation for $i$ is taken over the half-skyrmions non-reacting until $\ell \Delta t$, and $N_{\rm nr}(\ell\Delta t)$ denotes their number. The bracket represents the statistical and time averages.

\subsection{Elastic fields}
To find out the coupling between the half-skyrmion configuration and the elastic fields, we calculate local strain and stress fields~\cite{TakaeKawasaki}. The strain tensor for each particle is defined by
\begin{equation}
\tensor{\varepsilon}_i=\frac{2}{r_{\rm M}^2N_{{\rm b}i}}\sum_j\bi r_{ij}\bi r_{ij},
\end{equation}
where $r_{\rm M}$ is the first maximum of the radial distribution function required to normalize the strain tensor such that $\tensor{\varepsilon}=\tensor{1}$ for the perfect hexagonal crystal. The volumetric, uniaxial, and shear strains are defined as $\det[ \tensor{\varepsilon}]-1$, $\varepsilon_{xx}-\varepsilon_{yy}$, and $\varepsilon_{xy}$, respectively. The strain field displayed in the figures is obtained by coarse-graining the strain tensor as
\begin{equation}
\tensor{\varepsilon}(\bi r)=\int d\bi r' w(\bi r-\bi r')\sum_i\tensor{\varepsilon}_i\delta(\bi r' -\bi r_i),
\end{equation}
where $w(\bi r)=(1/2\pi \sigma^2)e^{-r^2/2\sigma^2}$ is the weight function.

The local stress field is calculated using the Irving-Kirkwood formula as~\cite{IrvingKirkwood}
\begin{align}
\tensor{\sigma}(\bi r)=&-\sum_i m_i \bi v_i \bi v_i\delta(\bi r -\bi r_i)\nonumber\\
&-\sum_{i<j}\bi f_{ij}\bi r_{ij}\int_0^1 ds\, \delta(s\bi r_i+(1-s)\bi r_j-\bi r),
\end{align}
where the first and second terms denote the kinetic and interaction (configuration) terms, respectively. $\bi f_{ij}$ represents interparticle forces arising from the pair interaction term (the first term in Eq.~(1)). In the figures, we also apply coarse-graining to the stress field using the weight function $w(\bi r)$.
The elastic fields for Fig.~\ref{fig:elementary}(a,b) at the inherent (zero temperature) states are displayed in Fig.~\ref{fig:fusionelastic}.



\section{Results}
\subsection{Elementary processes}
There are three elementary processes governing half-skyrmion dynamics.
The first one is the fusion and fission of half-skyrmions (Fig.~\ref{fig:elementary}(c)), as described in Sec.~IIC.
Half-skyrmion fusion-fission induces elastic deformation of the background crystal~\cite{TakaeKawasaki}, as displayed in Fig.~\ref{fig:fusionelastic}. Though large anisotropic strain is induced by the fusion and fission events, the deformation and translation of the surrounding half-skyrmions reduce the elastic energy. As a result, the elastic energy needed to excite fusion-fission becomes comparable to thermal energy, as shown in Fig.~\ref{fig:order}(c,d). Hence, the fusion-fission is excited thermally.
The other important elementary processes contributing to the structural relaxation are the bond-breaking and reconnection between adjacent half-skyrmions (Fig.~\ref{fig:elementary}(e)) and deformation of half-skyrmions (Fig.~\ref{fig:elementary}(f)). As explained later, the bond-breaking is induced by half-skyrmion displacement relative to the surrounding half-skyrmions, called cage-relative displacement~\cite{Keim}, as shown in Fig.~\ref{fig:elementary}(g). We will see below that the dominant contribution to the structural relaxation of half-skyrmions in our system is the fusion-fission as displayed in Fig.~\ref{fig:elementary}(c), which arises from the non-conserved nature of quasiparticles.

\begin{figure}[t!]
\centering
\includegraphics[width=8cm]{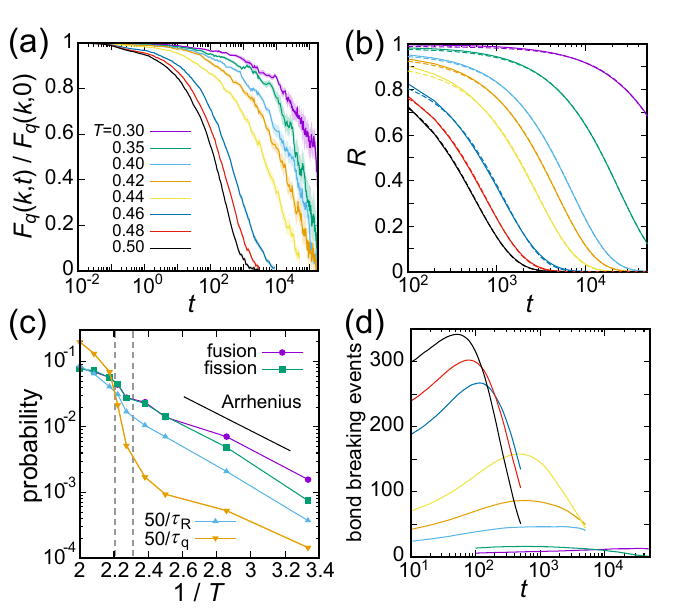}
\caption{
{Structural relaxation of half-skyrmions.}
(a) Time correlation function representing structural relaxation of half-skyrmion liquid, hexatic, and crystalline phases. The long-time behavior is described by stretched exponential functions, where the relaxation time $\tau_{\rm q}$ is displayed in (c). The shaded regions represent standard error.
(b) Fraction of non-reacting half-skyrmions $R$ with respect to time, obeying exponential relaxation (broken curves) with the relaxation time $\tau_{\rm R}$ displayed in (c).
(c) Fusion probability, fission probability, the non-reacting time $\tau_{\rm R}$, and the structural relaxation time $\tau_{\rm q}$ are plotted. The first three quantities exhibit the Arrhenius behavior. The broken Grey lines represent liquid-hexatic (left) and hexatic-crystal (right) phase boundaries.
(d) The number of bond-breaking events with respect to time interval. The bond-breaking is well defined only for the non-reacting half-skyrmions without fusion and fission. Hence, the number of the bond-breaking events decreases as $R$ decreases.
}
\label{fig:relaxation}
\end{figure}

\begin{figure}[t!]
\centering
\includegraphics[width=8cm]{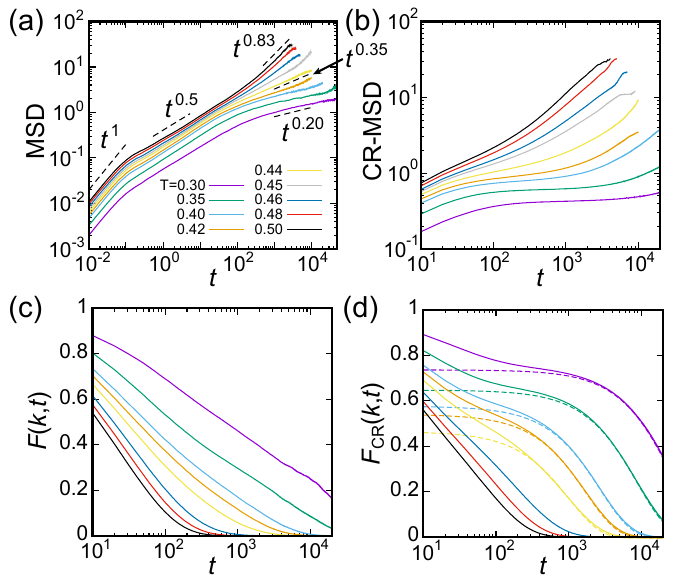}
\caption{
{Subdiffusive motion of half-skyrmions}
(a) The {mean-square-displacement (MSD)} of non-reacting half-skyrmions exhibits subdiffusion, where the late-stage subdiffusive exponent decreases as temperature decreases.
(b) The MSD calculated by the cage-relative displacement {(CR-MSD)} defined in Sec.~IIIC. In contrast to the MSD in (a), the CR-MSD exhibits a plateau region in half-skyrmion crystals. This implies that the subdiffusion in the late-stage of half-skyrmion MSD in the crystalline phase results from Mermin-Wagner fluctuation, characteristic in two-dimensional systems.
(c) Time correlation function of half-skyrmion displacement $F(k,t)$ {defined in Eq.~(\ref{eq:F})}. The wavenumber is chosen such that the structure factor with respect to the half-skyrmion positions exhibits maximum, representing the wavenumber of the adjacent half-skyrmions. Mermin-Wagner fluctuation rapidly relaxes $F(k,t)$ even in the half-skyrmion crystalline states.
(d) Time correlation function $F_{\rm CR}(k,t)$ calculated from the cage-relative displacement, {defined in Eq.~(\ref{eq:FCR})}. The broken curves denote the exponential fitting of the late-stage relaxation in the hexatic and crystalline phases, where the relaxation time is close to the bond-breaking time $\tau_{\rm B}$ in Sec.IIIB.
}
\label{fig:MSD}
\end{figure}

\subsection{Structural relaxation of half-skyrmions}
Because the deformation and translation of half-skyrmions are driven by the rotational motion of constituent molecules with $\bi n \leftrightarrow -\bi n$ symmetry, the structural relaxation of half-skyrmion liquid ($T\ge 0.46$), hexatic ($T=0.44$), and crystalline ($T\le 0.42$) phases are governed by cooperative rotational dynamics of the director field $\tensor{q}(\bi r,t)=\sum_i(\bi n_i(t)\bi n_i(t)-\tensor{1}/3)\delta(\bi r-\bi r_i(t))$. 
The time correlation function is defined by
\begin{equation}
F_q(k,t)=\frac{1}{N}\av{\tensor{q}(\bi k,t):\tensor{q}(-\bi k,0)},
\label{eq:Fq}
\end{equation}
where $\tensor{q}(\bi k,t)$ is the spatial Fourier transform of $\tensor{q}(\bi r,t)$ and the bracket denotes the angular, time, and statistical average. In Fig.~\ref{fig:relaxation}(a), we display the normalized time correlation function $F_q(k,t)/F_q(k,0)$, where $k$ is chosen to be the first peak of the static structure factor $\av{\tensor{q}(\bi k,t):\tensor{q}(-\bi k,t)}/N$, corresponding to the nearest neighbor wavenumber of half-skyrmions.

\begin{figure}[t!]
\centering
\includegraphics[width=8.5cm]{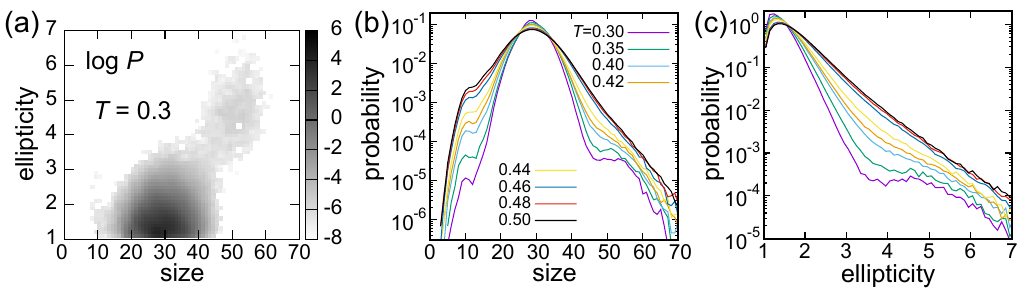}
\caption{
(a) Probability distribution of the size and the ellipticity of the half-skyrmions at temperature $T=0.3$. The secondary peak in the top right region represents fused half-skyrmions.
(b,c) The temperature dependence of the probability distribution. As temperature increases, the width of both the size (b) and the ellipticity (c) distribution becomes broad, implying that the deformation of half-skyrmions occurs more easily. At sufficiently high temperatures, the distributions exhibit exponential behavior. Thermal excitation of the orientation of each molecule deforms half-skyrmions continuously.
}
\label{fig:texture}
\end{figure}

In the late stage, the time correlation functions do not obey a single exponential function but are fitted by stretched exponential functions $e^{-(t/\tau_{\rm q})^\beta}$, where $\tau_{\rm q}$ and $\beta$ represent the relaxation time and the stretch exponent, respectively. The temperature dependence of $\tau_{\rm q}$ is displayed in Fig.~\ref{fig:relaxation}(c). This behavior is reminiscent of the structural relaxation of glass forming liquids~\cite{Glasses,Hansen}, but occurs by a qualitatively different mechanism: The relaxation also occurs in the half-skyrmion crystalline phase. The relaxation arises from three relaxation processes described in Fig.~\ref{fig:elementary}(c-f), occurring heterogeneously as discussed later in Fig.~\ref{fig:heterogeneity}.
The primary (slowest) relaxation is mainly governed by the fusion-fission of half-skyrmions in Fig.~\ref{fig:elementary}(c). To see this, we calculate the fraction $R$ of non-reacting half-skyrmions in Fig.~\ref{fig:relaxation}(b). $R$ decreases to zero exponentially at every temperature with the relaxation time $\tau_{\rm R}$, exhibiting Arrhenius behavior ($\ln{\tau_{\rm R}}\propto 1/T$) as displayed in Fig.~\ref{fig:relaxation}(c) with a small change in the hexatic phase. As decreasing temperature, $\tau_{\rm R}$ becomes close to the structural relaxation time $\tau_{\rm q}$. This implies that thermally excited fusion and fission modes, which are characteristic of non-conserved systems, are responsible for the late-stage structural relaxation of half-skyrmions. It should be noted, however, that half-skyrmions undergoing fusion and fission can thermally recover their original configuration (see also Fig.~\ref{fig:MSD2}), which is not incorporated in the calculation of $\tau_{\rm R}$, whereas contributes to the relaxation of $F_q(k,t)$. Hence $\tau_{\rm q}>\tau_{\rm R}$ in the crystalline phase.

The secondary relaxation is induced by the bond-breaking arising from the rearrangement of half-skyrmions induced by the cage-relative displacement as described in Fig.~\ref{fig:elementary}(e) and (g). This is the primary relaxation mechanism in supercooled liquids and amorphous solids~\cite{Glasses,YamamotoOnukiPRE,Langer}, where the number of particles is conserved {and fusion-fission does not occur}. Fig.~\ref{fig:relaxation}(d) shows the number of bond-breaking events between the non-reacting half-skyrmions.
The bond-breaking is defined as follows.
To calculate the bond connection between half-skyrmions, the Delaunay triangulation method was adopted in our previous paper~\cite{TakaeKawasaki}. However, the distance between the core position of half-skyrmions easily fluctuates since half-skyrmions are deformable. Therefore, it is not appropriate to identify dynamical bond reconnection between half-skyrmions by the Delaunay method, which only uses $\bi r_{\rm v}$. Alternatively, in this paper, two half-skyrmions are regarded to be bonded when two adjacent particles belong to different half-skyrmions. Then, bond-breaking occurs when the bonded two half-skyrmions at time $t_0$ are not bonded at time $t_0+t$. Since half-skyrmions often undergo fusion and fission, bond-breaking is calculated only between the non-reacting half-skyrmions.
As time proceeds, the number of bond-breaking events increases. However, since the bond-breaking is well-defined only between the non-reacting half-skyrmions, the number of the bond-breaking decreases as $R$ decreases. Thus, it exhibits a maximum at a finite time $\tau_{\rm B}$. {$\tau_{\rm B} \lesssim 100$ in the liquid phase, $\tau_{\rm B} \lesssim 1000$ in the hexatic phase, and $\tau_{\rm B} \lesssim 2000$ in the crystalline phases at $T=0.42$, respectively, as shown in Fig.~\ref{fig:relaxation}(d).} $\tau_{\rm B}$ is close to the relaxation time of the cage-relative displacement $F_{\rm CR}(k,t)$ in Fig.~\ref{fig:MSD}(d).
Although $\tau_{\rm B}$ grows as lowering the temperature, $\tau_{\rm B}<\tau_{\rm R}$ always holds, implying that the bond-breaking is relevant in the intermediate-stage, {not the late stage}.

\begin{figure}[t!]
\centering
\includegraphics[width=8.5cm]{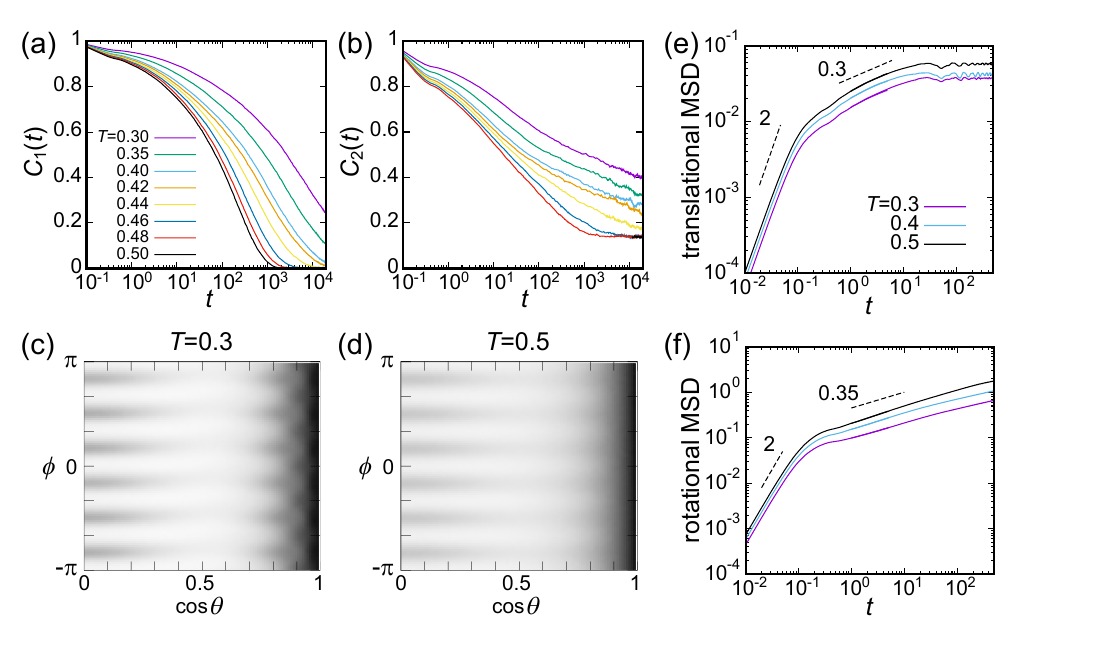}
\caption{
The incoherent (the self-part) time correlation function.
(a,b) two rotational autocorrelation functions $C_1(t)=\av{\sum_i\bi n_i(t)\cdot\bi n_i(0)}/N$ (a) and $C_2(t)=\av{\sum_i 3(\bi n_i(t)\cdot\bi n_i(0))^2-1}/2N$ (b) are displayed. The former decays by configurational relaxation and molecular flipping without orientational relaxation, resulting in a small relaxation time. On the other hand, $C_2(t)$ decays by configurational relaxation and cooperative molecular rotation, leading to rigid translational motion of half-skyrmions. The latter is the origin of why $C_2(t)$ decays faster than $F_q(k,t)$ {in Eq.~(\ref{eq:Fq})}. However, $C_2(t)$ does not decay to zero because the probability distribution of the molecular angle is not homogeneous, as shown in (c,d). 
(c,d) Probability distribution of the particle angle in half-skyrmion crystal at $T=0.3$ (c) and half-skyrmion liquid at $T=0.5$ (d). Here, $\cos{\theta}=n_z$ stands for the tilt angle, and $\phi$ represents the molecular angle in the $xy$ plane.
(e) {Mean-square-displacement} (MSD) regarding the particle translation exhibits subdiffusion ($\sim t^{0.3}$) in the intermediate stage ($0.1\lesssim t \lesssim 30$) between the ballistic ($\sim t^2$) and vibrating ($\sim t^0$) stages.
(f) Rotational MSD continues growing after the ballistic ($\sim t^2$) stage, giving rise to subdiffusion in the intermediate stage in (e) due to translation-rotation coupling.
}
\label{fig:incoherent}
\end{figure}

\begin{figure}[t!]
\centering
\includegraphics[width=8cm]{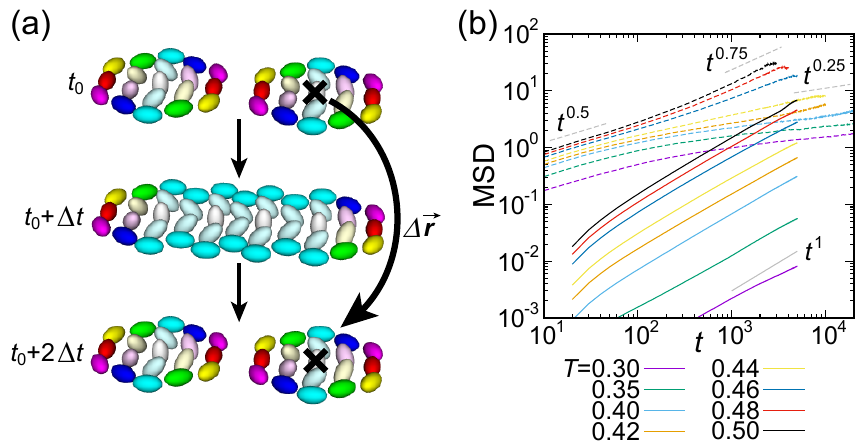}
\caption{
{Mean square displacement {(MSD)} of re-dividing half-skyrmions.}
(a) Definition of re-dividing half-skyrmions. A half-skyrmion is regarded to re-divide when two half-skyrmions undergo fusion between $t_0$ and $t_0+\Delta t$, and subsequently undergo fission into two original half-skyrmions between $t_0+\Delta t$ and $t_0+2\Delta t$. By comparing the position of the half-skyrmion at $t$ and $t+2\Delta t$, one can calculate the displacement of re-dividing half-skyrmions as $\Delta \bi r_{\rm v}(2\Delta t)=\bi r_{\rm v}(t+2\Delta t)-\bi r_{\rm v}(t)$.
(b) MSD of re-dividing half-skyrmions using $\Delta \bi r_{\rm v}$ defined in (a) with $\Delta t=10$ are displayed in solid curves.
The broken curves represent the MSD of non-reacting half-skyrmions displayed in Fig.~\ref{fig:relaxation}(f). Note that the precise value of the MSD in this definition depends on the choice of $\Delta t$, but the qualitative feature is not affected by changing $\Delta t$.
}
\label{fig:MSD2}
\end{figure}

Thirdly, the deformation of half-skyrmions, shown in Fig.~\ref{fig:elementary}(f), is described in Fig.~\ref{fig:texture}. The half-skyrmion morphology fluctuates around the primary peak of the size and ellipticity ($\sim 30$ and 1), while the secondary peak ($\sim 50$ and 4.5) indicates fused half-skyrmions. As discussed in the next subsection, the deformation process contributes to the early-stage, resulting in the gentle relaxation of the time correlation function for $t\lesssim 30$.

To conclude this subsection, structural relaxation is governed by early-stage half-skyrmion deformation (relaxation of the internal degree of freedom), intermediate-stage bond-breaking (conserved dynamics), and late-stage fusion-fission (non-conserved dynamics).
Here, it should be noted that the incoherent (the self-part) rotational time correlation in Fig.~\ref{fig:incoherent}(a,b), which is often calculated in supercooled liquids to examine structural relaxation, decouples with structural relaxation because cooperative particle rotation induces rigid translation of half-skyrmions as a whole.

\begin{figure*}[t!]
\centering
\includegraphics[width=14cm]{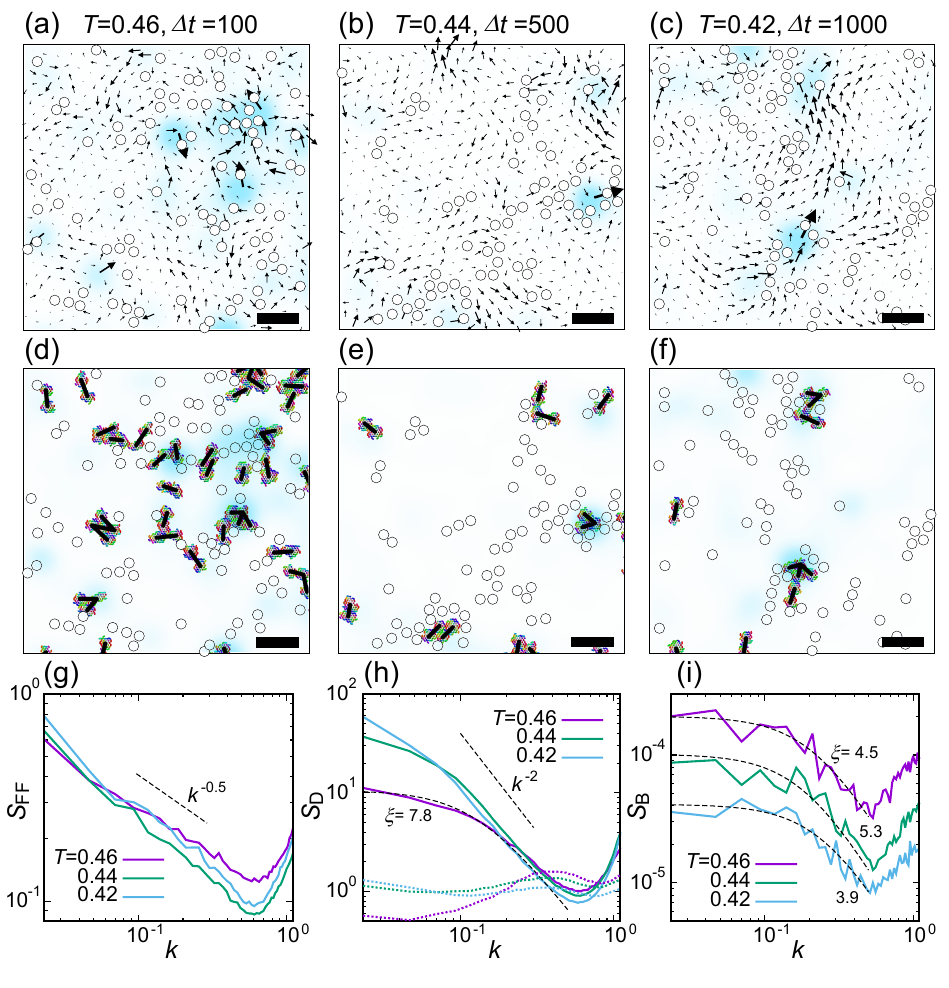}
\caption{
{Dynamic heterogeneity of half-skyrmion motions.}
(a-c) Heterogeneous displacement of half-skyrmions during the time interval $\Delta t$ in liquid (a), hexatic (b), and crystalline (c) phases. Open circles represent the half-skyrmions undergoing fusion and fission. The arrows represent the displacement vector of non-reacting half-skyrmions where the length of each arrow is 2 times the actual displacement. Background color denotes the magnitude of the cage-relative displacement.
(d-f) Spatial distribution of bond-breaking (in black lines) in liquid (d), hexatic (e), and crystalline (f) phases. The molecules constituting the bond-broken half-skyrmions are also displayed. Open circles and the background color are drawn in the same manner as (a-c).
(g-i) Structure factors of half-skyrmions undergoing fusion-fission (g), half-skyrmion displacements (h), and bond-breaking (i). Broken curves in (h,i) denote fitting by the Ornstein-Zernike function to yield the correlation length $\xi$.
In (h), the structure factor of cage-relative displacement is shown by dotted curves.
The scale bar in (a-f) denotes $20\sigma$.
}
\label{fig:heterogeneity}
\end{figure*}

\begin{figure*}[t!]
\centering
\includegraphics[width=15cm]{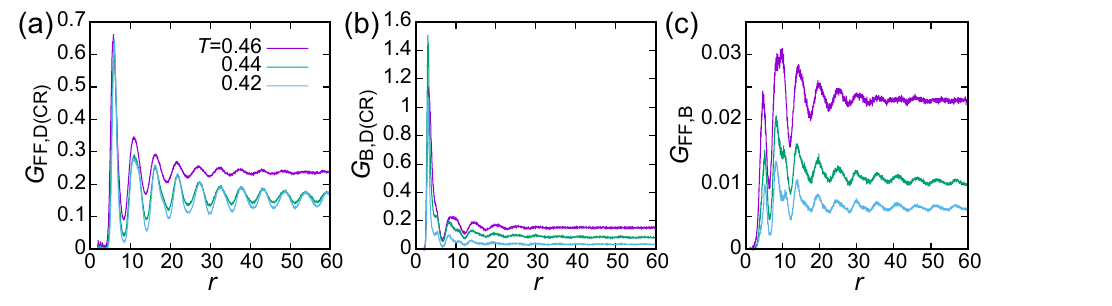}
\caption{
Cross spatial correlation functions of the fusion-fission events $\rho_{\rm FF}$, cage-relative displacement density $\rho_{\rm D}^{\rm CR}$, and bond-breaking events $\rho_{\rm B}$, {which are defined in Sec.~IIID}. (a) Cross-correlation between the fusion-fission and cage-relative displacement $G_{\rm FF,D(CR)}(r,t)$. (b) Cross-correlation between the bond breaking and cage-relative displacement $G_{\rm B,D(CR)}(r,t)$. (c) Cross-correlation between the fusion-fission and bond breaking $G_{\rm FF,B}(r,t)$.
They exhibit short-range spatial correlation, implying that these events positively affect each other. $G_{\rm FF,D(CR)}(r,t)$ and $G_{\rm FF,B}(r,t)$ exhibit spatial correlation up to $r\sim 30$ (several times half-skyrmion diameter).
The first peak of $G_{\rm B,D(CR)}(r,t)$ corresponds to the half-skyrmions undergoing bond-breaking. Thus, bond-breaking induces large translational motion of half-skyrmions, resulting in normal diffusive motion of half-skyrmions when re-division occurs, as shown in Fig.~\ref{fig:MSD2}.
}
\label{fig:cross}
\end{figure*}

\subsection{Subdiffusive motion of half-skyrmions}
In condensed matter, configurational relaxation strongly couples with molecular diffusion~\cite{Hansen,ChaikinLubensky}. To examine their relationship in half-skyrmions, the MSD $\av{|\Delta \bi r_{\rm v}|^2}$ of half-skyrmions is presented in Fig.~\ref{fig:MSD}(a) (see Sec.IIC for its definition ). The MSD of half-skyrmions exhibits three stages.
i) In the initial stage ($t\lesssim 0.1$), $\av{|\Delta \bi r_{\rm v}|^2} \sim t^1$ where each molecule exhibits the inertial translational motion $\av{|\Delta \bi r|^2} \sim t^2$ (see Fig.~\ref{fig:incoherent}(e,f)). A half-skyrmion consists of approximately 30 molecules, and the orientation of their motion is not correlated in this stage. Hence $\av{|\Delta \bi r_{\rm v}|^2}\sim t^1$ is realized by averaging over the molecules in the same half-skyrmion.
ii) In the intermediate stage ($0.1\lesssim t\lesssim 30$), $\av{|\Delta \bi r_{\rm v}|^2}\sim t^{0.5}$ regardless the temperature. In this stage, particle rotation exhibits subdiffusive growth (Fig.~\ref{fig:incoherent}(f)), which induces particle translation (Fig.~\ref{fig:incoherent}(e)) because steric anisotropy of the particles indicates anisotropic free space for the translational motion. When molecular rotation occurs, the anisotropic free space also changes, resulting in the enhancement of particle translation. These translational and rotational motions are responsible for half-skyrmion translation and deformation. For the latter, the joint probability distribution of the size and ellipticity exhibits a broad distribution, implying that half-skyrmion texture changes frequently, as shown in Fig.~\ref{fig:texture}(a) (see supplementary Videos 1,2,3 for texture fluctuation in half-skyrmion liquid, hexatic, and crystalline phases, respectively).
iii) In the late-stage ($t\gtrsim 100$), $\av{|\Delta \bi r_{\rm v}|^2}\sim t^\alpha$ with $\alpha < 1$ is found, indicating subdiffusion of half-skyrmions. 
The exponent $\alpha$ decreases as decreasing temperature.

To investigate why subdiffusion emerges and normal diffusion is absent, we calculate the MSD of re-dividing half-skyrmions. A half-skyrmion is defined to re-divide when a fused half-skyrmion undergoes fission to divide into the original two half-skyrmions in a short time range, as displayed in Fig.~\ref{fig:MSD2}(a). In Fig.~\ref{fig:MSD2}(b), the MSD of re-dividing half-skyrmions obeys normal diffusion $t^1$ irrespective of temperature, implying that large displacements of half-skyrmions are induced by successive fusion and fission.
In the liquid phase, their contribution becomes large as time proceeds, leading to the crossover from subdiffusion to normal diffusion. However, the lifetime of half-skyrmions is shorter than the crossover time. Hence, the normal diffusion is not apparent. In the hexatic and crystalline phases, their contribution remains small compared to the MSD of non-reacting half-skyrmions. Thus, the enhancement of the MSD by fusion and fission is not apparent.

Alternatively, there is another cooperative motion of half-skyrmions in two dimensions, leading to subdiffusion in the crystalline phase. To see this, we divide the displacement into Mermin-Wagner fluctuation and cage-relative displacement, where the latter describes the particle displacement relative to the streaming displacement~\cite{Keim}, as described in Fig.~\ref{fig:elementary}(g). 
We note the lifetime of a half-skyrmion as $\tau_{{\rm v}i}$, at which fusion and fission of the half-skyrmion occur.
Then the cage displacement is defined by $\Delta \bi r^{\rm C}_{{\rm v}i}(t)=\sum_j \Delta \bi r_{{\rm v}j}(t) / N_{{\rm b}i}$, where the summation is taken over the adjacent half-skyrmions at $t=0$, and $N_{{\rm b}i}$ is their number ($\Delta \bi r_{{\rm v}j}(t)$ reads $\Delta \bi r_{{\rm v}j}(\tau_{{\rm v}j})$ if $t>\tau_{{\rm v}j}$). The cage-relative displacement is defined by $\Delta \bi r^{\rm CR}_{{\rm v}i}(t)= \Delta \bi r_{{\rm v}i}(t)-\Delta \bi r^{\rm C}_{{\rm v}i}(t)$.
{Using $\Delta \bi r_{\rm v}$ and $\Delta \bi r^{\rm CR}_{\rm v}$, the time correlation function of the displacement and cage-relative displacement read 
\begin{align}
F(k,t)&=\av{\sum_{j=1}^{N_{\rm v}}\exp[-i\bi k\cdot\Delta\bi r_{{\rm v}j}(t)]/N_{\rm nr}(t)},
\label{eq:F}
\\
F_{\rm CR}(k,t)&=\av{\sum_{j=1}^{N_{\rm v}}\exp[-i\bi k\cdot\Delta\bi r^{\rm CR}_{{\rm v}j}(t)]/N_{\rm nr}(t)},
\label{eq:FCR}
\end{align}
}where the bracket represents the statistical, time, and angular average.
{The MSD calculated from the cage-relative displacement (CR-MSD) is displayed in Fig.~\ref{fig:MSD}(b), and $F(k,t)$ and $F_{\rm CR}(k,t)$ are displayed in Fig.~\ref{fig:MSD}(c,d), respectively.} It is found that the CR-MSD exhibits a plateau region. Accordingly, $F_{\rm CR}(k,t)$ exhibits two-step relaxation, in stark contrast to $F(k,t)$. The difference between the MSD and cage-relative MSD implies that the subdiffusive motion results from Mermin-Wagner fluctuation. The emergence of the plateau in CR-MSD and two-step relaxation in $F_{\rm CR}(k,t)$ is reminiscent of the dynamics of glass-forming liquids, where particle diffusion is suppressed by cages formed by surrounding particles with characteristic cage lifetime~\cite{Glasses}. In glass-forming liquids, the late-stage relaxation of $F_{\rm CR}(k,t)$ is described by stretch exponential, implying the broad distribution of the cage lifetime. In our system, however, the late-stage relaxation becomes single exponential, implying a homogeneous cage lifetime since no intrinsic disorder exists. 

Thus, the MSD is divided into Mermin-Wagner fluctuation and cage-relative displacements. The former represents cooperative particle motion and decouples with structural relaxation, resulting in subdiffusion. The latter, which induces bond-breaking between adjacent half-skyrmions, contributes to the structural relaxation in the intermediate stage. However, the CR-MSD is much smaller than the MSD arising from Mermin-Wagner fluctuation, resulting in apparent decoupling between the MSD and structural relaxation. Because the former is enhanced in two dimensions~\cite{Shiba,Keim}, this decoupling should be noticeable as the sample thickness decreases in experiments.



\subsection{Heterogeneous dynamics}
The correlation between the fusion-fission, half-skyrmion cage-relative displacement, and bond-breaking can be understood by analyzing spatially heterogeneous dynamics. Their spatially heterogeneous distribution at a given time interval is displayed in Fig.~\ref{fig:heterogeneity}(a-f). The time interval is chosen to be close to $\tau_{\rm B}$ so that heterogeneous dynamics manifest. Correlations between fusion-fission, bond-breaking, and cage-relative displacement are conspicuous in liquid, hexatic, and crystalline phases. More quantitatively, we calculate their structure factors in Fig.~\ref{fig:heterogeneity}(g-i). 
They are defined by $S_{\alpha,k}=\av{\rho_{\alpha,k}\rho_{\alpha,-k}}/\bar{N_{\rm v}}$,
where $k$ stands for the wavenumber, and $\alpha={\rm FF}$, ${\rm D}$, and ${\rm B}$ represent the structure factor of fusion-fission, displacement, and bond-breaking, respectively. The bracket denotes angular, time, and statistical average and $\bar{N_{\rm v}}$ is the average of $N_{\rm v}$.
$\rho_{\alpha,k}$ is the spatial Fourier transform of the density fields given by
$\rho_{\rm FF}(\bi r,t)=\sum_{j\in {\rm FF}(t)}\delta(\bi r-\bi r_{{\rm v}j}(0))$,
$\rho_{\rm D}(\bi r,t)=\sum_j|\Delta \bi r_{{\rm v}j}(t)|^2\delta(\bi r-\bi r_{{\rm v}j}(0))$,
$\rho_{\rm D}^{\rm CR}(\bi r,t)=\sum_j|\Delta \bi r_{{\rm v}j}^{\rm CR}(t)|^2\delta(\bi r-\bi r_{{\rm v}j}(0))$,
and $\rho_{\rm B}(\bi r,t)=\sum_{(i,j)\in {\rm B}(t)}\delta(\bi r-\bi r_{{\rm v}ij{\rm B}}(0))$, where $r_{{\rm v}ij{\rm B}}(t)=(\bi r_{{\rm v}i}(t)+\bi r_{{\rm v}j}(t))/2$ is the middle point of the two adjacent half-skyrmions $i$ and $j$.
In Fig.~\ref{fig:heterogeneity}(g), the structure factor of the half-skyrmions undergoing fusion and fission $S_{\rm FF}$ exhibits anomalous increase $\sim k^{-0.5}$ in low wavenumber regime. This implies that fusion-fission events, and conversely, half-skyrmions robust to fusion-fission excitation, distribute heterogeneously regardless of the temperature. The structure factor of half-skyrmion displacement $S_{\rm D}$ in Fig.~\ref{fig:heterogeneity}(h) grows at low wavenumber, reminiscent of order parameter fluctuations in critical phenomena~\cite{ChaikinLubensky,Onukibook}. Similar to the critical fluctuation, it is fitted by the Ornstein-Zernike function $\sim 1/(1+(\xi k)^2)$, where $\xi$ denotes the correlation length. $\xi$ grows as decreasing temperature, though it is not possible to obtain the precise value of the correlation length in the hexatic and the crystalline phase due to the limitation of the system size. The growing correlation length arises from Mermin-Wagner fluctuation, which represents streaming displacement~\cite{Keim}. By calculating the structure factor with respect to the cage-relative displacement, long wavelength fluctuation is suppressed, implying that the cage-relative displacement is excited locally.

The cross-correlation function of fusion-fission and cage-relative displacement, which is defined by
$G_{\rm FF,D(CR)}(r,t)=\av{\rho_{\rm FF}(\bi r,t)\rho_{\rm D}^{\rm CR}(\bi 0,t)}$, is presented in Fig.~\ref{fig:cross}(a).
It exhibits medium-range correlation (approximately several times half-skyrmion diameter), implying that fusion and fission induce large displacement of half-skyrmions nearby via the emergent elastic fields: The elastic fields vary due to the elastic deformation of the background crystal induced by the fusion-fission, which in turn influence the motion of surrounding half-skyrmions via the elastic interaction. In Fig.~\ref{fig:heterogeneity}(i), the structure factor of the bond-breaking has a small correlation length $\xi_{\rm B}$, which is close to half-skyrmion diameter and does not grow even at low temperatures. However, bond-breaking induces a large displacement of surrounding half-skyrmions via the emergent elastic fields. This is apparent in the cross-spatial correlation function shown in Fig.~\ref{fig:cross}(b), which is defined by
$G_{{\rm B},{\rm D (CR)}}(r,t)=\av{\rho_{\rm B}(\bi r,t)\rho_{\rm D}^{\rm CR}(\bi 0,t)}$. Thus, large cage-relative displacement of half-skyrmions is locally excited by fusion-fission and bond-breaking, while the latter two events $G_{{\rm FF},{\rm B}}(r,t)=\av{\rho_{\rm FF}(\bi r,t)\rho_{\rm B}(\bi 0,t)}$ are also spatially correlated, as displayed in Fig.~\ref{fig:cross}(c).

\begin{figure}[t!]
\centering
\includegraphics[width=8cm]{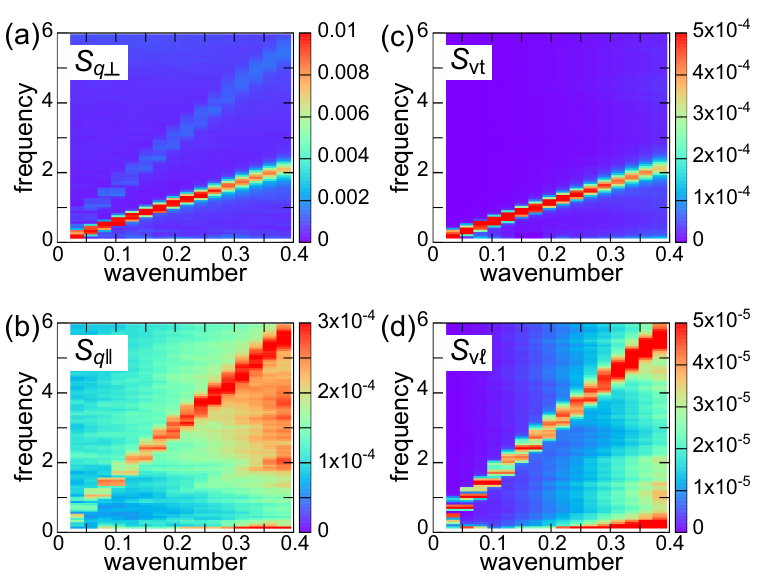}
\caption{
Dynamic structure factors with respect to molecular orientation {$S_{q\perp}$ and $S_{q\parallel}$} (a,b) and half-skyrmion velocity autocorrelation {$S_{{\rm v}t}$ and $S_{{\rm v}\ell}$} (c,d) represent transverse (a,c) and longitudinal (b,d) sound propagation, where the latter exhibits strong sound attenuation. Reorientation of molecules at the edge of half-skyrmions couples with transverse phonon. $T=0.01$ in this figure.
}
\label{fig:Skw}
\end{figure}


\subsection{Manipulation of half-skyrmions by sound waves}
The relationship between half-skyrmion dynamics and emergent elastic fields becomes clear when we examine sound wave excitation, {because sound wave induces elastic deformation of the background crystal.}
In magnetic systems, sound waves couple with magnetic degrees of freedom~\cite{Tokura2014elastic,YangSchmidt}. {Examples include strong coupling between transverse phonon and magnon resulting in anti-crossing~\cite{berk2019strongly,hioki2022coherent}, shear-horizontal surface acoustic waves and magnon~\cite{hwang2024strongly}, and creation and} manipulation of magnetic skyrmions by surface acoustic waves~\cite{nepal2018magnetic,Yokouchi-acoustic,yang2024acoustic}. It is also proposed in another elastic system that applied sound waves produce acoustic skyrmions~\cite{acoustic}. In our systems, the coupling between sound wave excitation and half-skyrmion dynamics arises because the deformation and translation of half-skyrmions are induced by elastic deformation of the background crystal, that is, the translation and rotation of each molecule. To see this, we calculate the dynamic structure factors with respect to molecular rotation and half-skyrmion translation. They are defined by the Fourier-Laplace transformation of the autocorrelation function of $q$ and $\bi v_{\rm v}$. The perpendicular and parallel components of $F_q(k,t)$ are defined by
\begin{align}
F_{q\perp}(k,t)&=\frac{1}{N}\langle q_{xx}(\bi k,t)q_{xx}(-\bi k,0)+q_{yy}(\bi k,t)q_{yy}(-\bi k,0)
\nonumber\\
&+2q_{xy}(\bi k,t)q_{xy}(-\bi k,0)\rangle,\\
F_{q\parallel}(k,t)&=\frac{1}{N}\av{q_{zz}(\bi k,t)q_{zz}(-\bi k,0)}.
\label{eq:Fq2}
\end{align}
Their Fourier-Laplace transformation yields the transverse and longitudinal dynamic structure factors with respect to molecular orientation $S_{q\perp}$ and $S_{q\parallel}$, as shown in Fig.~\ref{fig:Skw}(a,b), respectively.

\begin{figure}[t!]
\centering
\includegraphics[width=8cm]{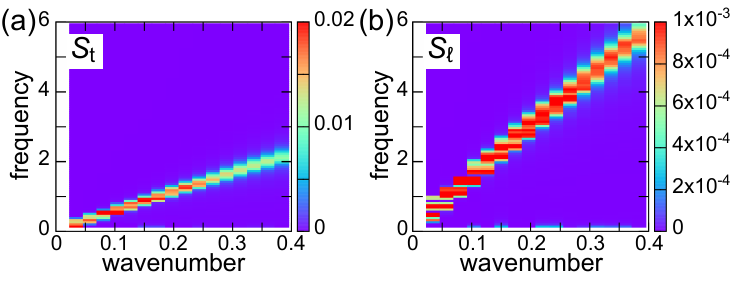}
\caption{
Dynamic structure factors with respect to transverse {$S_t$} (a) and longitudinal {$S_\ell$} (b) molecular motion, representing transverse and longitudinal sound propagation. $T=0.01$ in this figure.
The main peak of $S_t$ ($S_\ell$) is the same as that of $S_{q\perp}$ and $S_{{\rm v}t}$ ($S_{q\parallel}$ and $S_{{\rm v}\ell}$) {in Fig.~\ref{fig:Skw}}. The transverse sound velocity $c_t\simeq 5.8$ is approximately $1/3$ of the longitudinal sound velocity $c_\ell\simeq 15$.
}
\label{fig:phonon}
\end{figure}

\begin{figure*}[t!]
\centering
\includegraphics[width=14cm]{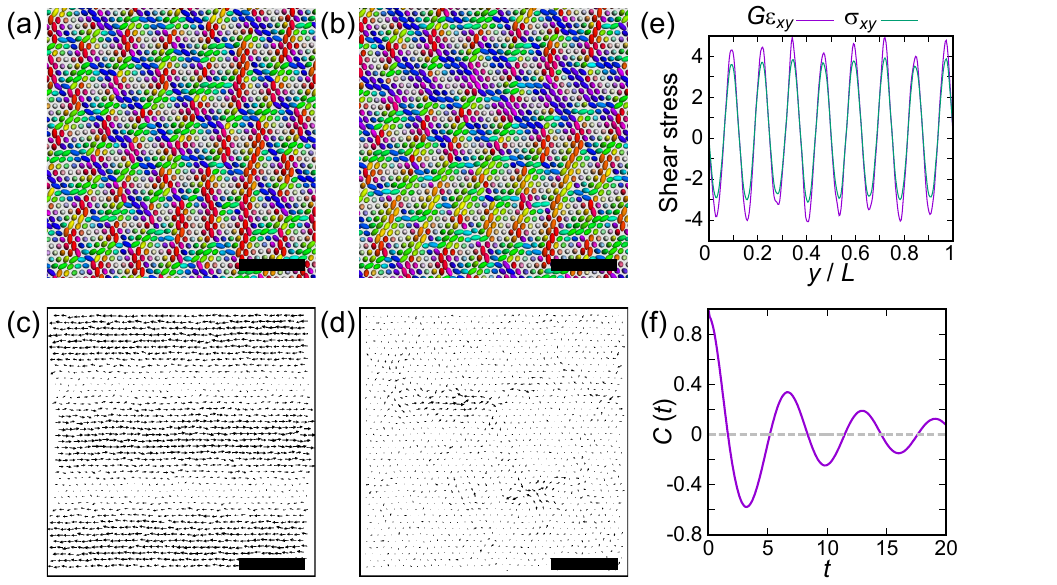}
\caption{
{Response to transverse sound excitation at low temperature.}
(a-d) Molecular orientation (a,b) and the velocity distribution (c,d). Spatially modulated transverse sound wave with wavelength $\lambda=L/8\simeq 33.8$ {($L$ is the lateral length of the system)} is applied at $t=0$ (a,c), which induces molecular reorientation at $t=1.7\simeq t_{\omega t}/4$ as shown in (b) with non-affine velocity distribution in (d), where $t_{\omega t} \sim 6.8$ is the period of {transverse} sound propagation. $T=0.01$ in the figure, and
the scale bar denotes $10\sigma$.
(e) One-dimensional average of the spatial modulation of the shear strain and stress. $G=\rho c_t^2$ is the shear modulus of two-dimensional hexagonal crystals, where $\rho$ is the mass density. {The shear stress} $\sigma_{xy}$ becomes slightly smaller than $G\varepsilon_{xy}$ {($\varepsilon_{xy}$ is the shear strain)}, implying that internal deformation of half-skyrmions reduces shear stress.
(f) Velocity autocorrelation function after transverse sound excitation at $t=0$.
$C(t)=\sum_j\bi v_j(t)\cdot\bi v_j(0)/\sum_j\bi v_j(0)\cdot\bi v_j(0)$ {($\bi v$ is the velocity of each molecule)} after transverse sound excitation at $t=0$ indicates oscillatory molecular motion, exhibiting rapid attenuation.
}
\label{fig:sound}
\end{figure*}

\begin{figure}[t!]
\centering
\includegraphics[width=8cm]{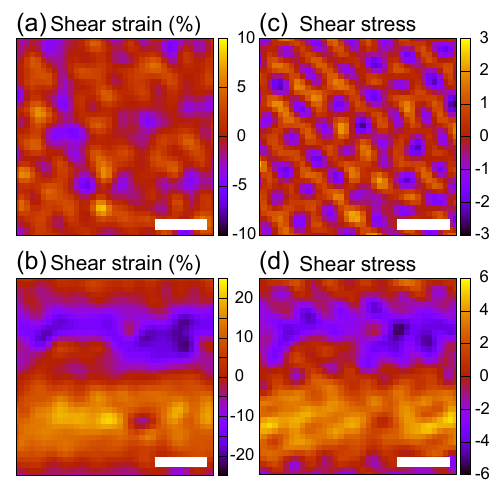}
\caption{
{Elastic field response under transverse sound excitation.}
Shear strain $\varepsilon_{xy}$ (a,b) and stress $\sigma_{xy}$ (c,d) at the initial (a,c) and $t_{\omega t}/4=1.7$ (b,d){, where $t_{\omega t} \sim 6.8$ is the period of transverse sound propagation}. 
{Temperature} $T=0.01$ in this figure. The scale bar in (a-d) denotes $10\sigma$.
}
\label{fig:soundelastic}
\end{figure}

\begin{figure*}[t!]
\centering
\includegraphics[width=15cm]{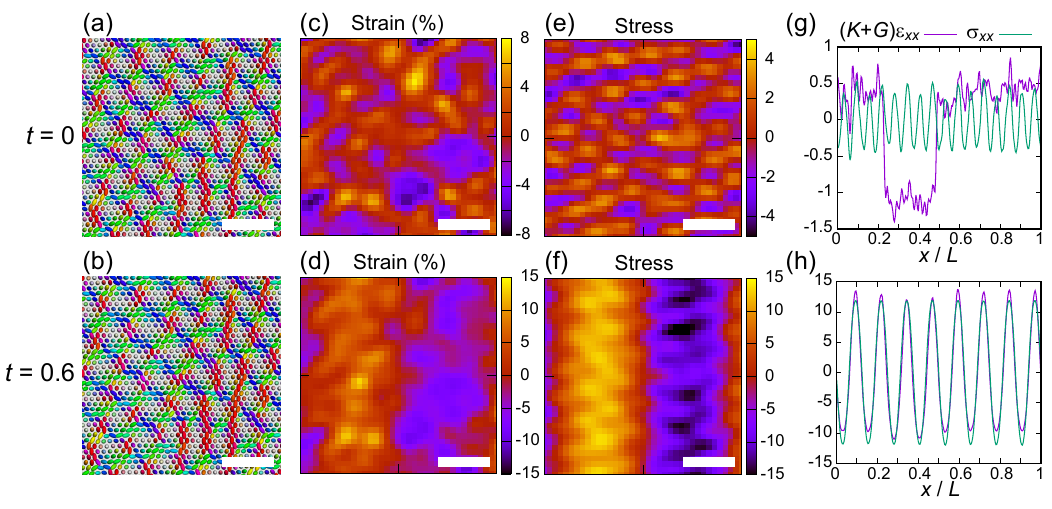}
\caption{
{Longitudinal sound excitation.}
(a,b) Snapshot of particle configuration at the initial (a) and $t_{\omega\ell}/4=0.6$ {($t_{\omega\ell}$ is the period of the longitudinal sound wave)} (b). 
(c-f) Longitudinal strain $\varepsilon_{xx}$ (c,d) and stress $\sigma_{xx}$ (e,f) are spatially modulated by sound excitation.
(g,h) One-dimensional average of the spatial modulation of the strain and stress at $t=0$ (g) and $t=0.6$ (h), where $K$ and $G$ are the bulk and shear modulus, respectively. $K+G=\rho c_\ell^2$ represents the longitudinal modulus of two-dimensional hexagonal crystals, where $\rho$ is the mass density and $c_\ell$ is the longitudinal sound velocity.
{Temperature} $T=0.01$ in this figure. The scale bar in (a-f) denotes $10\sigma$.
}
\label{fig:longitudinal}
\end{figure*}

Similarly, the dynamic structure factors with respect to half-skyrmion translation in Fig.~\ref{fig:Skw}(c,d) are defined by
\begin{align}
S_{{\rm v}t}(k,\omega)&=\frac{k^2}{N_{\rm v}\omega^2}\int_0^\infty dt e^{i\omega t}\av{\bi v_{{\rm v}\perp}(\bi k,t)\cdot \bi v_{{\rm v}\perp}(-\bi k,0)},\\
S_{{\rm v}\ell}(k,\omega)&=\frac{k^2}{N_{\rm v}\omega^2}\int_0^\infty dt e^{i\omega t}\av{\bi v_{{\rm v}\parallel}(\bi k,t)\cdot \bi v_{{\rm v}\parallel}(-\bi k,0)},
\label{eq:Sv}
\end{align}
where $\bi v_{{\rm v}\perp}(\bi k,t)=(\tensor{1}-\bi k\bi k/k^2)\cdot \bi v_{\rm v}(\bi k,t)$ and $\bi v_{{\rm v}\parallel}(\bi k,t)=(\bi k\bi k/k^2) \cdot \bi v_{\rm v}(\bi k,t)$ are the transverse and longitudinal velocity component with $\bi v_{\rm v}(\bi k,t)=\sum_j \bi v_{{\rm v}j}(t)\exp[-i\bi k\cdot \bi r_{{\rm v}j}(t)]$ being the spatial Fourier transformation of the half-skyrmion velocity field $\bi v_{\rm v}(\bi r,t)=\sum_j\bi v_{{\rm v}j}(t)\delta(\bi r-\bi r_{{\rm v}j}(t))$. Here, it should be noted that $N_{\rm v}$ varies by fusion and fission, whose excitation becomes more frequent as increasing temperature, as shown in Fig.~\ref{fig:relaxation}(c). To suppress the variation in $N_{\rm v}$, we calculate $S_{{\rm v}t}$ and $S_{{\rm v}\ell}$ at very low temperature $T=0.01$ in Fig.~\ref{fig:Skw}, where the fusion and fission modes do not appear. In this condition, $N_{\rm v}$ is conserved in time so that $S_{{\rm v}\ell}$ becomes identical with the Fourier-Laplace transformation of density autocorrelation function $S_\rho(k,t)=\frac{1}{N_{\rm v}}\int_0^\infty dt e^{i\omega t}\av{\rho_{\rm v}(\bi k,t)\rho_{\rm v}(-\bi k,0)}$ where $\rho_{\rm v}(\bi k,t)=\sum_j\exp[-i\bi k\cdot \bi r_{{\rm v}j}(t)]$ is the spatial Fourier transform of the density distribution function of half-skyrmions.

$S_q$ and $S_{\rm v}$ are presented in Fig.~\ref{fig:Skw}. The major branches of the transverse (a,c) and longitudinal (b,d) modes coincide with the transverse and longitudinal branches of particle translation, respectively, which are displayed in Fig.~\ref{fig:phonon}, though there is a weak coupling between the transverse and longitudinal branches in $S_q$ in Fig.~\ref{fig:Skw}(a,b). {The similarity between $S_{q\perp}$, $S_{{\rm v}t}$, and $S_t$ ($S_{q\parallel}$, $S_{{\rm v}\ell}$, and $S_\ell$)} implies that molecular rotation and half-skyrmion translation occur in the same manner as molecular translation. The difference between molecular rotation and half-skyrmion translation arises from the deformation of half-skyrmions, incorporated only in the former. The deformation of half-skyrmions, which is induced by particle rotation, weakly couples with sound wave propagation, as shown in Fig.~\ref{fig:sound}(a-d) (Supplementary Video 4). In Fig.~\ref{fig:sound}(b), the particle configuration at $t=t_{\omega t}/4$ is displayed, where $t_{\omega t}$ is the period of transverse sound propagation shown in Fig.~\ref{fig:sound}(f). Deformation of half-skyrmions associated with cooperative molecular rotation is induced by the transverse sound excitation, resulting in the weak coupling between transverse and longitudinal branches in Fig.~\ref{fig:Skw}(a,b). In contrast, the deformation of half-skyrmions has little effect on the translational motion of half-skyrmions, resulting in the disappearance of the coupling in $S_{\rm v}$, as shown in Fig.~\ref{fig:Skw}(c,d).
Thus, applied transverse sound wave induces both the translation and deformation of half-skyrmions, leading to a softer elastic response than the linear elastic theory of uniform hexagonal crystals. The reduction of the shear stress is displayed in Fig.~\ref{fig:sound}(e). {Here, one-dimensional profiles along $y$ direction are shown by taking the average over $x$ direction}. The linear shear modulus $G$ is calculated by $G=\rho c_t^2$, where $\rho$ is the mass density and $c_t$ is the transverse sound velocity. $c_t$ is determined as the slope of the transverse dynamic structure factor in Fig.~\ref{fig:phonon}(a). The linear elastic theory predicts $\sigma_{xy}=G\ve_{xy}$~\cite{Landau7}. In the figure, however, $\sigma_{xy}$ is smaller than $G\ve_{xy}$, implying that the deformation of half-skyrmions reduces the shear stress. This is analogous to the stress reduction induced by cooperative molecular rotation~\cite{Takae2014origlass}.

The deformation of half-skyrmions is also responsible for the emergence of a non-affine velocity response, as shown in Fig.~\ref{fig:sound}(c,d), which implies heterogeneous distribution of soft elastic regions~\cite{Takae2014origlass}. The existence of soft elastic region, which is called elastic heterogeneity~\cite{Onukibook,Barrat2018review,Wagner2011local,Mitsumoto}, results from the elastic fields associated with half-skyrmion deformation, as shown in Fig.~\ref{fig:soundelastic}. The regions with large non-affine velocity distribution in Fig.~\ref{fig:sound}(d) correspond to elastically heterogeneous regions with large shear strain and stress gradient. In amorphous systems, it is known that elastic heterogeneity causes acoustic scattering~\cite{Mizuno2014,MizunoIkeda}. Similarly, elastic heterogeneity in half-skyrmion condensed phases results in transverse sound attenuation, shown in Fig.~\ref{fig:sound}(f), via non-affine strain responses.
Thus, the coupling between sound excitation and half-skyrmion deformation leads to strong acoustic attenuation via the translation-rotation coupling and elastic heterogeneity.

\begin{figure}[t!]
\centering
\includegraphics[width=8.5cm]{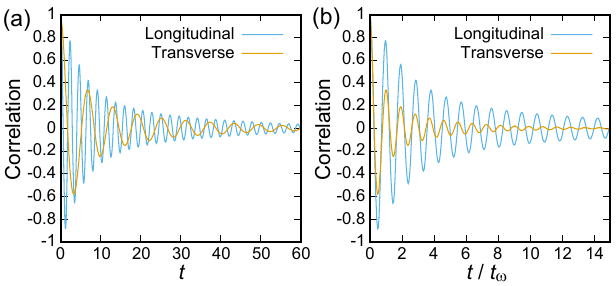}
\caption{
{Sound attenuation.}
Attenuation of the transverse and longitudinal sound waves with respect to time $t$ (a) and time divided by the period of sound propagation $t/t_{\omega}$ (b). Attenuation of the transverse mode is faster than the longitudinal mode because the transverse mode couples with half-skyrmion deformation.
}
\label{fig:attenuation}
\end{figure}

When the longitudinal sound wave is applied, the deformation of half-skyrmions is not apparent as shown in Fig.~\ref{fig:longitudinal}, where the strain and stress heterogeneity at $t=0$ and $t=t_{\omega\ell}/4$ are shown in Fig.~\ref{fig:longitudinal}(c,e,g) and Fig.~\ref{fig:longitudinal}(d,f,h), respectively (see Supplementary Video 5).
According to the linear elastic theory in two dimensions~\cite{Landau7}, the longitudinal stress $\sigma_{xx}$ is given by $(K+G)\ve_{xx}$, where $K+G=\rho c_\ell^2$ is the longitudinal elastic modulus with $c_\ell$ be the longitudinal sound velocity. $c_\ell$ is determined as the slope of the longitudinal dynamic structure factor in Fig.~\ref{fig:phonon}(b). {Their spatial profiles along $x$ direction averaged over $y$ direction} are presented in Fig.~\ref{fig:longitudinal}(g,h). In (g), there is embedded strain heterogeneity, which is responsible for the slight discrepancy between $\sigma_{xx}$ and  $(K+G)\ve_{xx}$ in (h). 
However, their amplitudes coincide, implying that the strain and stress obey the linear elastic relationship of uniform hexagonal crystals with embedded strain and stress heterogeneity.
This agreement implies that longitudinal sound waves do not couple with half-skyrmion deformation, resulting in a slower attenuation of the longitudinal sound waves than the transverse sound waves, as shown in Fig.~\ref{fig:attenuation}.
Thus, we may conclude that transverse and longitudinal sound perturbations can control the motion of half-skyrmions in a different manner, utilizing a non-affine strain response arising from emergent elastic fields induced by half-skyrmion deformation.
In this paper, low-temperature dynamics is examined, where fusion and fission do not occur.
At finite temperatures, however, transverse sound waves should couple with fusion-fission modes, providing another origin of acoustic attenuation.

\section{Discussion}
Here, we discuss the relationship between half-skyrmion dynamics in anisotropic molecular systems and skyrmion dynamics in magnetic systems.
Major differences between these systems arise from differences in symmetry and dynamical equations.
In anisotropic molecular crystals and liquid crystals, molecules are often regarded as {uniaxial molecules} possessing head-to-tail rotation symmetry, which results in the formation of half-skyrmions in many cases.
{The equation of motion for each molecule is given by Eqs.~(\ref{eq:eomt}) and (\ref{eq:eomr}).}
Magnetic skyrmions, on the other hand, skyrmions are typically observed because magnetic spins have polar symmetry, {though half-skyrmions (merons) have also been observed~\cite{Tokura2018N}. The equation of motion is Landau-Lifshitz-Gilbert equation.}

{The qualitative difference between molecular and magnetic systems arising from the difference in the equation of motion becomes apparent by deriving the equation of motion for the collective coordinates.
As well known, the motion of an isolated magnetic skyrmion is described by Thiele equation~\cite{thiele1973steady,ohki2025fundamental}, which reads
\begin{equation}
\bi G\times \bi V +\alpha D \bi V =\bi F,
\label{eq:thiele_mag}
\end{equation}
where $\bi V$ is the drift velocity of the skyrmion, $\bi G$ is the gyrocoupling vector, $\alpha D$ represents the frictional motion arising from the Gilbert damping, and $\bi F$ is the force exerted on the skyrmion. Here, we assume rigid, undeformable skyrmions, indicating massless skyrmion motion.
By performing the same procedure, we can derive the equation of motion for a single, isolated, undeformable half-skyrmion as follows.
For simplicity, we neglect the translational motion of each particle, focusing on the rotational motion. Then, the equation of motion is given by Eq.~(\ref{eq:eomr}).
To make a comparison with the Thiele equation, we add a phenomenological friction term $-\zeta \dot{\bi n}_i$ arising from the translation-rotation coupling.
By taking the inner product with $\nabla_\alpha \bi n$, this equation can be rewritten as
\begin{equation}
I(\nabla_\alpha \bi n)\cdot \ddot{\bi n}=-\zeta (\nabla_\alpha \bi n)\cdot \dot{\bi n} - (\nabla_\alpha \bi n)\cdot\frac{\p U}{\p \bi n}.
\end{equation}
We then take the spatial integration of this equation, substituting $\bi n(\bi r,t)=\bi n(\bi r-\bi R(t))$ and its polar coordinates $(\theta,\phi)=(\pi r/2R_{\rm sk}, \tan^{-1}(y/x))$ for $r<R_{\rm sk}$, where $\bi R$ is the center-of-mass position of the half-skyrmion and $R_{\rm sk}$ is the radius. Some calculation yields
\begin{equation}
ID \dot{\bi V} + \zeta D \bi V = \bi F,
\label{eq:thiele_mol}
\end{equation}
where use has been made of $\int dx dy (\nabla_\alpha \bi n)\cdot(\nabla_\beta \bi n)=D\delta_{\alpha\beta}$ with $D$ be a function of $R_{\rm sk}$, and $\bi F=-\int dxdy (d U / d \bi R)$ is the force exerted on the half-skyrmion.
Thus, the equation of motion for the collective coordinate includes the effective mass and dissipation terms, whereas the gyrocoupling term is absent.
The friction term has the same form as the Gilbert damping term. Unlike magnetic skyrmions, the Magnus force (or odd viscosity~\cite{Vitelli}) is not relevant in molecular half-skyrmion dynamics, though the assumption of rigid and dilute half-skyrmions in the derivation of Eq.(\ref{eq:thiele_mol}) is an over-simplification. However, investigating the quantitative difference in cooperative dynamics of magnetic skyrmions and molecular half-skyrmions is beyond the scope of this study.}

{Thus, both the static structures and dynamics are different between molecular half-skyrmions and magnetic skyrmions.
Nevertheless, concerning the late-stage structural relaxation and diffusion,} these differences will be less relevant in the condensed phase because elementary processes of structural relaxation---fusion-fission, bond-breaking, and Mermin-Wagner fluctuation---are common to skyrmions and half-skyrmions. {They are governed by the non-conserved nature of the quasiparticle number density, intermolecular rearrangement, and thermal fluctuation peculiar to two-dimensional systems, respectively,} regardless of the specific equation of motion for each quasiparticle.
{When fusion-fission of magnetic skyrmions occurs, it can also be shown that the total vorticity and skyrmion number do not change.}
Therefore, qualitatively similar anomalous transport is also expected for skyrmions in magnetic systems.

However, the relevance of the fusion-fission needs to be examined carefully.
In this study, the excitation energy of fusion and fission is comparable to the thermal energy, resulting in diffusionless structural relaxation of half-skyrmions. This is because the numerical parameters in this study are close to the phase boundary between half-skyrmion and helical phases.
In magnetic skyrmions, on the other hand, skyrmion melting can occur far from the phase boundary between the skyrmion and helical phases~\cite{Nishikawa,Zazvorka,Huang}. Far from the helical-skyrmion phase boundary, skyrmions have a robust structure against thermal fluctuation. Then, the fusion-fission becomes less relevant in structural relaxation.
It can occur in the vicinity of the helical-skyrmion phase boundary.
Thus, to examine the role of fusion-fission on the diffusion of magnetic skyrmions in crystalline and liquid phases, one needs to measure diffusion close to the triple point between helical, skyrmion crystal, and skyrmion liquid phases.

{Next, we discuss sound wave excitation in molecular and magnetic systems.}
The half-skyrmions studied in this study are Bloch-type half-skyrmions, where the change in molecular orientation is perpendicular to the intermolecular displacement (twist).
This is why the transverse sound wave can excite molecular rotation and, accordingly, the deformation of half-skyrmions. 
In N\'eel-type half-skyrmion systems, on the other hand, the change in molecular orientation is parallel to the intermolecular displacement (splay). 
In this system, the deformation of half-skyrmions should be excited by the longitudinal sound wave rather than the transverse sound wave, resulting in strong attenuation of the longitudinal sound wave. For magnetic systems, it is known that N\'eel-type skyrmions are observed in polar magnets~\cite{Tokura2021review}. 
Similarly, polar liquid crystals, which are known to exhibit the splay-nematic phase in the bulk~\cite{splaynematic}, will be candidates to realize a N\'eel-type half-skyrmion phase when confined in thin films, though orientation reconfiguration in the third dimension may destabilize N\'eel-type half-skyrmions~\cite{fukuda2022liquid,deGennesProst}.
{
Another candidate for examining molecular (half-)skyrmion dynamics is anisotropic colloidal systems~\cite{Zerrouki,Han,Sacanna2021}. Colloidal crystals exhibit the transverse and longitudinal sound dispersion consistent with harmonic lattice theory~\cite{keim2004harmonic,yunker2014physics}. When colloids possess an anisotropic shape, coupling between the rotational and translational motion arises. Because studies on the formation of (half-)skyrmion crystals have been sparse, it will be challenging to explore the possibility of the creation and manipulation of (half-)skyrmions in molecular and colloidal crystals~\cite{Dogic,Dijkstra2016,Skyrmion-MOF,subert2024achiral}.
}

{
In magnetic systems, coupling between surface acoustic waves and single magnetic skyrmion has been examined~\cite{nepal2018magnetic,Yokouchi-acoustic,yang2024acoustic}. Here, surface acoustic waves include Rayleigh wave (shear-vertical and longitudinal strains are excited) and Love wave (shear-horizontal strain is excited)~\cite{hwang2024strongly,yang2024acoustic}. It is theoretically suggested that longitudinal acoustic wave induces translational motion of magnetic bubbles~\cite{nepal2018magnetic}. Experimentally, however, shear-vertical component in Rayleigh waves may disturb translation of N\'eel-type skyrmion~\cite{Yokouchi-acoustic,yang2024acoustic}.
Rather, shear-horizontal wave induces their translation~\cite{yang2024acoustic}. Thus, it should be important to separate the role of shear-vertical and longitudinal strains.
For skyrmion condensed states, collective skyrmion translation corresponds to magnon excitation with a particular wave number~\cite{mohanta2020signatures}. While experiments on skyrmion collective translation by surface acoustic waves are lacking, it has been known that the transverse and longitudinal acoustic wave couples with magnon in a different manner~\cite{kikkawa2016magnon,hioki2022coherent}. Therefore, there may be a possibility to control collective skyrmion translation by the longitudinal and transverse acoustic waves differently.
}

{Finally, }
it should also be noted that the relevance of the non-conserved dynamics is also recognized in foams~\cite{Foam}, chemical reactions~\cite{Kuramoto}, cell divisions~\cite{loose2008spatial}, and ecosystems~\cite{ecology}, where the number of particles varies in time under the strong influence of surrounding environments. These systems are suitable for examining the coupling between non-conserved (non-diffusive) and conserved (diffusive) dynamics.

\section{Conclusion}
In conclusion, we presented the decoupling between the primary structural relaxation and the MSD of half-skyrmions in condensed phases. In contrast to conserved particle dynamics, fusion and fission govern the primary structural relaxation. Bond-breaking induced by the cage-relative displacement contributes to the secondary structural relaxation. The diffusion of half-skyrmions is suppressed by dynamic cages formed by surrounding half-skyrmions having finite lifetime, whereas enhanced by Mermin-Wagner fluctuation characteristic to two-dimensional systems. The displacement of half-skyrmions is correlated to fusion-fission and bond-breaking via the emergent elastic field, leading to spatially heterogeneous dynamics. In our Bloch-type half-skyrmion systems, transverse sound waves couple with the deformation of half-skyrmions, resulting in stronger acoustic attenuation than longitudinal sound waves.



We treated half-skyrmion dynamics in a two-dimensional monolayer. Experimentally, however, layer thickness plays an essential role in the formation and dynamics of skyrmions and half-skyrmions~\cite{Milde,Tokura2014three,Fukuda2017,nagase2019smectic,Tokura2020-propagation,blue}. 
Examining the role of the Mermin-Wagner fluctuation and cage-relative displacements on structural relaxation in finite-thickness systems should be a further research direction.




\vspace{2mm}
\noindent 
{\bf Acknowledgments} The authors would like to thank Dr. Rei Kurita for valuable discussions. This work was supported by {Inamori Research Grants and} JSPS KAKENHI Grant No. JP20H05619 and JP24K00594.





\begin{thebibliography}{95}%
\makeatletter
\providecommand \@ifxundefined [1]{%
 \@ifx{#1\undefined}
}%
\providecommand \@ifnum [1]{%
 \ifnum #1\expandafter \@firstoftwo
 \else \expandafter \@secondoftwo
 \fi
}%
\providecommand \@ifx [1]{%
 \ifx #1\expandafter \@firstoftwo
 \else \expandafter \@secondoftwo
 \fi
}%
\providecommand \natexlab [1]{#1}%
\providecommand \enquote  [1]{``#1''}%
\providecommand \bibnamefont  [1]{#1}%
\providecommand \bibfnamefont [1]{#1}%
\providecommand \citenamefont [1]{#1}%
\providecommand \href@noop [0]{\@secondoftwo}%
\providecommand \href [0]{\begingroup \@sanitize@url \@href}%
\providecommand \@href[1]{\@@startlink{#1}\@@href}%
\providecommand \@@href[1]{\endgroup#1\@@endlink}%
\providecommand \@sanitize@url [0]{\catcode `\\12\catcode `\$12\catcode
  `\&12\catcode `\#12\catcode `\^12\catcode `\_12\catcode `\%12\relax}%
\providecommand \@@startlink[1]{}%
\providecommand \@@endlink[0]{}%
\providecommand \url  [0]{\begingroup\@sanitize@url \@url }%
\providecommand \@url [1]{\endgroup\@href {#1}{\urlprefix }}%
\providecommand \urlprefix  [0]{URL }%
\providecommand \Eprint [0]{\href }%
\providecommand \doibase [0]{https://doi.org/}%
\providecommand \selectlanguage [0]{\@gobble}%
\providecommand \bibinfo  [0]{\@secondoftwo}%
\providecommand \bibfield  [0]{\@secondoftwo}%
\providecommand \translation [1]{[#1]}%
\providecommand \BibitemOpen [0]{}%
\providecommand \bibitemStop [0]{}%
\providecommand \bibitemNoStop [0]{.\EOS\space}%
\providecommand \EOS [0]{\spacefactor3000\relax}%
\providecommand \BibitemShut  [1]{\csname bibitem#1\endcsname}%
\let\auto@bib@innerbib\@empty
\bibitem [{\citenamefont {H\"{o}fling}\ and\ \citenamefont
  {Franosch}(2013)}]{Franosch}%
  \BibitemOpen
  \bibfield  {author} {\bibinfo {author} {\bibfnamefont {F.}~\bibnamefont
  {H\"{o}fling}}\ and\ \bibinfo {author} {\bibfnamefont {T.}~\bibnamefont
  {Franosch}},\ }\bibfield  {title} {\bibinfo {title} {Anomalous transport in
  the crowded world of biological cells},\ }\href
  {https://doi.org/10.1088/0034-4885/76/4/046602} {\bibfield  {journal}
  {\bibinfo  {journal} {Rep. Prog. Phys.}\ }\textbf {\bibinfo {volume} {76}},\
  \bibinfo {pages} {046602} (\bibinfo {year} {2013})}\BibitemShut {NoStop}%
\bibitem [{\citenamefont {Lee}\ \emph {et~al.}(2021)\citenamefont {Lee},
  \citenamefont {Wingreen},\ and\ \citenamefont {Brangwynne}}]{Brangwynne}%
  \BibitemOpen
  \bibfield  {author} {\bibinfo {author} {\bibfnamefont {D.~S.}\ \bibnamefont
  {Lee}}, \bibinfo {author} {\bibfnamefont {N.~S.}\ \bibnamefont {Wingreen}},\
  and\ \bibinfo {author} {\bibfnamefont {C.~P.}\ \bibnamefont {Brangwynne}},\
  }\bibfield  {title} {\bibinfo {title} {Chromatin mechanics dictates
  subdiffusion and coarsening dynamics of embedded condensates},\ }\href
  {https://doi.org/10.1038/s41567-020-01125-8} {\bibfield  {journal} {\bibinfo
  {journal} {Nature Phys.}\ }\textbf {\bibinfo {volume} {17}},\ \bibinfo
  {pages} {531} (\bibinfo {year} {2021})}\BibitemShut {NoStop}%
\bibitem [{\citenamefont {Amblard}\ \emph {et~al.}(1996)\citenamefont
  {Amblard}, \citenamefont {Maggs}, \citenamefont {Yurke}, \citenamefont
  {Pargellis},\ and\ \citenamefont {Leibler}}]{Amblard}%
  \BibitemOpen
  \bibfield  {author} {\bibinfo {author} {\bibfnamefont {F.}~\bibnamefont
  {Amblard}}, \bibinfo {author} {\bibfnamefont {A.~C.}\ \bibnamefont {Maggs}},
  \bibinfo {author} {\bibfnamefont {B.}~\bibnamefont {Yurke}}, \bibinfo
  {author} {\bibfnamefont {A.~N.}\ \bibnamefont {Pargellis}},\ and\ \bibinfo
  {author} {\bibfnamefont {S.}~\bibnamefont {Leibler}},\ }\bibfield  {title}
  {\bibinfo {title} {Subdiffusion and anomalous local viscoelasticity in actin
  networks},\ }\href {https://doi.org/10.1103/PhysRevLett.77.4470} {\bibfield
  {journal} {\bibinfo  {journal} {Phys. Rev. Lett.}\ }\textbf {\bibinfo
  {volume} {77}},\ \bibinfo {pages} {4470} (\bibinfo {year}
  {1996})}\BibitemShut {NoStop}%
\bibitem [{\citenamefont {Gittes}\ \emph {et~al.}(1997)\citenamefont {Gittes},
  \citenamefont {Schnurr}, \citenamefont {Olmsted}, \citenamefont
  {MacKintosh},\ and\ \citenamefont {Schmidt}}]{Schmidt}%
  \BibitemOpen
  \bibfield  {author} {\bibinfo {author} {\bibfnamefont {F.}~\bibnamefont
  {Gittes}}, \bibinfo {author} {\bibfnamefont {B.}~\bibnamefont {Schnurr}},
  \bibinfo {author} {\bibfnamefont {P.~D.}\ \bibnamefont {Olmsted}}, \bibinfo
  {author} {\bibfnamefont {F.~C.}\ \bibnamefont {MacKintosh}},\ and\ \bibinfo
  {author} {\bibfnamefont {C.~F.}\ \bibnamefont {Schmidt}},\ }\bibfield
  {title} {\bibinfo {title} {Microscopic viscoelasticity: Shear moduli of soft
  materials determined from thermal fluctuations},\ }\href
  {https://doi.org/10.1103/PhysRevLett.79.3286} {\bibfield  {journal} {\bibinfo
   {journal} {Phys. Rev. Lett.}\ }\textbf {\bibinfo {volume} {79}},\ \bibinfo
  {pages} {3286} (\bibinfo {year} {1997})}\BibitemShut {NoStop}%
\bibitem [{\citenamefont {K{\"a}rger}\ \emph {et~al.}(2012)\citenamefont
  {K{\"a}rger}, \citenamefont {Ruthven}, \citenamefont {Theodorou} \emph
  {et~al.}}]{Karger}%
  \BibitemOpen
  \bibfield  {author} {\bibinfo {author} {\bibfnamefont {J.}~\bibnamefont
  {K{\"a}rger}}, \bibinfo {author} {\bibfnamefont {D.~M.}\ \bibnamefont
  {Ruthven}}, \bibinfo {author} {\bibfnamefont {D.~N.}\ \bibnamefont
  {Theodorou}}, \emph {et~al.},\ }\href {https://doi.org/10.1002/9783527651276}
  {\emph {\bibinfo {title} {Diffusion in Nanoporous Materials}}},\
  Vol.~\bibinfo {volume} {48}\ (\bibinfo  {publisher} {Wiley Online Library},\
  \bibinfo {year} {2012})\BibitemShut {NoStop}%
\bibitem [{\citenamefont {Berthier}\ \emph {et~al.}(2011)\citenamefont
  {Berthier}, \citenamefont {Biroli}, \citenamefont {Bouchaud}, \citenamefont
  {Cipelletti},\ and\ \citenamefont {van Saarloos}}]{Glasses}%
  \BibitemOpen
  \bibfield  {author} {\bibinfo {author} {\bibfnamefont {L.}~\bibnamefont
  {Berthier}}, \bibinfo {author} {\bibfnamefont {G.}~\bibnamefont {Biroli}},
  \bibinfo {author} {\bibfnamefont {J.-P.}\ \bibnamefont {Bouchaud}}, \bibinfo
  {author} {\bibfnamefont {L.}~\bibnamefont {Cipelletti}},\ and\ \bibinfo
  {author} {\bibfnamefont {W.}~\bibnamefont {van Saarloos}},\ }\href
  {https://doi.org/10.1093/acprof:oso/9780199691470.001.0001} {\emph {\bibinfo
  {title} {Dynamical Heterogeneities in Glasses, Colloids, and Granular
  media}}}\ (\bibinfo  {publisher} {Oxford University Press},\ \bibinfo {year}
  {2011})\BibitemShut {NoStop}%
\bibitem [{\citenamefont {Helbing}\ \emph {et~al.}(2005)\citenamefont
  {Helbing}, \citenamefont {Buzna}, \citenamefont {Johansson},\ and\
  \citenamefont {Werner}}]{Crowd}%
  \BibitemOpen
  \bibfield  {author} {\bibinfo {author} {\bibfnamefont {D.}~\bibnamefont
  {Helbing}}, \bibinfo {author} {\bibfnamefont {L.}~\bibnamefont {Buzna}},
  \bibinfo {author} {\bibfnamefont {A.}~\bibnamefont {Johansson}},\ and\
  \bibinfo {author} {\bibfnamefont {T.}~\bibnamefont {Werner}},\ }\bibfield
  {title} {\bibinfo {title} {Self-organized pedestrian crowd dynamics:
  Experiments, simulations, and design solutions},\ }\href
  {https://doi.org/10.1287/trsc.1040.0108} {\bibfield  {journal} {\bibinfo
  {journal} {Transp. Sci.}\ }\textbf {\bibinfo {volume} {39}},\ \bibinfo
  {pages} {1} (\bibinfo {year} {2005})}\BibitemShut {NoStop}%
\bibitem [{\citenamefont {Hansen}\ and\ \citenamefont
  {McDonald}(2013)}]{Hansen}%
  \BibitemOpen
  \bibfield  {author} {\bibinfo {author} {\bibfnamefont {J.-P.}\ \bibnamefont
  {Hansen}}\ and\ \bibinfo {author} {\bibfnamefont {I.~R.}\ \bibnamefont
  {McDonald}},\ }\href {https://doi.org/10.1016/C2010-0-66723-X} {\emph
  {\bibinfo {title} {{Theory of Simple Liquids}}}}\ (\bibinfo  {publisher}
  {Academic Press, Amsterdam},\ \bibinfo {year} {2013})\BibitemShut {NoStop}%
\bibitem [{\citenamefont {Bouchaud}\ and\ \citenamefont
  {Georges}(1990)}]{Bouchaud}%
  \BibitemOpen
  \bibfield  {author} {\bibinfo {author} {\bibfnamefont {J.-P.}\ \bibnamefont
  {Bouchaud}}\ and\ \bibinfo {author} {\bibfnamefont {A.}~\bibnamefont
  {Georges}},\ }\bibfield  {title} {\bibinfo {title} {Anomalous diffusion in
  disordered media: Statistical mechanisms, models and physical applications},\
  }\href {https://doi.org/https://doi.org/10.1016/0370-1573(90)90099-N}
  {\bibfield  {journal} {\bibinfo  {journal} {Phys. Rep.}\ }\textbf {\bibinfo
  {volume} {195}},\ \bibinfo {pages} {127} (\bibinfo {year}
  {1990})}\BibitemShut {NoStop}%
\bibitem [{\citenamefont {Nagaosa}\ and\ \citenamefont
  {Tokura}(2013)}]{NagaosaTokura2013}%
  \BibitemOpen
  \bibfield  {author} {\bibinfo {author} {\bibfnamefont {N.}~\bibnamefont
  {Nagaosa}}\ and\ \bibinfo {author} {\bibfnamefont {Y.}~\bibnamefont
  {Tokura}},\ }\bibfield  {title} {\bibinfo {title} {Topological properties and
  dynamics of magnetic skyrmions},\ }\href
  {https://doi.org/10.1038/nnano.2013.243} {\bibfield  {journal} {\bibinfo
  {journal} {Nature Nanotech.}\ }\textbf {\bibinfo {volume} {8}},\ \bibinfo
  {pages} {899} (\bibinfo {year} {2013})}\BibitemShut {NoStop}%
\bibitem [{\citenamefont {Tokura}\ and\ \citenamefont
  {Kanazawa}(2021)}]{Tokura2021review}%
  \BibitemOpen
  \bibfield  {author} {\bibinfo {author} {\bibfnamefont {Y.}~\bibnamefont
  {Tokura}}\ and\ \bibinfo {author} {\bibfnamefont {N.}~\bibnamefont
  {Kanazawa}},\ }\bibfield  {title} {\bibinfo {title} {Magnetic skyrmion
  materials},\ }\href {https://doi.org/10.1021/acs.chemrev.0c00297} {\bibfield
  {journal} {\bibinfo  {journal} {Chem. Rev.}\ }\textbf {\bibinfo {volume}
  {121}},\ \bibinfo {pages} {2857} (\bibinfo {year} {2021})}\BibitemShut
  {NoStop}%
\bibitem [{\citenamefont {Okubo}\ \emph {et~al.}(2012)\citenamefont {Okubo},
  \citenamefont {Chung},\ and\ \citenamefont {Kawamura}}]{Kawamura}%
  \BibitemOpen
  \bibfield  {author} {\bibinfo {author} {\bibfnamefont {T.}~\bibnamefont
  {Okubo}}, \bibinfo {author} {\bibfnamefont {S.}~\bibnamefont {Chung}},\ and\
  \bibinfo {author} {\bibfnamefont {H.}~\bibnamefont {Kawamura}},\ }\bibfield
  {title} {\bibinfo {title} {Multiple-$q$ states and the skyrmion lattice of
  the triangular-lattice heisenberg antiferromagnet under magnetic fields},\
  }\href {https://doi.org/10.1103/PhysRevLett.108.017206} {\bibfield  {journal}
  {\bibinfo  {journal} {Phys. Rev. Lett.}\ }\textbf {\bibinfo {volume} {108}},\
  \bibinfo {pages} {017206} (\bibinfo {year} {2012})}\BibitemShut {NoStop}%
\bibitem [{\citenamefont {Fukuda}\ and\ \citenamefont
  {{\v{Z}}umer}(2011)}]{Fukuda2011}%
  \BibitemOpen
  \bibfield  {author} {\bibinfo {author} {\bibfnamefont {J.-i.}\ \bibnamefont
  {Fukuda}}\ and\ \bibinfo {author} {\bibfnamefont {S.}~\bibnamefont
  {{\v{Z}}umer}},\ }\bibfield  {title} {\bibinfo {title} {Quasi-two-dimensional
  skyrmion lattices in a chiral nematic liquid crystal},\ }\href
  {https://doi.org/10.1038/ncomms1250} {\bibfield  {journal} {\bibinfo
  {journal} {Nature Commun.}\ }\textbf {\bibinfo {volume} {2}},\ \bibinfo
  {pages} {246} (\bibinfo {year} {2011})}\BibitemShut {NoStop}%
\bibitem [{\citenamefont {Nych}\ \emph {et~al.}(2017)\citenamefont {Nych},
  \citenamefont {Fukuda}, \citenamefont {Ognysta}, \citenamefont
  {{\v{Z}}umer},\ and\ \citenamefont {Mu{\v{s}}evi{\v{c}}}}]{Fukuda2017}%
  \BibitemOpen
  \bibfield  {author} {\bibinfo {author} {\bibfnamefont {A.}~\bibnamefont
  {Nych}}, \bibinfo {author} {\bibfnamefont {J.-i.}\ \bibnamefont {Fukuda}},
  \bibinfo {author} {\bibfnamefont {U.}~\bibnamefont {Ognysta}}, \bibinfo
  {author} {\bibfnamefont {S.}~\bibnamefont {{\v{Z}}umer}},\ and\ \bibinfo
  {author} {\bibfnamefont {I.}~\bibnamefont {Mu{\v{s}}evi{\v{c}}}},\ }\bibfield
   {title} {\bibinfo {title} {Spontaneous formation and dynamics of
  half-skyrmions in a chiral liquid-crystal film},\ }\href
  {https://doi.org/10.1038/nphys4245} {\bibfield  {journal} {\bibinfo
  {journal} {Nature Phys.}\ }\textbf {\bibinfo {volume} {13}},\ \bibinfo
  {pages} {1215} (\bibinfo {year} {2017})}\BibitemShut {NoStop}%
\bibitem [{\citenamefont {Ohmi}\ and\ \citenamefont
  {Machida}(1998)}]{OhmiMachida}%
  \BibitemOpen
  \bibfield  {author} {\bibinfo {author} {\bibfnamefont {T.}~\bibnamefont
  {Ohmi}}\ and\ \bibinfo {author} {\bibfnamefont {K.}~\bibnamefont {Machida}},\
  }\bibfield  {title} {\bibinfo {title} {{Bose-Einstein} condensation with
  internal degrees of freedom in alkali atom gases},\ }\href
  {https://doi.org/10.1143/jpsj.67.1822} {\bibfield  {journal} {\bibinfo
  {journal} {J. Phys. Soc. Jpn.}\ }\textbf {\bibinfo {volume} {67}},\ \bibinfo
  {pages} {1822} (\bibinfo {year} {1998})}\BibitemShut {NoStop}%
\bibitem [{\citenamefont {Ho}(1998)}]{Ho-BEC}%
  \BibitemOpen
  \bibfield  {author} {\bibinfo {author} {\bibfnamefont {T.-L.}\ \bibnamefont
  {Ho}},\ }\bibfield  {title} {\bibinfo {title} {Spinor {Bose} condensates in
  optical traps},\ }\href {https://doi.org/10.1103/PhysRevLett.81.742}
  {\bibfield  {journal} {\bibinfo  {journal} {Phys. Rev. Lett.}\ }\textbf
  {\bibinfo {volume} {81}},\ \bibinfo {pages} {742} (\bibinfo {year}
  {1998})}\BibitemShut {NoStop}%
\bibitem [{\citenamefont {Das}\ \emph {et~al.}(2019)\citenamefont {Das},
  \citenamefont {Tang}, \citenamefont {Hong}, \citenamefont {Gon{\c{c}}alves},
  \citenamefont {McCarter}, \citenamefont {Klewe}, \citenamefont {Nguyen},
  \citenamefont {G{\'o}mez-Ortiz}, \citenamefont {Shafer}, \citenamefont
  {Arenholz} \emph {et~al.}}]{Ramesh-polar2019}%
  \BibitemOpen
  \bibfield  {author} {\bibinfo {author} {\bibfnamefont {S.}~\bibnamefont
  {Das}}, \bibinfo {author} {\bibfnamefont {Y.}~\bibnamefont {Tang}}, \bibinfo
  {author} {\bibfnamefont {Z.}~\bibnamefont {Hong}}, \bibinfo {author}
  {\bibfnamefont {M.}~\bibnamefont {Gon{\c{c}}alves}}, \bibinfo {author}
  {\bibfnamefont {M.}~\bibnamefont {McCarter}}, \bibinfo {author}
  {\bibfnamefont {C.}~\bibnamefont {Klewe}}, \bibinfo {author} {\bibfnamefont
  {K.}~\bibnamefont {Nguyen}}, \bibinfo {author} {\bibfnamefont
  {F.}~\bibnamefont {G{\'o}mez-Ortiz}}, \bibinfo {author} {\bibfnamefont
  {P.}~\bibnamefont {Shafer}}, \bibinfo {author} {\bibfnamefont
  {E.}~\bibnamefont {Arenholz}}, \emph {et~al.},\ }\bibfield  {title} {\bibinfo
  {title} {Observation of room-temperature polar skyrmions},\ }\href
  {https://doi.org/10.1038/s41586-019-1092-8} {\bibfield  {journal} {\bibinfo
  {journal} {Nature}\ }\textbf {\bibinfo {volume} {568}},\ \bibinfo {pages}
  {368} (\bibinfo {year} {2019})}\BibitemShut {NoStop}%
\bibitem [{\citenamefont {Das}\ \emph {et~al.}(2021)\citenamefont {Das},
  \citenamefont {Hong}, \citenamefont {Stoica}, \citenamefont
  {Gon{\c{c}}alves}, \citenamefont {Shao}, \citenamefont {Parsonnet},
  \citenamefont {Marksz}, \citenamefont {Saremi}, \citenamefont {McCarter},
  \citenamefont {Reynoso} \emph {et~al.}}]{Ramesh-polar2021}%
  \BibitemOpen
  \bibfield  {author} {\bibinfo {author} {\bibfnamefont {S.}~\bibnamefont
  {Das}}, \bibinfo {author} {\bibfnamefont {Z.}~\bibnamefont {Hong}}, \bibinfo
  {author} {\bibfnamefont {V.}~\bibnamefont {Stoica}}, \bibinfo {author}
  {\bibfnamefont {M.}~\bibnamefont {Gon{\c{c}}alves}}, \bibinfo {author}
  {\bibfnamefont {Y.-T.}\ \bibnamefont {Shao}}, \bibinfo {author}
  {\bibfnamefont {E.}~\bibnamefont {Parsonnet}}, \bibinfo {author}
  {\bibfnamefont {E.~J.}\ \bibnamefont {Marksz}}, \bibinfo {author}
  {\bibfnamefont {S.}~\bibnamefont {Saremi}}, \bibinfo {author} {\bibfnamefont
  {M.}~\bibnamefont {McCarter}}, \bibinfo {author} {\bibfnamefont
  {A.}~\bibnamefont {Reynoso}}, \emph {et~al.},\ }\bibfield  {title} {\bibinfo
  {title} {Local negative permittivity and topological phase transition in
  polar skyrmions},\ }\href {https://doi.org/10.1038/s41563-020-00818-y}
  {\bibfield  {journal} {\bibinfo  {journal} {Nature Mater.}\ }\textbf
  {\bibinfo {volume} {20}},\ \bibinfo {pages} {194} (\bibinfo {year}
  {2021})}\BibitemShut {NoStop}%
\bibitem [{\citenamefont {Sch\"utte}\ \emph {et~al.}(2014)\citenamefont
  {Sch\"utte}, \citenamefont {Iwasaki}, \citenamefont {Rosch},\ and\
  \citenamefont {Nagaosa}}]{Nagaosa2014}%
  \BibitemOpen
  \bibfield  {author} {\bibinfo {author} {\bibfnamefont {C.}~\bibnamefont
  {Sch\"utte}}, \bibinfo {author} {\bibfnamefont {J.}~\bibnamefont {Iwasaki}},
  \bibinfo {author} {\bibfnamefont {A.}~\bibnamefont {Rosch}},\ and\ \bibinfo
  {author} {\bibfnamefont {N.}~\bibnamefont {Nagaosa}},\ }\bibfield  {title}
  {\bibinfo {title} {Inertia, diffusion, and dynamics of a driven skyrmion},\
  }\href {https://doi.org/10.1103/PhysRevB.90.174434} {\bibfield  {journal}
  {\bibinfo  {journal} {Phys. Rev. B}\ }\textbf {\bibinfo {volume} {90}},\
  \bibinfo {pages} {174434} (\bibinfo {year} {2014})}\BibitemShut {NoStop}%
\bibitem [{\citenamefont {Z{\'a}zvorka}\ \emph {et~al.}(2019)\citenamefont
  {Z{\'a}zvorka}, \citenamefont {Jakobs}, \citenamefont {Heinze}, \citenamefont
  {Keil}, \citenamefont {Kromin}, \citenamefont {Jaiswal}, \citenamefont
  {Litzius}, \citenamefont {Jakob}, \citenamefont {Virnau}, \citenamefont
  {Pinna} \emph {et~al.}}]{diffusion}%
  \BibitemOpen
  \bibfield  {author} {\bibinfo {author} {\bibfnamefont {J.}~\bibnamefont
  {Z{\'a}zvorka}}, \bibinfo {author} {\bibfnamefont {F.}~\bibnamefont
  {Jakobs}}, \bibinfo {author} {\bibfnamefont {D.}~\bibnamefont {Heinze}},
  \bibinfo {author} {\bibfnamefont {N.}~\bibnamefont {Keil}}, \bibinfo {author}
  {\bibfnamefont {S.}~\bibnamefont {Kromin}}, \bibinfo {author} {\bibfnamefont
  {S.}~\bibnamefont {Jaiswal}}, \bibinfo {author} {\bibfnamefont
  {K.}~\bibnamefont {Litzius}}, \bibinfo {author} {\bibfnamefont
  {G.}~\bibnamefont {Jakob}}, \bibinfo {author} {\bibfnamefont
  {P.}~\bibnamefont {Virnau}}, \bibinfo {author} {\bibfnamefont
  {D.}~\bibnamefont {Pinna}}, \emph {et~al.},\ }\bibfield  {title} {\bibinfo
  {title} {Thermal skyrmion diffusion used in a reshuffler device},\ }\href
  {https://doi.org/10.1038/s41565-019-0436-8} {\bibfield  {journal} {\bibinfo
  {journal} {Nature Nanotech.}\ }\textbf {\bibinfo {volume} {14}},\ \bibinfo
  {pages} {658} (\bibinfo {year} {2019})}\BibitemShut {NoStop}%
\bibitem [{\citenamefont {Reichhardt}\ \emph {et~al.}(2022)\citenamefont
  {Reichhardt}, \citenamefont {Reichhardt},\ and\ \citenamefont
  {Milo\v{s}evi\'{c}}}]{pinning}%
  \BibitemOpen
  \bibfield  {author} {\bibinfo {author} {\bibfnamefont {C.}~\bibnamefont
  {Reichhardt}}, \bibinfo {author} {\bibfnamefont {C.~J.~O.}\ \bibnamefont
  {Reichhardt}},\ and\ \bibinfo {author} {\bibfnamefont {M.~V.}\ \bibnamefont
  {Milo\v{s}evi\'{c}}},\ }\bibfield  {title} {\bibinfo {title} {Statics and
  dynamics of skyrmions interacting with disorder and nanostructures},\ }\href
  {https://doi.org/10.1103/RevModPhys.94.035005} {\bibfield  {journal}
  {\bibinfo  {journal} {Rev. Mod. Phys.}\ }\textbf {\bibinfo {volume} {94}},\
  \bibinfo {pages} {035005} (\bibinfo {year} {2022})}\BibitemShut {NoStop}%
\bibitem [{\citenamefont {Mochizuki}(2012)}]{Mochizuki}%
  \BibitemOpen
  \bibfield  {author} {\bibinfo {author} {\bibfnamefont {M.}~\bibnamefont
  {Mochizuki}},\ }\bibfield  {title} {\bibinfo {title} {Spin-wave modes and
  their intense excitation effects in skyrmion crystals},\ }\href
  {https://doi.org/10.1103/PhysRevLett.108.017601} {\bibfield  {journal}
  {\bibinfo  {journal} {Phys. Rev. Lett.}\ }\textbf {\bibinfo {volume} {108}},\
  \bibinfo {pages} {017601} (\bibinfo {year} {2012})}\BibitemShut {NoStop}%
\bibitem [{\citenamefont {B{\"u}ttner}\ \emph {et~al.}(2015)\citenamefont
  {B{\"u}ttner}, \citenamefont {Moutafis}, \citenamefont {Schneider},
  \citenamefont {Kr{\"u}ger}, \citenamefont {G{\"u}nther}, \citenamefont
  {Geilhufe}, \citenamefont {Schmising}, \citenamefont {Mohanty}, \citenamefont
  {Pfau}, \citenamefont {Schaffert} \emph {et~al.}}]{Buttner}%
  \BibitemOpen
  \bibfield  {author} {\bibinfo {author} {\bibfnamefont {F.}~\bibnamefont
  {B{\"u}ttner}}, \bibinfo {author} {\bibfnamefont {C.}~\bibnamefont
  {Moutafis}}, \bibinfo {author} {\bibfnamefont {M.}~\bibnamefont {Schneider}},
  \bibinfo {author} {\bibfnamefont {B.}~\bibnamefont {Kr{\"u}ger}}, \bibinfo
  {author} {\bibfnamefont {C.}~\bibnamefont {G{\"u}nther}}, \bibinfo {author}
  {\bibfnamefont {J.}~\bibnamefont {Geilhufe}}, \bibinfo {author}
  {\bibfnamefont {C.~v.~K.}\ \bibnamefont {Schmising}}, \bibinfo {author}
  {\bibfnamefont {J.}~\bibnamefont {Mohanty}}, \bibinfo {author} {\bibfnamefont
  {B.}~\bibnamefont {Pfau}}, \bibinfo {author} {\bibfnamefont {S.}~\bibnamefont
  {Schaffert}}, \emph {et~al.},\ }\bibfield  {title} {\bibinfo {title}
  {Dynamics and inertia of skyrmionic spin structures},\ }\href
  {https://doi.org/10.1038/nphys3234} {\bibfield  {journal} {\bibinfo
  {journal} {Nature Phys.}\ }\textbf {\bibinfo {volume} {11}},\ \bibinfo
  {pages} {225} (\bibinfo {year} {2015})}\BibitemShut {NoStop}%
\bibitem [{\citenamefont {M\"uller}\ \emph {et~al.}(2017)\citenamefont
  {M\"uller}, \citenamefont {Rajeswari}, \citenamefont {Huang}, \citenamefont
  {Murooka}, \citenamefont {R\o{}nnow}, \citenamefont {Carbone},\ and\
  \citenamefont {Rosch}}]{Muller}%
  \BibitemOpen
  \bibfield  {author} {\bibinfo {author} {\bibfnamefont {J.}~\bibnamefont
  {M\"uller}}, \bibinfo {author} {\bibfnamefont {J.}~\bibnamefont {Rajeswari}},
  \bibinfo {author} {\bibfnamefont {P.}~\bibnamefont {Huang}}, \bibinfo
  {author} {\bibfnamefont {Y.}~\bibnamefont {Murooka}}, \bibinfo {author}
  {\bibfnamefont {H.~M.}\ \bibnamefont {R\o{}nnow}}, \bibinfo {author}
  {\bibfnamefont {F.}~\bibnamefont {Carbone}},\ and\ \bibinfo {author}
  {\bibfnamefont {A.}~\bibnamefont {Rosch}},\ }\bibfield  {title} {\bibinfo
  {title} {Magnetic skyrmions and skyrmion clusters in the helical phase of
  {${\mathrm{Cu}}_{2}{\mathrm{OSeO}}_{3}$}},\ }\href
  {https://doi.org/10.1103/PhysRevLett.119.137201} {\bibfield  {journal}
  {\bibinfo  {journal} {Phys. Rev. Lett.}\ }\textbf {\bibinfo {volume} {119}},\
  \bibinfo {pages} {137201} (\bibinfo {year} {2017})}\BibitemShut {NoStop}%
\bibitem [{\citenamefont {Yu}\ \emph {et~al.}(2018{\natexlab{a}})\citenamefont
  {Yu}, \citenamefont {Morikawa}, \citenamefont {Yokouchi}, \citenamefont
  {Shibata}, \citenamefont {Kanazawa}, \citenamefont {Kagawa}, \citenamefont
  {Arima},\ and\ \citenamefont {Tokura}}]{Tokura2018NP}%
  \BibitemOpen
  \bibfield  {author} {\bibinfo {author} {\bibfnamefont {X.}~\bibnamefont
  {Yu}}, \bibinfo {author} {\bibfnamefont {D.}~\bibnamefont {Morikawa}},
  \bibinfo {author} {\bibfnamefont {T.}~\bibnamefont {Yokouchi}}, \bibinfo
  {author} {\bibfnamefont {K.}~\bibnamefont {Shibata}}, \bibinfo {author}
  {\bibfnamefont {N.}~\bibnamefont {Kanazawa}}, \bibinfo {author}
  {\bibfnamefont {F.}~\bibnamefont {Kagawa}}, \bibinfo {author} {\bibfnamefont
  {T.-h.}\ \bibnamefont {Arima}},\ and\ \bibinfo {author} {\bibfnamefont
  {Y.}~\bibnamefont {Tokura}},\ }\bibfield  {title} {\bibinfo {title}
  {Aggregation and collapse dynamics of skyrmions in a non-equilibrium state},\
  }\href {https://doi.org/10.1038/s41567-018-0155-3} {\bibfield  {journal}
  {\bibinfo  {journal} {Nature Phys.}\ }\textbf {\bibinfo {volume} {14}},\
  \bibinfo {pages} {832} (\bibinfo {year} {2018}{\natexlab{a}})}\BibitemShut
  {NoStop}%
\bibitem [{\citenamefont {Litzius}\ \emph {et~al.}(2020)\citenamefont
  {Litzius}, \citenamefont {Leliaert}, \citenamefont {Bassirian}, \citenamefont
  {Rodrigues}, \citenamefont {Kromin}, \citenamefont {Lemesh}, \citenamefont
  {Zazvorka}, \citenamefont {Lee}, \citenamefont {Mulkers}, \citenamefont
  {Kerber} \emph {et~al.}}]{Litzius}%
  \BibitemOpen
  \bibfield  {author} {\bibinfo {author} {\bibfnamefont {K.}~\bibnamefont
  {Litzius}}, \bibinfo {author} {\bibfnamefont {J.}~\bibnamefont {Leliaert}},
  \bibinfo {author} {\bibfnamefont {P.}~\bibnamefont {Bassirian}}, \bibinfo
  {author} {\bibfnamefont {D.}~\bibnamefont {Rodrigues}}, \bibinfo {author}
  {\bibfnamefont {S.}~\bibnamefont {Kromin}}, \bibinfo {author} {\bibfnamefont
  {I.}~\bibnamefont {Lemesh}}, \bibinfo {author} {\bibfnamefont
  {J.}~\bibnamefont {Zazvorka}}, \bibinfo {author} {\bibfnamefont {K.-J.}\
  \bibnamefont {Lee}}, \bibinfo {author} {\bibfnamefont {J.}~\bibnamefont
  {Mulkers}}, \bibinfo {author} {\bibfnamefont {N.}~\bibnamefont {Kerber}},
  \emph {et~al.},\ }\bibfield  {title} {\bibinfo {title} {The role of
  temperature and drive current in skyrmion dynamics},\ }\href
  {https://doi.org/10.1038/s41928-019-0359-2} {\bibfield  {journal} {\bibinfo
  {journal} {Nature Electron.}\ }\textbf {\bibinfo {volume} {3}},\ \bibinfo
  {pages} {30} (\bibinfo {year} {2020})}\BibitemShut {NoStop}%
\bibitem [{\citenamefont {Shimojima}\ \emph {et~al.}(2021)\citenamefont
  {Shimojima}, \citenamefont {Nakamura}, \citenamefont {Yu}, \citenamefont
  {Karube}, \citenamefont {Taguchi}, \citenamefont {Tokura},\ and\
  \citenamefont {Ishizaka}}]{Tokura2021}%
  \BibitemOpen
  \bibfield  {author} {\bibinfo {author} {\bibfnamefont {T.}~\bibnamefont
  {Shimojima}}, \bibinfo {author} {\bibfnamefont {A.}~\bibnamefont {Nakamura}},
  \bibinfo {author} {\bibfnamefont {X.}~\bibnamefont {Yu}}, \bibinfo {author}
  {\bibfnamefont {K.}~\bibnamefont {Karube}}, \bibinfo {author} {\bibfnamefont
  {Y.}~\bibnamefont {Taguchi}}, \bibinfo {author} {\bibfnamefont
  {Y.}~\bibnamefont {Tokura}},\ and\ \bibinfo {author} {\bibfnamefont
  {K.}~\bibnamefont {Ishizaka}},\ }\bibfield  {title} {\bibinfo {title}
  {Nano-to-micro spatiotemporal imaging of magnetic skyrmion's life cycle},\
  }\href {https://doi.org/10.1126/sciadv.abg1322} {\bibfield  {journal}
  {\bibinfo  {journal} {Sci. Adv.}\ }\textbf {\bibinfo {volume} {7}},\ \bibinfo
  {pages} {eabg1322} (\bibinfo {year} {2021})}\BibitemShut {NoStop}%
\bibitem [{\citenamefont {Fert}\ \emph {et~al.}(2017)\citenamefont {Fert},
  \citenamefont {Reyren},\ and\ \citenamefont {Cros}}]{Fert}%
  \BibitemOpen
  \bibfield  {author} {\bibinfo {author} {\bibfnamefont {A.}~\bibnamefont
  {Fert}}, \bibinfo {author} {\bibfnamefont {N.}~\bibnamefont {Reyren}},\ and\
  \bibinfo {author} {\bibfnamefont {V.}~\bibnamefont {Cros}},\ }\bibfield
  {title} {\bibinfo {title} {Magnetic skyrmions: advances in physics and
  potential applications},\ }\href {https://doi.org/10.1038/natrevmats.2017.31}
  {\bibfield  {journal} {\bibinfo  {journal} {Nature Rev. Mater.}\ }\textbf
  {\bibinfo {volume} {2}},\ \bibinfo {pages} {1} (\bibinfo {year}
  {2017})}\BibitemShut {NoStop}%
\bibitem [{\citenamefont {Everschor-Sitte}\ \emph {et~al.}(2018)\citenamefont
  {Everschor-Sitte}, \citenamefont {Masell}, \citenamefont {Reeve},\ and\
  \citenamefont {Kl{\"{a}}ui}}]{Klaui-review}%
  \BibitemOpen
  \bibfield  {author} {\bibinfo {author} {\bibfnamefont {K.}~\bibnamefont
  {Everschor-Sitte}}, \bibinfo {author} {\bibfnamefont {J.}~\bibnamefont
  {Masell}}, \bibinfo {author} {\bibfnamefont {R.~M.}\ \bibnamefont {Reeve}},\
  and\ \bibinfo {author} {\bibfnamefont {M.}~\bibnamefont {Kl{\"{a}}ui}},\
  }\bibfield  {title} {\bibinfo {title} {Perspective: Magnetic
  skyrmions--overview of recent progress in an active research field},\ }\href
  {https://doi.org/10.1063/1.5048972} {\bibfield  {journal} {\bibinfo
  {journal} {J. Appl. Phys.}\ }\textbf {\bibinfo {volume} {124}},\ \bibinfo
  {pages} {240901} (\bibinfo {year} {2018})}\BibitemShut {NoStop}%
\bibitem [{\citenamefont {Foster}\ \emph {et~al.}(2019)\citenamefont {Foster},
  \citenamefont {Kind}, \citenamefont {Ackerman}, \citenamefont {Tai},
  \citenamefont {Dennis},\ and\ \citenamefont {Smalyukh}}]{Smalyukh}%
  \BibitemOpen
  \bibfield  {author} {\bibinfo {author} {\bibfnamefont {D.}~\bibnamefont
  {Foster}}, \bibinfo {author} {\bibfnamefont {C.}~\bibnamefont {Kind}},
  \bibinfo {author} {\bibfnamefont {P.~J.}\ \bibnamefont {Ackerman}}, \bibinfo
  {author} {\bibfnamefont {J.-S.~B.}\ \bibnamefont {Tai}}, \bibinfo {author}
  {\bibfnamefont {M.~R.}\ \bibnamefont {Dennis}},\ and\ \bibinfo {author}
  {\bibfnamefont {I.~I.}\ \bibnamefont {Smalyukh}},\ }\bibfield  {title}
  {\bibinfo {title} {Two-dimensional skyrmion bags in liquid crystals and
  ferromagnets},\ }\href {https://doi.org/10.1038/s41567-019-0476-x} {\bibfield
   {journal} {\bibinfo  {journal} {Nature Phys.}\ }\textbf {\bibinfo {volume}
  {15}},\ \bibinfo {pages} {655} (\bibinfo {year} {2019})}\BibitemShut
  {NoStop}%
\bibitem [{\citenamefont {Pi\v{s}ljar}\ \emph {et~al.}(2022)\citenamefont
  {Pi\v{s}ljar}, \citenamefont {Ghosh}, \citenamefont {Turlapati},
  \citenamefont {Rao}, \citenamefont {\v{S}karabot}, \citenamefont {Mertelj},
  \citenamefont {Petelin}, \citenamefont {Nych}, \citenamefont
  {Marin\v{c}i\v{c}}, \citenamefont {Pusovnik}, \citenamefont {Ravnik},\ and\
  \citenamefont {Mu\v{s}evi\v{c}}}]{blue}%
  \BibitemOpen
  \bibfield  {author} {\bibinfo {author} {\bibfnamefont {J.}~\bibnamefont
  {Pi\v{s}ljar}}, \bibinfo {author} {\bibfnamefont {S.}~\bibnamefont {Ghosh}},
  \bibinfo {author} {\bibfnamefont {S.}~\bibnamefont {Turlapati}}, \bibinfo
  {author} {\bibfnamefont {N.~V.~S.}\ \bibnamefont {Rao}}, \bibinfo {author}
  {\bibfnamefont {M.}~\bibnamefont {\v{S}karabot}}, \bibinfo {author}
  {\bibfnamefont {A.}~\bibnamefont {Mertelj}}, \bibinfo {author} {\bibfnamefont
  {A.}~\bibnamefont {Petelin}}, \bibinfo {author} {\bibfnamefont
  {A.}~\bibnamefont {Nych}}, \bibinfo {author} {\bibfnamefont {M.}~\bibnamefont
  {Marin\v{c}i\v{c}}}, \bibinfo {author} {\bibfnamefont {A.}~\bibnamefont
  {Pusovnik}}, \bibinfo {author} {\bibfnamefont {M.}~\bibnamefont {Ravnik}},\
  and\ \bibinfo {author} {\bibfnamefont {I.}~\bibnamefont {Mu\v{s}evi\v{c}}},\
  }\bibfield  {title} {\bibinfo {title} {Blue phase {III}: Topological fluid of
  skyrmions},\ }\href {https://doi.org/10.1103/PhysRevX.12.011003} {\bibfield
  {journal} {\bibinfo  {journal} {Phys. Rev. X}\ }\textbf {\bibinfo {volume}
  {12}},\ \bibinfo {pages} {011003} (\bibinfo {year} {2022})}\BibitemShut
  {NoStop}%
\bibitem [{\citenamefont {Pi\v{s}ljar}\ \emph {et~al.}(2024)\citenamefont
  {Pi\v{s}ljar}, \citenamefont {Nych}, \citenamefont {Ognysta}, \citenamefont
  {Petelin}, \citenamefont {Kralj},\ and\ \citenamefont
  {Mu\v{s}evi\v{c}}}]{Musevic}%
  \BibitemOpen
  \bibfield  {author} {\bibinfo {author} {\bibfnamefont {J.}~\bibnamefont
  {Pi\v{s}ljar}}, \bibinfo {author} {\bibfnamefont {A.}~\bibnamefont {Nych}},
  \bibinfo {author} {\bibfnamefont {U.}~\bibnamefont {Ognysta}}, \bibinfo
  {author} {\bibfnamefont {A.}~\bibnamefont {Petelin}}, \bibinfo {author}
  {\bibfnamefont {S.}~\bibnamefont {Kralj}},\ and\ \bibinfo {author}
  {\bibfnamefont {I.}~\bibnamefont {Mu\v{s}evi\v{c}}},\ }\bibfield  {title}
  {\bibinfo {title} {Dynamics and topology of symmetry breaking with
  skyrmions},\ }\href {https://doi.org/10.1103/PhysRevLett.132.178101}
  {\bibfield  {journal} {\bibinfo  {journal} {Phys. Rev. Lett.}\ }\textbf
  {\bibinfo {volume} {132}},\ \bibinfo {pages} {178101} (\bibinfo {year}
  {2024})}\BibitemShut {NoStop}%
\bibitem [{\citenamefont {M{\"u}hlbauer}\ \emph {et~al.}(2009)\citenamefont
  {M{\"u}hlbauer}, \citenamefont {Binz}, \citenamefont {Jonietz}, \citenamefont
  {Pfleiderer}, \citenamefont {Rosch}, \citenamefont {Neubauer}, \citenamefont
  {Georgii},\ and\ \citenamefont {B{\"o}ni}}]{Muhlbauer}%
  \BibitemOpen
  \bibfield  {author} {\bibinfo {author} {\bibfnamefont {S.}~\bibnamefont
  {M{\"u}hlbauer}}, \bibinfo {author} {\bibfnamefont {B.}~\bibnamefont {Binz}},
  \bibinfo {author} {\bibfnamefont {F.}~\bibnamefont {Jonietz}}, \bibinfo
  {author} {\bibfnamefont {C.}~\bibnamefont {Pfleiderer}}, \bibinfo {author}
  {\bibfnamefont {A.}~\bibnamefont {Rosch}}, \bibinfo {author} {\bibfnamefont
  {A.}~\bibnamefont {Neubauer}}, \bibinfo {author} {\bibfnamefont
  {R.}~\bibnamefont {Georgii}},\ and\ \bibinfo {author} {\bibfnamefont
  {P.}~\bibnamefont {B{\"o}ni}},\ }\bibfield  {title} {\bibinfo {title}
  {Skyrmion lattice in a chiral magnet},\ }\href
  {https://doi.org/10.1126/science.1166767} {\bibfield  {journal} {\bibinfo
  {journal} {Science}\ }\textbf {\bibinfo {volume} {323}},\ \bibinfo {pages}
  {915} (\bibinfo {year} {2009})}\BibitemShut {NoStop}%
\bibitem [{\citenamefont {Yu}\ \emph {et~al.}(2010)\citenamefont {Yu},
  \citenamefont {Onose}, \citenamefont {Kanazawa}, \citenamefont {Park},
  \citenamefont {Han}, \citenamefont {Matsui}, \citenamefont {Nagaosa},\ and\
  \citenamefont {Tokura}}]{Tokura2010}%
  \BibitemOpen
  \bibfield  {author} {\bibinfo {author} {\bibfnamefont {X.}~\bibnamefont
  {Yu}}, \bibinfo {author} {\bibfnamefont {Y.}~\bibnamefont {Onose}}, \bibinfo
  {author} {\bibfnamefont {N.}~\bibnamefont {Kanazawa}}, \bibinfo {author}
  {\bibfnamefont {J.~H.}\ \bibnamefont {Park}}, \bibinfo {author}
  {\bibfnamefont {J.}~\bibnamefont {Han}}, \bibinfo {author} {\bibfnamefont
  {Y.}~\bibnamefont {Matsui}}, \bibinfo {author} {\bibfnamefont
  {N.}~\bibnamefont {Nagaosa}},\ and\ \bibinfo {author} {\bibfnamefont
  {Y.}~\bibnamefont {Tokura}},\ }\bibfield  {title} {\bibinfo {title}
  {Real-space observation of a two-dimensional skyrmion crystal},\ }\href
  {https://doi.org/10.1038/nature09124} {\bibfield  {journal} {\bibinfo
  {journal} {Nature}\ }\textbf {\bibinfo {volume} {465}},\ \bibinfo {pages}
  {901} (\bibinfo {year} {2010})}\BibitemShut {NoStop}%
\bibitem [{\citenamefont {Huang}\ \emph {et~al.}(2020)\citenamefont {Huang},
  \citenamefont {Sch{\"o}nenberger}, \citenamefont {Cantoni}, \citenamefont
  {Heinen}, \citenamefont {Magrez}, \citenamefont {Rosch}, \citenamefont
  {Carbone},\ and\ \citenamefont {R{\o}nnow}}]{Huang}%
  \BibitemOpen
  \bibfield  {author} {\bibinfo {author} {\bibfnamefont {P.}~\bibnamefont
  {Huang}}, \bibinfo {author} {\bibfnamefont {T.}~\bibnamefont
  {Sch{\"o}nenberger}}, \bibinfo {author} {\bibfnamefont {M.}~\bibnamefont
  {Cantoni}}, \bibinfo {author} {\bibfnamefont {L.}~\bibnamefont {Heinen}},
  \bibinfo {author} {\bibfnamefont {A.}~\bibnamefont {Magrez}}, \bibinfo
  {author} {\bibfnamefont {A.}~\bibnamefont {Rosch}}, \bibinfo {author}
  {\bibfnamefont {F.}~\bibnamefont {Carbone}},\ and\ \bibinfo {author}
  {\bibfnamefont {H.~M.}\ \bibnamefont {R{\o}nnow}},\ }\bibfield  {title}
  {\bibinfo {title} {Melting of a skyrmion lattice to a skyrmion liquid via a
  hexatic phase},\ }\href {https://doi.org/10.1038/s41565-020-0716-3}
  {\bibfield  {journal} {\bibinfo  {journal} {Nature Nanotech.}\ }\textbf
  {\bibinfo {volume} {15}},\ \bibinfo {pages} {761} (\bibinfo {year}
  {2020})}\BibitemShut {NoStop}%
\bibitem [{\citenamefont {Z{\'{a}}zvorka}\ \emph {et~al.}(2020)\citenamefont
  {Z{\'{a}}zvorka}, \citenamefont {Dittrich}, \citenamefont {Ge}, \citenamefont
  {Kerber}, \citenamefont {Raab}, \citenamefont {Winkler}, \citenamefont
  {Litzius}, \citenamefont {Veis}, \citenamefont {Virnau},\ and\ \citenamefont
  {Kl{\"{a}}ui}}]{Zazvorka}%
  \BibitemOpen
  \bibfield  {author} {\bibinfo {author} {\bibfnamefont {J.}~\bibnamefont
  {Z{\'{a}}zvorka}}, \bibinfo {author} {\bibfnamefont {F.}~\bibnamefont
  {Dittrich}}, \bibinfo {author} {\bibfnamefont {Y.}~\bibnamefont {Ge}},
  \bibinfo {author} {\bibfnamefont {N.}~\bibnamefont {Kerber}}, \bibinfo
  {author} {\bibfnamefont {K.}~\bibnamefont {Raab}}, \bibinfo {author}
  {\bibfnamefont {T.}~\bibnamefont {Winkler}}, \bibinfo {author} {\bibfnamefont
  {K.}~\bibnamefont {Litzius}}, \bibinfo {author} {\bibfnamefont
  {M.}~\bibnamefont {Veis}}, \bibinfo {author} {\bibfnamefont {P.}~\bibnamefont
  {Virnau}},\ and\ \bibinfo {author} {\bibfnamefont {M.}~\bibnamefont
  {Kl{\"{a}}ui}},\ }\bibfield  {title} {\bibinfo {title} {Skyrmion lattice
  phases in thin film multilayer},\ }\href
  {https://doi.org/https://doi.org/10.1002/adfm.202004037} {\bibfield
  {journal} {\bibinfo  {journal} {Adv. Funct. Mater.}\ }\textbf {\bibinfo
  {volume} {30}},\ \bibinfo {pages} {2004037} (\bibinfo {year}
  {2020})}\BibitemShut {NoStop}%
\bibitem [{\citenamefont {Nishikawa}\ \emph {et~al.}(2019)\citenamefont
  {Nishikawa}, \citenamefont {Hukushima},\ and\ \citenamefont
  {Krauth}}]{Nishikawa}%
  \BibitemOpen
  \bibfield  {author} {\bibinfo {author} {\bibfnamefont {Y.}~\bibnamefont
  {Nishikawa}}, \bibinfo {author} {\bibfnamefont {K.}~\bibnamefont
  {Hukushima}},\ and\ \bibinfo {author} {\bibfnamefont {W.}~\bibnamefont
  {Krauth}},\ }\bibfield  {title} {\bibinfo {title} {Solid-liquid transition of
  skyrmions in a two-dimensional chiral magnet},\ }\href
  {https://doi.org/10.1103/PhysRevB.99.064435} {\bibfield  {journal} {\bibinfo
  {journal} {Phys. Rev. B}\ }\textbf {\bibinfo {volume} {99}},\ \bibinfo
  {pages} {064435} (\bibinfo {year} {2019})}\BibitemShut {NoStop}%
\bibitem [{\citenamefont {Meisenheimer}\ \emph {et~al.}(2023)\citenamefont
  {Meisenheimer}, \citenamefont {Zhang}, \citenamefont {Raftrey}, \citenamefont
  {Chen}, \citenamefont {Shao}, \citenamefont {Chan}, \citenamefont {Yalisove},
  \citenamefont {Chen}, \citenamefont {Yao}, \citenamefont {Scott} \emph
  {et~al.}}]{meisenheimer2023ordering}%
  \BibitemOpen
  \bibfield  {author} {\bibinfo {author} {\bibfnamefont {P.}~\bibnamefont
  {Meisenheimer}}, \bibinfo {author} {\bibfnamefont {H.}~\bibnamefont {Zhang}},
  \bibinfo {author} {\bibfnamefont {D.}~\bibnamefont {Raftrey}}, \bibinfo
  {author} {\bibfnamefont {X.}~\bibnamefont {Chen}}, \bibinfo {author}
  {\bibfnamefont {Y.-T.}\ \bibnamefont {Shao}}, \bibinfo {author}
  {\bibfnamefont {Y.-T.}\ \bibnamefont {Chan}}, \bibinfo {author}
  {\bibfnamefont {R.}~\bibnamefont {Yalisove}}, \bibinfo {author}
  {\bibfnamefont {R.}~\bibnamefont {Chen}}, \bibinfo {author} {\bibfnamefont
  {J.}~\bibnamefont {Yao}}, \bibinfo {author} {\bibfnamefont {M.~C.}\
  \bibnamefont {Scott}}, \emph {et~al.},\ }\bibfield  {title} {\bibinfo {title}
  {Ordering of room-temperature magnetic skyrmions in a polar van der waals
  magnet},\ }\href {https://doi.org/10.1038/s41467-023-39442-0} {\bibfield
  {journal} {\bibinfo  {journal} {Nature Commun.}\ }\textbf {\bibinfo {volume}
  {14}},\ \bibinfo {pages} {3744} (\bibinfo {year} {2023})}\BibitemShut
  {NoStop}%
\bibitem [{\citenamefont {Kosterlitz}\ and\ \citenamefont
  {Thouless}(1973)}]{KT1973}%
  \BibitemOpen
  \bibfield  {author} {\bibinfo {author} {\bibfnamefont {J.~M.}\ \bibnamefont
  {Kosterlitz}}\ and\ \bibinfo {author} {\bibfnamefont {D.~J.}\ \bibnamefont
  {Thouless}},\ }\bibfield  {title} {\bibinfo {title} {Ordering, metastability
  and phase transitions in two-dimensional systems},\ }\href
  {https://doi.org/10.1088/0022-3719/6/7/010} {\bibfield  {journal} {\bibinfo
  {journal} {J. Phys. C: Solid State Phys.}\ }\textbf {\bibinfo {volume} {6}},\
  \bibinfo {pages} {1181} (\bibinfo {year} {1973})}\BibitemShut {NoStop}%
\bibitem [{\citenamefont {Nelson}\ and\ \citenamefont
  {Halperin}(1979)}]{NelsonHalperin}%
  \BibitemOpen
  \bibfield  {author} {\bibinfo {author} {\bibfnamefont {D.~R.}\ \bibnamefont
  {Nelson}}\ and\ \bibinfo {author} {\bibfnamefont {B.~I.}\ \bibnamefont
  {Halperin}},\ }\bibfield  {title} {\bibinfo {title} {Dislocation-mediated
  melting in two dimensions},\ }\href
  {https://doi.org/10.1103/PhysRevB.19.2457} {\bibfield  {journal} {\bibinfo
  {journal} {Phys. Rev. B}\ }\textbf {\bibinfo {volume} {19}},\ \bibinfo
  {pages} {2457} (\bibinfo {year} {1979})}\BibitemShut {NoStop}%
\bibitem [{\citenamefont {Young}(1979)}]{Young}%
  \BibitemOpen
  \bibfield  {author} {\bibinfo {author} {\bibfnamefont {A.~P.}\ \bibnamefont
  {Young}},\ }\bibfield  {title} {\bibinfo {title} {Melting and the vector
  coulomb gas in two dimensions},\ }\href
  {https://doi.org/10.1103/PhysRevB.19.1855} {\bibfield  {journal} {\bibinfo
  {journal} {Phys. Rev. B}\ }\textbf {\bibinfo {volume} {19}},\ \bibinfo
  {pages} {1855} (\bibinfo {year} {1979})}\BibitemShut {NoStop}%
\bibitem [{\citenamefont {Kapfer}\ and\ \citenamefont
  {Krauth}(2015)}]{Krauth2015}%
  \BibitemOpen
  \bibfield  {author} {\bibinfo {author} {\bibfnamefont {S.~C.}\ \bibnamefont
  {Kapfer}}\ and\ \bibinfo {author} {\bibfnamefont {W.}~\bibnamefont
  {Krauth}},\ }\bibfield  {title} {\bibinfo {title} {Two-dimensional melting:
  From liquid-hexatic coexistence to continuous transitions},\ }\href
  {https://doi.org/10.1103/PhysRevLett.114.035702} {\bibfield  {journal}
  {\bibinfo  {journal} {Phys. Rev. Lett.}\ }\textbf {\bibinfo {volume} {114}},\
  \bibinfo {pages} {035702} (\bibinfo {year} {2015})}\BibitemShut {NoStop}%
\bibitem [{\citenamefont {Shiba}\ \emph {et~al.}(2012)\citenamefont {Shiba},
  \citenamefont {Kawasaki},\ and\ \citenamefont {Onuki}}]{Shiba}%
  \BibitemOpen
  \bibfield  {author} {\bibinfo {author} {\bibfnamefont {H.}~\bibnamefont
  {Shiba}}, \bibinfo {author} {\bibfnamefont {T.}~\bibnamefont {Kawasaki}},\
  and\ \bibinfo {author} {\bibfnamefont {A.}~\bibnamefont {Onuki}},\ }\bibfield
   {title} {\bibinfo {title} {Relationship between bond-breakage correlations
  and four-point correlations in heterogeneous glassy dynamics: Configuration
  changes and vibration modes},\ }\href
  {https://doi.org/10.1103/PhysRevE.86.041504} {\bibfield  {journal} {\bibinfo
  {journal} {Phys. Rev. E}\ }\textbf {\bibinfo {volume} {86}},\ \bibinfo
  {pages} {041504} (\bibinfo {year} {2012})}\BibitemShut {NoStop}%
\bibitem [{\citenamefont {Illing}\ \emph {et~al.}(2017)\citenamefont {Illing},
  \citenamefont {Fritschi}, \citenamefont {Kaiser}, \citenamefont {Klix},
  \citenamefont {Maret},\ and\ \citenamefont {Keim}}]{Keim}%
  \BibitemOpen
  \bibfield  {author} {\bibinfo {author} {\bibfnamefont {B.}~\bibnamefont
  {Illing}}, \bibinfo {author} {\bibfnamefont {S.}~\bibnamefont {Fritschi}},
  \bibinfo {author} {\bibfnamefont {H.}~\bibnamefont {Kaiser}}, \bibinfo
  {author} {\bibfnamefont {C.~L.}\ \bibnamefont {Klix}}, \bibinfo {author}
  {\bibfnamefont {G.}~\bibnamefont {Maret}},\ and\ \bibinfo {author}
  {\bibfnamefont {P.}~\bibnamefont {Keim}},\ }\bibfield  {title} {\bibinfo
  {title} {Mermin--{Wagner} fluctuations in {2D} amorphous solids},\ }\href
  {https://doi.org/10.1073/pnas.1612964114} {\bibfield  {journal} {\bibinfo
  {journal} {Proc. Natl. Acad. Sci.}\ }\textbf {\bibinfo {volume} {114}},\
  \bibinfo {pages} {1856} (\bibinfo {year} {2017})}\BibitemShut {NoStop}%
\bibitem [{\citenamefont {Takae}\ and\ \citenamefont
  {Kawasaki}(2022)}]{TakaeKawasaki}%
  \BibitemOpen
  \bibfield  {author} {\bibinfo {author} {\bibfnamefont {K.}~\bibnamefont
  {Takae}}\ and\ \bibinfo {author} {\bibfnamefont {T.}~\bibnamefont
  {Kawasaki}},\ }\bibfield  {title} {\bibinfo {title} {Emergent elastic fields
  induced by topological phase transitions: Impact of molecular chirality and
  steric anisotropy},\ }\href {https://doi.org/10.1073/pnas.2118492119}
  {\bibfield  {journal} {\bibinfo  {journal} {Proc. Natl. Acad. Sci.}\ }\textbf
  {\bibinfo {volume} {119}},\ \bibinfo {pages} {e2118492119} (\bibinfo {year}
  {2022})}\BibitemShut {NoStop}%
\bibitem [{\citenamefont {Takae}\ and\ \citenamefont
  {Onuki}(2014)}]{Takae2014origlass}%
  \BibitemOpen
  \bibfield  {author} {\bibinfo {author} {\bibfnamefont {K.}~\bibnamefont
  {Takae}}\ and\ \bibinfo {author} {\bibfnamefont {A.}~\bibnamefont {Onuki}},\
  }\bibfield  {title} {\bibinfo {title} {Orientational glass in mixtures of
  elliptic and circular particles: Structural heterogeneities, rotational
  dynamics, and rheology},\ }\href {https://doi.org/10.1103/PhysRevE.89.022308}
  {\bibfield  {journal} {\bibinfo  {journal} {Phys. Rev. E}\ }\textbf {\bibinfo
  {volume} {89}},\ \bibinfo {pages} {022308} (\bibinfo {year}
  {2014})}\BibitemShut {NoStop}%
\bibitem [{\citenamefont {Duzgun}\ \emph {et~al.}(2018)\citenamefont {Duzgun},
  \citenamefont {Selinger},\ and\ \citenamefont {Saxena}}]{Duzgun2018}%
  \BibitemOpen
  \bibfield  {author} {\bibinfo {author} {\bibfnamefont {A.}~\bibnamefont
  {Duzgun}}, \bibinfo {author} {\bibfnamefont {J.~V.}\ \bibnamefont
  {Selinger}},\ and\ \bibinfo {author} {\bibfnamefont {A.}~\bibnamefont
  {Saxena}},\ }\bibfield  {title} {\bibinfo {title} {Comparing skyrmions and
  merons in chiral liquid crystals and magnets},\ }\href
  {https://doi.org/10.1103/PhysRevE.97.062706} {\bibfield  {journal} {\bibinfo
  {journal} {Phys. Rev. E}\ }\textbf {\bibinfo {volume} {97}},\ \bibinfo
  {pages} {062706} (\bibinfo {year} {2018})}\BibitemShut {NoStop}%
\bibitem [{\citenamefont {Allen}\ and\ \citenamefont
  {Tildesley}(1989)}]{Allen}%
  \BibitemOpen
  \bibfield  {author} {\bibinfo {author} {\bibfnamefont {M.~P.}\ \bibnamefont
  {Allen}}\ and\ \bibinfo {author} {\bibfnamefont {D.~J.}\ \bibnamefont
  {Tildesley}},\ }\href@noop {} {\emph {\bibinfo {title} {{Computer Simulation
  of Liquids}}}}\ (\bibinfo  {publisher} {Oxford university press},\ \bibinfo
  {year} {1989})\BibitemShut {NoStop}%
\bibitem [{\citenamefont {De~Gennes}\ and\ \citenamefont
  {Prost}(1993)}]{deGennesProst}%
  \BibitemOpen
  \bibfield  {author} {\bibinfo {author} {\bibfnamefont {P.-G.}\ \bibnamefont
  {De~Gennes}}\ and\ \bibinfo {author} {\bibfnamefont {J.}~\bibnamefont
  {Prost}},\ }\href {https://doi.org/10.1093/oso/9780198520245.001.0001} {\emph
  {\bibinfo {title} {{The Physics of Liquid Crystals}}}}\ (\bibinfo
  {publisher} {Oxford university press},\ \bibinfo {year} {1993})\BibitemShut
  {NoStop}%
\bibitem [{\citenamefont {Aoyama}\ and\ \citenamefont
  {Kawamura}(2020)}]{Aoyama}%
  \BibitemOpen
  \bibfield  {author} {\bibinfo {author} {\bibfnamefont {K.}~\bibnamefont
  {Aoyama}}\ and\ \bibinfo {author} {\bibfnamefont {H.}~\bibnamefont
  {Kawamura}},\ }\bibfield  {title} {\bibinfo {title} {Spin current as a probe
  of the $\mathbb{Z}_2$-vortex topological transition in the classical
  heisenberg antiferromagnet on the triangular lattice},\ }\href
  {https://doi.org/10.1103/PhysRevLett.124.047202} {\bibfield  {journal}
  {\bibinfo  {journal} {Phys. Rev. Lett.}\ }\textbf {\bibinfo {volume} {124}},\
  \bibinfo {pages} {047202} (\bibinfo {year} {2020})}\BibitemShut {NoStop}%
\bibitem [{\citenamefont {Irving}\ and\ \citenamefont
  {Kirkwood}(1950)}]{IrvingKirkwood}%
  \BibitemOpen
  \bibfield  {author} {\bibinfo {author} {\bibfnamefont {J.~H.}\ \bibnamefont
  {Irving}}\ and\ \bibinfo {author} {\bibfnamefont {J.~G.}\ \bibnamefont
  {Kirkwood}},\ }\bibfield  {title} {\bibinfo {title} {The statistical
  mechanical theory of transport processes. {IV.} {The} equations of
  hydrodynamics},\ }\href {https://doi.org/10.1063/1.1747782} {\bibfield
  {journal} {\bibinfo  {journal} {J. Chem. Phys.}\ }\textbf {\bibinfo {volume}
  {18}},\ \bibinfo {pages} {817} (\bibinfo {year} {1950})}\BibitemShut
  {NoStop}%
\bibitem [{\citenamefont {Yamamoto}\ and\ \citenamefont
  {Onuki}(1998)}]{YamamotoOnukiPRE}%
  \BibitemOpen
  \bibfield  {author} {\bibinfo {author} {\bibfnamefont {R.}~\bibnamefont
  {Yamamoto}}\ and\ \bibinfo {author} {\bibfnamefont {A.}~\bibnamefont
  {Onuki}},\ }\bibfield  {title} {\bibinfo {title} {Dynamics of highly
  supercooled liquids: Heterogeneity, rheology, and diffusion},\ }\href
  {https://doi.org/10.1103/PhysRevE.58.3515} {\bibfield  {journal} {\bibinfo
  {journal} {Phys. Rev. E}\ }\textbf {\bibinfo {volume} {58}},\ \bibinfo
  {pages} {3515} (\bibinfo {year} {1998})}\BibitemShut {NoStop}%
\bibitem [{\citenamefont {Falk}\ and\ \citenamefont {Langer}(1998)}]{Langer}%
  \BibitemOpen
  \bibfield  {author} {\bibinfo {author} {\bibfnamefont {M.~L.}\ \bibnamefont
  {Falk}}\ and\ \bibinfo {author} {\bibfnamefont {J.~S.}\ \bibnamefont
  {Langer}},\ }\bibfield  {title} {\bibinfo {title} {Dynamics of viscoplastic
  deformation in amorphous solids},\ }\href
  {https://doi.org/10.1103/PhysRevE.57.7192} {\bibfield  {journal} {\bibinfo
  {journal} {Phys. Rev. E}\ }\textbf {\bibinfo {volume} {57}},\ \bibinfo
  {pages} {7192} (\bibinfo {year} {1998})}\BibitemShut {NoStop}%
\bibitem [{\citenamefont {Chaikin}\ and\ \citenamefont
  {Lubensky}(1995)}]{ChaikinLubensky}%
  \BibitemOpen
  \bibfield  {author} {\bibinfo {author} {\bibfnamefont {P.~M.}\ \bibnamefont
  {Chaikin}}\ and\ \bibinfo {author} {\bibfnamefont {T.~C.}\ \bibnamefont
  {Lubensky}},\ }\href {https://doi.org/10.1017/CBO9780511813467} {\emph
  {\bibinfo {title} {{Principles of Condensed Matter Physics}}}}\ (\bibinfo
  {publisher} {Cambridge University Press},\ \bibinfo {year}
  {1995})\BibitemShut {NoStop}%
\bibitem [{\citenamefont {Onuki}(2002)}]{Onukibook}%
  \BibitemOpen
  \bibfield  {author} {\bibinfo {author} {\bibfnamefont {A.}~\bibnamefont
  {Onuki}},\ }\href {https://doi.org/10.1017/CBO9780511534874} {\emph {\bibinfo
  {title} {{Phase Transition Dynamics}}}}\ (\bibinfo  {publisher} {Cambridge
  University Press},\ \bibinfo {year} {2002})\BibitemShut {NoStop}%
\bibitem [{\citenamefont {Nii}\ \emph {et~al.}(2014)\citenamefont {Nii},
  \citenamefont {Kikkawa}, \citenamefont {Taguchi}, \citenamefont {Tokura},\
  and\ \citenamefont {Iwasa}}]{Tokura2014elastic}%
  \BibitemOpen
  \bibfield  {author} {\bibinfo {author} {\bibfnamefont {Y.}~\bibnamefont
  {Nii}}, \bibinfo {author} {\bibfnamefont {A.}~\bibnamefont {Kikkawa}},
  \bibinfo {author} {\bibfnamefont {Y.}~\bibnamefont {Taguchi}}, \bibinfo
  {author} {\bibfnamefont {Y.}~\bibnamefont {Tokura}},\ and\ \bibinfo {author}
  {\bibfnamefont {Y.}~\bibnamefont {Iwasa}},\ }\bibfield  {title} {\bibinfo
  {title} {Elastic stiffness of a skyrmion crystal},\ }\href
  {https://doi.org/10.1103/PhysRevLett.113.267203} {\bibfield  {journal}
  {\bibinfo  {journal} {Phys. Rev. Lett.}\ }\textbf {\bibinfo {volume} {113}},\
  \bibinfo {pages} {267203} (\bibinfo {year} {2014})}\BibitemShut {NoStop}%
\bibitem [{\citenamefont {Yang}\ and\ \citenamefont
  {Schmidt}(2021)}]{YangSchmidt}%
  \BibitemOpen
  \bibfield  {author} {\bibinfo {author} {\bibfnamefont {W.-G.}\ \bibnamefont
  {Yang}}\ and\ \bibinfo {author} {\bibfnamefont {H.}~\bibnamefont {Schmidt}},\
  }\bibfield  {title} {\bibinfo {title} {Acoustic control of magnetism toward
  energy-efficient applications},\ }\href {https://doi.org/10.1063/5.0042138}
  {\bibfield  {journal} {\bibinfo  {journal} {Appl. Phys. Rev.}\ }\textbf
  {\bibinfo {volume} {8}},\ \bibinfo {pages} {021304} (\bibinfo {year}
  {2021})}\BibitemShut {NoStop}%
\bibitem [{\citenamefont {Berk}\ \emph {et~al.}(2019)\citenamefont {Berk},
  \citenamefont {Jaris}, \citenamefont {Yang}, \citenamefont {Dhuey},
  \citenamefont {Cabrini},\ and\ \citenamefont {Schmidt}}]{berk2019strongly}%
  \BibitemOpen
  \bibfield  {author} {\bibinfo {author} {\bibfnamefont {C.}~\bibnamefont
  {Berk}}, \bibinfo {author} {\bibfnamefont {M.}~\bibnamefont {Jaris}},
  \bibinfo {author} {\bibfnamefont {W.}~\bibnamefont {Yang}}, \bibinfo {author}
  {\bibfnamefont {S.}~\bibnamefont {Dhuey}}, \bibinfo {author} {\bibfnamefont
  {S.}~\bibnamefont {Cabrini}},\ and\ \bibinfo {author} {\bibfnamefont
  {H.}~\bibnamefont {Schmidt}},\ }\bibfield  {title} {\bibinfo {title}
  {Strongly coupled magnon--phonon dynamics in a single nanomagnet},\ }\href
  {https://doi.org/10.1038/s41467-019-10545-x} {\bibfield  {journal} {\bibinfo
  {journal} {Nature Commun.}\ }\textbf {\bibinfo {volume} {10}},\ \bibinfo
  {pages} {2652} (\bibinfo {year} {2019})}\BibitemShut {NoStop}%
\bibitem [{\citenamefont {Hioki}\ \emph {et~al.}(2022)\citenamefont {Hioki},
  \citenamefont {Hashimoto},\ and\ \citenamefont {Saitoh}}]{hioki2022coherent}%
  \BibitemOpen
  \bibfield  {author} {\bibinfo {author} {\bibfnamefont {T.}~\bibnamefont
  {Hioki}}, \bibinfo {author} {\bibfnamefont {Y.}~\bibnamefont {Hashimoto}},\
  and\ \bibinfo {author} {\bibfnamefont {E.}~\bibnamefont {Saitoh}},\
  }\bibfield  {title} {\bibinfo {title} {Coherent oscillation between phonons
  and magnons},\ }\href {https://doi.org/10.1038/s42005-022-00888-1} {\bibfield
   {journal} {\bibinfo  {journal} {Commun. Phys.}\ }\textbf {\bibinfo {volume}
  {5}},\ \bibinfo {pages} {115} (\bibinfo {year} {2022})}\BibitemShut {NoStop}%
\bibitem [{\citenamefont {Hwang}\ \emph {et~al.}(2024)\citenamefont {Hwang},
  \citenamefont {Puebla}, \citenamefont {Kondou}, \citenamefont
  {Gonzalez-Ballestero}, \citenamefont {Isshiki}, \citenamefont {Mu\~noz},
  \citenamefont {Liao}, \citenamefont {Chen}, \citenamefont {Luo},
  \citenamefont {Maekawa},\ and\ \citenamefont {Otani}}]{hwang2024strongly}%
  \BibitemOpen
  \bibfield  {author} {\bibinfo {author} {\bibfnamefont {Y.}~\bibnamefont
  {Hwang}}, \bibinfo {author} {\bibfnamefont {J.}~\bibnamefont {Puebla}},
  \bibinfo {author} {\bibfnamefont {K.}~\bibnamefont {Kondou}}, \bibinfo
  {author} {\bibfnamefont {C.}~\bibnamefont {Gonzalez-Ballestero}}, \bibinfo
  {author} {\bibfnamefont {H.}~\bibnamefont {Isshiki}}, \bibinfo {author}
  {\bibfnamefont {C.~S.}\ \bibnamefont {Mu\~noz}}, \bibinfo {author}
  {\bibfnamefont {L.}~\bibnamefont {Liao}}, \bibinfo {author} {\bibfnamefont
  {F.}~\bibnamefont {Chen}}, \bibinfo {author} {\bibfnamefont {W.}~\bibnamefont
  {Luo}}, \bibinfo {author} {\bibfnamefont {S.}~\bibnamefont {Maekawa}},\ and\
  \bibinfo {author} {\bibfnamefont {Y.}~\bibnamefont {Otani}},\ }\bibfield
  {title} {\bibinfo {title} {Strongly coupled spin waves and surface acoustic
  waves at room temperature},\ }\href
  {https://doi.org/10.1103/PhysRevLett.132.056704} {\bibfield  {journal}
  {\bibinfo  {journal} {Phys. Rev. Lett.}\ }\textbf {\bibinfo {volume} {132}},\
  \bibinfo {pages} {056704} (\bibinfo {year} {2024})}\BibitemShut {NoStop}%
\bibitem [{\citenamefont {Nepal}\ \emph {et~al.}(2018)\citenamefont {Nepal},
  \citenamefont {Güngördü},\ and\ \citenamefont
  {Kovalev}}]{nepal2018magnetic}%
  \BibitemOpen
  \bibfield  {author} {\bibinfo {author} {\bibfnamefont {R.}~\bibnamefont
  {Nepal}}, \bibinfo {author} {\bibfnamefont {U.}~\bibnamefont {Güngördü}},\
  and\ \bibinfo {author} {\bibfnamefont {A.~A.}\ \bibnamefont {Kovalev}},\
  }\bibfield  {title} {\bibinfo {title} {{Magnetic skyrmion bubble motion
  driven by surface acoustic waves}},\ }\href
  {https://doi.org/10.1063/1.5013620} {\bibfield  {journal} {\bibinfo
  {journal} {Appl. Phys. Lett.}\ }\textbf {\bibinfo {volume} {112}},\ \bibinfo
  {pages} {112404} (\bibinfo {year} {2018})}\BibitemShut {NoStop}%
\bibitem [{\citenamefont {Yokouchi}\ \emph {et~al.}(2020)\citenamefont
  {Yokouchi}, \citenamefont {Sugimoto}, \citenamefont {Rana}, \citenamefont
  {Seki}, \citenamefont {Ogawa}, \citenamefont {Kasai},\ and\ \citenamefont
  {Otani}}]{Yokouchi-acoustic}%
  \BibitemOpen
  \bibfield  {author} {\bibinfo {author} {\bibfnamefont {T.}~\bibnamefont
  {Yokouchi}}, \bibinfo {author} {\bibfnamefont {S.}~\bibnamefont {Sugimoto}},
  \bibinfo {author} {\bibfnamefont {B.}~\bibnamefont {Rana}}, \bibinfo {author}
  {\bibfnamefont {S.}~\bibnamefont {Seki}}, \bibinfo {author} {\bibfnamefont
  {N.}~\bibnamefont {Ogawa}}, \bibinfo {author} {\bibfnamefont
  {S.}~\bibnamefont {Kasai}},\ and\ \bibinfo {author} {\bibfnamefont
  {Y.}~\bibnamefont {Otani}},\ }\bibfield  {title} {\bibinfo {title} {Creation
  of magnetic skyrmions by surface acoustic waves},\ }\href
  {https://doi.org/10.1038/s41565-020-0661-1} {\bibfield  {journal} {\bibinfo
  {journal} {Nature Nanotech.}\ }\textbf {\bibinfo {volume} {15}},\ \bibinfo
  {pages} {361} (\bibinfo {year} {2020})}\BibitemShut {NoStop}%
\bibitem [{\citenamefont {Yang}\ \emph {et~al.}(2024)\citenamefont {Yang},
  \citenamefont {Zhao}, \citenamefont {Yi}, \citenamefont {Xu}, \citenamefont
  {Chai}, \citenamefont {Zhang}, \citenamefont {Jiang}, \citenamefont {Ji},
  \citenamefont {Hou}, \citenamefont {Jiang} \emph
  {et~al.}}]{yang2024acoustic}%
  \BibitemOpen
  \bibfield  {author} {\bibinfo {author} {\bibfnamefont {Y.}~\bibnamefont
  {Yang}}, \bibinfo {author} {\bibfnamefont {L.}~\bibnamefont {Zhao}}, \bibinfo
  {author} {\bibfnamefont {D.}~\bibnamefont {Yi}}, \bibinfo {author}
  {\bibfnamefont {T.}~\bibnamefont {Xu}}, \bibinfo {author} {\bibfnamefont
  {Y.}~\bibnamefont {Chai}}, \bibinfo {author} {\bibfnamefont {C.}~\bibnamefont
  {Zhang}}, \bibinfo {author} {\bibfnamefont {D.}~\bibnamefont {Jiang}},
  \bibinfo {author} {\bibfnamefont {Y.}~\bibnamefont {Ji}}, \bibinfo {author}
  {\bibfnamefont {D.}~\bibnamefont {Hou}}, \bibinfo {author} {\bibfnamefont
  {W.}~\bibnamefont {Jiang}}, \emph {et~al.},\ }\bibfield  {title} {\bibinfo
  {title} {Acoustic-driven magnetic skyrmion motion},\ }\href
  {https://doi.org/10.1038/s41467-024-45316-w} {\bibfield  {journal} {\bibinfo
  {journal} {Nature Commun.}\ }\textbf {\bibinfo {volume} {15}},\ \bibinfo
  {pages} {1018} (\bibinfo {year} {2024})}\BibitemShut {NoStop}%
\bibitem [{\citenamefont {Ge}\ \emph {et~al.}(2021)\citenamefont {Ge},
  \citenamefont {Xu}, \citenamefont {Liu}, \citenamefont {Xu}, \citenamefont
  {Lin}, \citenamefont {Yu}, \citenamefont {Bao}, \citenamefont {Jiang},
  \citenamefont {Lu},\ and\ \citenamefont {Chen}}]{acoustic}%
  \BibitemOpen
  \bibfield  {author} {\bibinfo {author} {\bibfnamefont {H.}~\bibnamefont
  {Ge}}, \bibinfo {author} {\bibfnamefont {X.-Y.}\ \bibnamefont {Xu}}, \bibinfo
  {author} {\bibfnamefont {L.}~\bibnamefont {Liu}}, \bibinfo {author}
  {\bibfnamefont {R.}~\bibnamefont {Xu}}, \bibinfo {author} {\bibfnamefont
  {Z.-K.}\ \bibnamefont {Lin}}, \bibinfo {author} {\bibfnamefont {S.-Y.}\
  \bibnamefont {Yu}}, \bibinfo {author} {\bibfnamefont {M.}~\bibnamefont
  {Bao}}, \bibinfo {author} {\bibfnamefont {J.-H.}\ \bibnamefont {Jiang}},
  \bibinfo {author} {\bibfnamefont {M.-H.}\ \bibnamefont {Lu}},\ and\ \bibinfo
  {author} {\bibfnamefont {Y.-F.}\ \bibnamefont {Chen}},\ }\bibfield  {title}
  {\bibinfo {title} {Observation of acoustic skyrmions},\ }\href
  {https://doi.org/10.1103/PhysRevLett.127.144502} {\bibfield  {journal}
  {\bibinfo  {journal} {Phys. Rev. Lett.}\ }\textbf {\bibinfo {volume} {127}},\
  \bibinfo {pages} {144502} (\bibinfo {year} {2021})}\BibitemShut {NoStop}%
\bibitem [{\citenamefont {Landau}\ and\ \citenamefont
  {Lifshitz}(1986)}]{Landau7}%
  \BibitemOpen
  \bibfield  {author} {\bibinfo {author} {\bibfnamefont {L.~D.}\ \bibnamefont
  {Landau}}\ and\ \bibinfo {author} {\bibfnamefont {E.~M.}\ \bibnamefont
  {Lifshitz}},\ }\href {https://doi.org/10.1016/C2009-0-25521-8} {\emph
  {\bibinfo {title} {{Theory of Elasticity}}}},\ \bibinfo {edition} {3rd}\ ed.\
  (\bibinfo  {publisher} {Pergamon Press},\ \bibinfo {year} {1986})\BibitemShut
  {NoStop}%
\bibitem [{\citenamefont {Nicolas}\ \emph {et~al.}(2018)\citenamefont
  {Nicolas}, \citenamefont {Ferrero}, \citenamefont {Martens},\ and\
  \citenamefont {Barrat}}]{Barrat2018review}%
  \BibitemOpen
  \bibfield  {author} {\bibinfo {author} {\bibfnamefont {A.}~\bibnamefont
  {Nicolas}}, \bibinfo {author} {\bibfnamefont {E.~E.}\ \bibnamefont
  {Ferrero}}, \bibinfo {author} {\bibfnamefont {K.}~\bibnamefont {Martens}},\
  and\ \bibinfo {author} {\bibfnamefont {J.-L.}\ \bibnamefont {Barrat}},\
  }\bibfield  {title} {\bibinfo {title} {Deformation and flow of amorphous
  solids: Insights from elastoplastic models},\ }\href
  {https://doi.org/10.1103/RevModPhys.90.045006} {\bibfield  {journal}
  {\bibinfo  {journal} {Rev. Mod. Phys.}\ }\textbf {\bibinfo {volume} {90}},\
  \bibinfo {pages} {045006} (\bibinfo {year} {2018})}\BibitemShut {NoStop}%
\bibitem [{\citenamefont {Wagner}\ \emph {et~al.}(2011)\citenamefont {Wagner},
  \citenamefont {Bedorf}, \citenamefont {Kuechemann}, \citenamefont {Schwabe},
  \citenamefont {Zhang}, \citenamefont {Arnold},\ and\ \citenamefont
  {Samwer}}]{Wagner2011local}%
  \BibitemOpen
  \bibfield  {author} {\bibinfo {author} {\bibfnamefont {H.}~\bibnamefont
  {Wagner}}, \bibinfo {author} {\bibfnamefont {D.}~\bibnamefont {Bedorf}},
  \bibinfo {author} {\bibfnamefont {S.}~\bibnamefont {Kuechemann}}, \bibinfo
  {author} {\bibfnamefont {M.}~\bibnamefont {Schwabe}}, \bibinfo {author}
  {\bibfnamefont {B.}~\bibnamefont {Zhang}}, \bibinfo {author} {\bibfnamefont
  {W.}~\bibnamefont {Arnold}},\ and\ \bibinfo {author} {\bibfnamefont
  {K.}~\bibnamefont {Samwer}},\ }\bibfield  {title} {\bibinfo {title} {Local
  elastic properties of a metallic glass},\ }\href
  {https://doi.org/10.1038/nmat3024} {\bibfield  {journal} {\bibinfo  {journal}
  {Nature Mater.}\ }\textbf {\bibinfo {volume} {10}},\ \bibinfo {pages} {439}
  (\bibinfo {year} {2011})}\BibitemShut {NoStop}%
\bibitem [{\citenamefont {Mitsumoto}\ and\ \citenamefont
  {Takae}(2023)}]{Mitsumoto}%
  \BibitemOpen
  \bibfield  {author} {\bibinfo {author} {\bibfnamefont {K.}~\bibnamefont
  {Mitsumoto}}\ and\ \bibinfo {author} {\bibfnamefont {K.}~\bibnamefont
  {Takae}},\ }\bibfield  {title} {\bibinfo {title} {Elastic heterogeneity
  governs asymmetric adsorption–desorption in a soft porous crystal},\ }\href
  {https://doi.org/10.1073/pnas.2302561120} {\bibfield  {journal} {\bibinfo
  {journal} {Proc. Natl. Acad. Sci.}\ }\textbf {\bibinfo {volume} {120}},\
  \bibinfo {pages} {e2302561120} (\bibinfo {year} {2023})}\BibitemShut
  {NoStop}%
\bibitem [{\citenamefont {Mizuno}\ \emph {et~al.}(2014)\citenamefont {Mizuno},
  \citenamefont {Mossa},\ and\ \citenamefont {Barrat}}]{Mizuno2014}%
  \BibitemOpen
  \bibfield  {author} {\bibinfo {author} {\bibfnamefont {H.}~\bibnamefont
  {Mizuno}}, \bibinfo {author} {\bibfnamefont {S.}~\bibnamefont {Mossa}},\ and\
  \bibinfo {author} {\bibfnamefont {J.-L.}\ \bibnamefont {Barrat}},\ }\bibfield
   {title} {\bibinfo {title} {Acoustic excitations and elastic heterogeneities
  in disordered solids},\ }\href {https://doi.org/10.1073/pnas.1409490111}
  {\bibfield  {journal} {\bibinfo  {journal} {Proc. Natl. Acad. Sci.}\ }\textbf
  {\bibinfo {volume} {111}},\ \bibinfo {pages} {11949} (\bibinfo {year}
  {2014})}\BibitemShut {NoStop}%
\bibitem [{\citenamefont {Mizuno}\ and\ \citenamefont
  {Ikeda}(2018)}]{MizunoIkeda}%
  \BibitemOpen
  \bibfield  {author} {\bibinfo {author} {\bibfnamefont {H.}~\bibnamefont
  {Mizuno}}\ and\ \bibinfo {author} {\bibfnamefont {A.}~\bibnamefont {Ikeda}},\
  }\bibfield  {title} {\bibinfo {title} {Phonon transport and vibrational
  excitations in amorphous solids},\ }\href
  {https://doi.org/10.1103/PhysRevE.98.062612} {\bibfield  {journal} {\bibinfo
  {journal} {Phys. Rev. E}\ }\textbf {\bibinfo {volume} {98}},\ \bibinfo
  {pages} {062612} (\bibinfo {year} {2018})}\BibitemShut {NoStop}%
\bibitem [{\citenamefont {Yu}\ \emph {et~al.}(2018{\natexlab{b}})\citenamefont
  {Yu}, \citenamefont {Koshibae}, \citenamefont {Tokunaga}, \citenamefont
  {Shibata}, \citenamefont {Taguchi}, \citenamefont {Nagaosa},\ and\
  \citenamefont {Tokura}}]{Tokura2018N}%
  \BibitemOpen
  \bibfield  {author} {\bibinfo {author} {\bibfnamefont {X.}~\bibnamefont
  {Yu}}, \bibinfo {author} {\bibfnamefont {W.}~\bibnamefont {Koshibae}},
  \bibinfo {author} {\bibfnamefont {Y.}~\bibnamefont {Tokunaga}}, \bibinfo
  {author} {\bibfnamefont {K.}~\bibnamefont {Shibata}}, \bibinfo {author}
  {\bibfnamefont {Y.}~\bibnamefont {Taguchi}}, \bibinfo {author} {\bibfnamefont
  {N.}~\bibnamefont {Nagaosa}},\ and\ \bibinfo {author} {\bibfnamefont
  {Y.}~\bibnamefont {Tokura}},\ }\bibfield  {title} {\bibinfo {title}
  {Transformation between meron and skyrmion topological spin textures in a
  chiral magnet},\ }\href {https://doi.org/10.1038/s41586-018-0745-3}
  {\bibfield  {journal} {\bibinfo  {journal} {Nature}\ }\textbf {\bibinfo
  {volume} {564}},\ \bibinfo {pages} {95} (\bibinfo {year}
  {2018}{\natexlab{b}})}\BibitemShut {NoStop}%
\bibitem [{\citenamefont {Thiele}(1973)}]{thiele1973steady}%
  \BibitemOpen
  \bibfield  {author} {\bibinfo {author} {\bibfnamefont {A.~A.}\ \bibnamefont
  {Thiele}},\ }\bibfield  {title} {\bibinfo {title} {Steady-state motion of
  magnetic domains},\ }\href {https://doi.org/10.1103/PhysRevLett.30.230}
  {\bibfield  {journal} {\bibinfo  {journal} {Phys. Rev. Lett.}\ }\textbf
  {\bibinfo {volume} {30}},\ \bibinfo {pages} {230} (\bibinfo {year}
  {1973})}\BibitemShut {NoStop}%
\bibitem [{\citenamefont {Ohki}\ and\ \citenamefont
  {Mochizuki}(2025)}]{ohki2025fundamental}%
  \BibitemOpen
  \bibfield  {author} {\bibinfo {author} {\bibfnamefont {Y.}~\bibnamefont
  {Ohki}}\ and\ \bibinfo {author} {\bibfnamefont {M.}~\bibnamefont
  {Mochizuki}},\ }\bibfield  {title} {\bibinfo {title} {Fundamental theory of
  current-induced motion of magnetic skyrmions},\ }\href
  {https://doi.org/10.1088/1361-648X/ad861b} {\bibfield  {journal} {\bibinfo
  {journal} {J. Phys.: Condens. Matter}\ }\textbf {\bibinfo {volume} {37}},\
  \bibinfo {pages} {023003} (\bibinfo {year} {2025})}\BibitemShut {NoStop}%
\bibitem [{\citenamefont {Shankar}\ \emph {et~al.}(2022)\citenamefont
  {Shankar}, \citenamefont {Souslov}, \citenamefont {Bowick}, \citenamefont
  {Marchetti},\ and\ \citenamefont {Vitelli}}]{Vitelli}%
  \BibitemOpen
  \bibfield  {author} {\bibinfo {author} {\bibfnamefont {S.}~\bibnamefont
  {Shankar}}, \bibinfo {author} {\bibfnamefont {A.}~\bibnamefont {Souslov}},
  \bibinfo {author} {\bibfnamefont {M.~J.}\ \bibnamefont {Bowick}}, \bibinfo
  {author} {\bibfnamefont {M.~C.}\ \bibnamefont {Marchetti}},\ and\ \bibinfo
  {author} {\bibfnamefont {V.}~\bibnamefont {Vitelli}},\ }\bibfield  {title}
  {\bibinfo {title} {Topological active matter},\ }\href
  {https://doi.org/10.1038/s42254-022-00445-3} {\bibfield  {journal} {\bibinfo
  {journal} {Nature Rev. Phys.}\ }\textbf {\bibinfo {volume} {4}},\ \bibinfo
  {pages} {380} (\bibinfo {year} {2022})}\BibitemShut {NoStop}%
\bibitem [{\citenamefont {Mertelj}\ \emph {et~al.}(2018)\citenamefont
  {Mertelj}, \citenamefont {Cmok}, \citenamefont {Sebasti\'an}, \citenamefont
  {Mandle}, \citenamefont {Parker}, \citenamefont {Whitwood}, \citenamefont
  {Goodby},\ and\ \citenamefont {\v{C}opi\v{c}}}]{splaynematic}%
  \BibitemOpen
  \bibfield  {author} {\bibinfo {author} {\bibfnamefont {A.}~\bibnamefont
  {Mertelj}}, \bibinfo {author} {\bibfnamefont {L.}~\bibnamefont {Cmok}},
  \bibinfo {author} {\bibfnamefont {N.}~\bibnamefont {Sebasti\'an}}, \bibinfo
  {author} {\bibfnamefont {R.~J.}\ \bibnamefont {Mandle}}, \bibinfo {author}
  {\bibfnamefont {R.~R.}\ \bibnamefont {Parker}}, \bibinfo {author}
  {\bibfnamefont {A.~C.}\ \bibnamefont {Whitwood}}, \bibinfo {author}
  {\bibfnamefont {J.~W.}\ \bibnamefont {Goodby}},\ and\ \bibinfo {author}
  {\bibfnamefont {M.}~\bibnamefont {\v{C}opi\v{c}}},\ }\bibfield  {title}
  {\bibinfo {title} {Splay nematic phase},\ }\href
  {https://doi.org/10.1103/PhysRevX.8.041025} {\bibfield  {journal} {\bibinfo
  {journal} {Phys. Rev. X}\ }\textbf {\bibinfo {volume} {8}},\ \bibinfo {pages}
  {041025} (\bibinfo {year} {2018})}\BibitemShut {NoStop}%
\bibitem [{\citenamefont {Fukuda}\ \emph {et~al.}(2022)\citenamefont {Fukuda},
  \citenamefont {Nych}, \citenamefont {Ognysta}, \citenamefont {Žumer},\ and\
  \citenamefont {Muševič}}]{fukuda2022liquid}%
  \BibitemOpen
  \bibfield  {author} {\bibinfo {author} {\bibfnamefont {J.-i.}\ \bibnamefont
  {Fukuda}}, \bibinfo {author} {\bibfnamefont {A.}~\bibnamefont {Nych}},
  \bibinfo {author} {\bibfnamefont {U.}~\bibnamefont {Ognysta}}, \bibinfo
  {author} {\bibfnamefont {S.}~\bibnamefont {Žumer}},\ and\ \bibinfo {author}
  {\bibfnamefont {I.}~\bibnamefont {Muševič}},\ }\bibfield  {title} {\bibinfo
  {title} {Liquid crystalline half-skyrmions and their optical properties},\
  }\href {https://doi.org/https://doi.org/10.1002/andp.202100336} {\bibfield
  {journal} {\bibinfo  {journal} {Ann. der Phys.}\ }\textbf {\bibinfo {volume}
  {534}},\ \bibinfo {pages} {2100336} (\bibinfo {year} {2022})}\BibitemShut
  {NoStop}%
\bibitem [{\citenamefont {Zerrouki}\ \emph {et~al.}(2008)\citenamefont
  {Zerrouki}, \citenamefont {Baudry}, \citenamefont {Pine}, \citenamefont
  {Chaikin},\ and\ \citenamefont {Bibette}}]{Zerrouki}%
  \BibitemOpen
  \bibfield  {author} {\bibinfo {author} {\bibfnamefont {D.}~\bibnamefont
  {Zerrouki}}, \bibinfo {author} {\bibfnamefont {J.}~\bibnamefont {Baudry}},
  \bibinfo {author} {\bibfnamefont {D.}~\bibnamefont {Pine}}, \bibinfo {author}
  {\bibfnamefont {P.}~\bibnamefont {Chaikin}},\ and\ \bibinfo {author}
  {\bibfnamefont {J.}~\bibnamefont {Bibette}},\ }\bibfield  {title} {\bibinfo
  {title} {Chiral colloidal clusters},\ }\href
  {https://doi.org/10.1038/nature07237} {\bibfield  {journal} {\bibinfo
  {journal} {Nature}\ }\textbf {\bibinfo {volume} {455}},\ \bibinfo {pages}
  {380} (\bibinfo {year} {2008})}\BibitemShut {NoStop}%
\bibitem [{\citenamefont {Li}\ \emph {et~al.}(2016)\citenamefont {Li},
  \citenamefont {Zhou},\ and\ \citenamefont {Han}}]{Han}%
  \BibitemOpen
  \bibfield  {author} {\bibinfo {author} {\bibfnamefont {B.}~\bibnamefont
  {Li}}, \bibinfo {author} {\bibfnamefont {D.}~\bibnamefont {Zhou}},\ and\
  \bibinfo {author} {\bibfnamefont {Y.}~\bibnamefont {Han}},\ }\bibfield
  {title} {\bibinfo {title} {Assembly and phase transitions of colloidal
  crystals},\ }\href {https://doi.org/10.1038/natrevmats.2015.11} {\bibfield
  {journal} {\bibinfo  {journal} {Nature Rev. Mater.}\ }\textbf {\bibinfo
  {volume} {1}},\ \bibinfo {pages} {15011} (\bibinfo {year}
  {2016})}\BibitemShut {NoStop}%
\bibitem [{\citenamefont {Hueckel}\ \emph {et~al.}(2021)\citenamefont
  {Hueckel}, \citenamefont {Hocky},\ and\ \citenamefont
  {Sacanna}}]{Sacanna2021}%
  \BibitemOpen
  \bibfield  {author} {\bibinfo {author} {\bibfnamefont {T.}~\bibnamefont
  {Hueckel}}, \bibinfo {author} {\bibfnamefont {G.~M.}\ \bibnamefont {Hocky}},\
  and\ \bibinfo {author} {\bibfnamefont {S.}~\bibnamefont {Sacanna}},\
  }\bibfield  {title} {\bibinfo {title} {Total synthesis of colloidal matter},\
  }\href {https://doi.org/10.1038/s41578-021-00323-x} {\bibfield  {journal}
  {\bibinfo  {journal} {Nature Rev. Mater.}\ }\textbf {\bibinfo {volume} {6}},\
  \bibinfo {pages} {1053} (\bibinfo {year} {2021})}\BibitemShut {NoStop}%
\bibitem [{\citenamefont {Keim}\ \emph {et~al.}(2004)\citenamefont {Keim},
  \citenamefont {Maret}, \citenamefont {Herz},\ and\ \citenamefont {von
  Gr\"unberg}}]{keim2004harmonic}%
  \BibitemOpen
  \bibfield  {author} {\bibinfo {author} {\bibfnamefont {P.}~\bibnamefont
  {Keim}}, \bibinfo {author} {\bibfnamefont {G.}~\bibnamefont {Maret}},
  \bibinfo {author} {\bibfnamefont {U.}~\bibnamefont {Herz}},\ and\ \bibinfo
  {author} {\bibfnamefont {H.~H.}\ \bibnamefont {von Gr\"unberg}},\ }\bibfield
  {title} {\bibinfo {title} {Harmonic lattice behavior of two-dimensional
  colloidal crystals},\ }\href {https://doi.org/10.1103/PhysRevLett.92.215504}
  {\bibfield  {journal} {\bibinfo  {journal} {Phys. Rev. Lett.}\ }\textbf
  {\bibinfo {volume} {92}},\ \bibinfo {pages} {215504} (\bibinfo {year}
  {2004})}\BibitemShut {NoStop}%
\bibitem [{\citenamefont {Yunker}\ \emph {et~al.}(2014)\citenamefont {Yunker},
  \citenamefont {Chen}, \citenamefont {Gratale}, \citenamefont {Lohr},
  \citenamefont {Still},\ and\ \citenamefont {Yodh}}]{yunker2014physics}%
  \BibitemOpen
  \bibfield  {author} {\bibinfo {author} {\bibfnamefont {P.~J.}\ \bibnamefont
  {Yunker}}, \bibinfo {author} {\bibfnamefont {K.}~\bibnamefont {Chen}},
  \bibinfo {author} {\bibfnamefont {M.~D.}\ \bibnamefont {Gratale}}, \bibinfo
  {author} {\bibfnamefont {M.~A.}\ \bibnamefont {Lohr}}, \bibinfo {author}
  {\bibfnamefont {T.}~\bibnamefont {Still}},\ and\ \bibinfo {author}
  {\bibfnamefont {A.~G.}\ \bibnamefont {Yodh}},\ }\bibfield  {title} {\bibinfo
  {title} {Physics in ordered and disordered colloidal matter composed of
  poly(n-isopropylacrylamide) microgel particles},\ }\href
  {https://doi.org/10.1088/0034-4885/77/5/056601} {\bibfield  {journal}
  {\bibinfo  {journal} {Rep. Prog. Phys.}\ }\textbf {\bibinfo {volume} {77}},\
  \bibinfo {pages} {056601} (\bibinfo {year} {2014})}\BibitemShut {NoStop}%
\bibitem [{\citenamefont {Sharma}\ \emph {et~al.}(2014)\citenamefont {Sharma},
  \citenamefont {Ward}, \citenamefont {Gibaud}, \citenamefont {Hagan},\ and\
  \citenamefont {Dogic}}]{Dogic}%
  \BibitemOpen
  \bibfield  {author} {\bibinfo {author} {\bibfnamefont {P.}~\bibnamefont
  {Sharma}}, \bibinfo {author} {\bibfnamefont {A.}~\bibnamefont {Ward}},
  \bibinfo {author} {\bibfnamefont {T.}~\bibnamefont {Gibaud}}, \bibinfo
  {author} {\bibfnamefont {M.~F.}\ \bibnamefont {Hagan}},\ and\ \bibinfo
  {author} {\bibfnamefont {Z.}~\bibnamefont {Dogic}},\ }\bibfield  {title}
  {\bibinfo {title} {Hierarchical organization of chiral rafts in colloidal
  membranes},\ }\href {https://doi.org/10.1038/nature13694} {\bibfield
  {journal} {\bibinfo  {journal} {Nature}\ }\textbf {\bibinfo {volume} {513}},\
  \bibinfo {pages} {77} (\bibinfo {year} {2014})}\BibitemShut {NoStop}%
\bibitem [{\citenamefont {Dussi}\ and\ \citenamefont
  {Dijkstra}(2016)}]{Dijkstra2016}%
  \BibitemOpen
  \bibfield  {author} {\bibinfo {author} {\bibfnamefont {S.}~\bibnamefont
  {Dussi}}\ and\ \bibinfo {author} {\bibfnamefont {M.}~\bibnamefont
  {Dijkstra}},\ }\bibfield  {title} {\bibinfo {title} {Entropy-driven formation
  of chiral nematic phases by computer simulations},\ }\href
  {https://doi.org/10.1038/ncomms11175} {\bibfield  {journal} {\bibinfo
  {journal} {Nature Commun.}\ }\textbf {\bibinfo {volume} {7}},\ \bibinfo
  {pages} {11175} (\bibinfo {year} {2016})}\BibitemShut {NoStop}%
\bibitem [{\citenamefont {Wolpert}\ \emph {et~al.}(2020)\citenamefont
  {Wolpert}, \citenamefont {Coudert},\ and\ \citenamefont
  {Goodwin}}]{Skyrmion-MOF}%
  \BibitemOpen
  \bibfield  {author} {\bibinfo {author} {\bibfnamefont {E.}~\bibnamefont
  {Wolpert}}, \bibinfo {author} {\bibfnamefont {F.-X.}\ \bibnamefont
  {Coudert}},\ and\ \bibinfo {author} {\bibfnamefont {A.}~\bibnamefont
  {Goodwin}},\ }\href {https://doi.org/10.26434/chemrxiv.12515594.v1} {\bibinfo
  {title} {Skyrmion lattices in chiral metal-organic frameworks}} (\bibinfo
  {year} {2020}),\ \bibinfo {note} {{ChemRxiv.12515594}}\BibitemShut {NoStop}%
\bibitem [{\citenamefont {Subert}\ \emph {et~al.}(2024)\citenamefont {Subert},
  \citenamefont {Campos-Villalobos},\ and\ \citenamefont
  {Dijkstra}}]{subert2024achiral}%
  \BibitemOpen
  \bibfield  {author} {\bibinfo {author} {\bibfnamefont {R.}~\bibnamefont
  {Subert}}, \bibinfo {author} {\bibfnamefont {G.}~\bibnamefont
  {Campos-Villalobos}},\ and\ \bibinfo {author} {\bibfnamefont
  {M.}~\bibnamefont {Dijkstra}},\ }\bibfield  {title} {\bibinfo {title}
  {Achiral hard bananas assemble double-twist skyrmions and blue phases},\
  }\href {https://doi.org/10.1038/s41467-024-50935-4} {\bibfield  {journal}
  {\bibinfo  {journal} {Nature Commun.}\ }\textbf {\bibinfo {volume} {15}},\
  \bibinfo {pages} {6780} (\bibinfo {year} {2024})}\BibitemShut {NoStop}%
\bibitem [{\citenamefont {Mohanta}\ \emph {et~al.}(2020)\citenamefont
  {Mohanta}, \citenamefont {Christianson}, \citenamefont {Okamoto},\ and\
  \citenamefont {Dagotto}}]{mohanta2020signatures}%
  \BibitemOpen
  \bibfield  {author} {\bibinfo {author} {\bibfnamefont {N.}~\bibnamefont
  {Mohanta}}, \bibinfo {author} {\bibfnamefont {A.~D.}\ \bibnamefont
  {Christianson}}, \bibinfo {author} {\bibfnamefont {S.}~\bibnamefont
  {Okamoto}},\ and\ \bibinfo {author} {\bibfnamefont {E.}~\bibnamefont
  {Dagotto}},\ }\bibfield  {title} {\bibinfo {title} {Signatures of a
  liquid-crystal transition in spin-wave excitations of skyrmions},\ }\href
  {https://doi.org/10.1038/s42005-020-00489-w} {\bibfield  {journal} {\bibinfo
  {journal} {Commun. Phys.}\ }\textbf {\bibinfo {volume} {3}},\ \bibinfo
  {pages} {229} (\bibinfo {year} {2020})}\BibitemShut {NoStop}%
\bibitem [{\citenamefont {Kikkawa}\ \emph {et~al.}(2016)\citenamefont
  {Kikkawa}, \citenamefont {Shen}, \citenamefont {Flebus}, \citenamefont
  {Duine}, \citenamefont {Uchida}, \citenamefont {Qiu}, \citenamefont {Bauer},\
  and\ \citenamefont {Saitoh}}]{kikkawa2016magnon}%
  \BibitemOpen
  \bibfield  {author} {\bibinfo {author} {\bibfnamefont {T.}~\bibnamefont
  {Kikkawa}}, \bibinfo {author} {\bibfnamefont {K.}~\bibnamefont {Shen}},
  \bibinfo {author} {\bibfnamefont {B.}~\bibnamefont {Flebus}}, \bibinfo
  {author} {\bibfnamefont {R.~A.}\ \bibnamefont {Duine}}, \bibinfo {author}
  {\bibfnamefont {K.-i.}\ \bibnamefont {Uchida}}, \bibinfo {author}
  {\bibfnamefont {Z.}~\bibnamefont {Qiu}}, \bibinfo {author} {\bibfnamefont
  {G.~E.~W.}\ \bibnamefont {Bauer}},\ and\ \bibinfo {author} {\bibfnamefont
  {E.}~\bibnamefont {Saitoh}},\ }\bibfield  {title} {\bibinfo {title} {Magnon
  polarons in the spin seebeck effect},\ }\href
  {https://doi.org/10.1103/PhysRevLett.117.207203} {\bibfield  {journal}
  {\bibinfo  {journal} {Phys. Rev. Lett.}\ }\textbf {\bibinfo {volume} {117}},\
  \bibinfo {pages} {207203} (\bibinfo {year} {2016})}\BibitemShut {NoStop}%
\bibitem [{\citenamefont {Weaire}\ and\ \citenamefont {Hutzler}(2000)}]{Foam}%
  \BibitemOpen
  \bibfield  {author} {\bibinfo {author} {\bibfnamefont {D.~L.}\ \bibnamefont
  {Weaire}}\ and\ \bibinfo {author} {\bibfnamefont {S.}~\bibnamefont
  {Hutzler}},\ }\href {https://doi.org/10.1093/oso/9780198505518.001.0001}
  {\emph {\bibinfo {title} {{The Physics of Foams}}}}\ (\bibinfo  {publisher}
  {Oxford University Press},\ \bibinfo {year} {2000})\BibitemShut {NoStop}%
\bibitem [{\citenamefont {Kuramoto}(1984)}]{Kuramoto}%
  \BibitemOpen
  \bibfield  {author} {\bibinfo {author} {\bibfnamefont {Y.}~\bibnamefont
  {Kuramoto}},\ }\href {https://doi.org/10.1007/978-3-642-69689-3} {\emph
  {\bibinfo {title} {{Chemical Oscillations, Waves, and Turbulence}}}}\
  (\bibinfo  {publisher} {Springer},\ \bibinfo {year} {1984})\BibitemShut
  {NoStop}%
\bibitem [{\citenamefont {Loose}\ \emph {et~al.}(2008)\citenamefont {Loose},
  \citenamefont {Fischer-Friedrich}, \citenamefont {Ries}, \citenamefont
  {Kruse},\ and\ \citenamefont {Schwille}}]{loose2008spatial}%
  \BibitemOpen
  \bibfield  {author} {\bibinfo {author} {\bibfnamefont {M.}~\bibnamefont
  {Loose}}, \bibinfo {author} {\bibfnamefont {E.}~\bibnamefont
  {Fischer-Friedrich}}, \bibinfo {author} {\bibfnamefont {J.}~\bibnamefont
  {Ries}}, \bibinfo {author} {\bibfnamefont {K.}~\bibnamefont {Kruse}},\ and\
  \bibinfo {author} {\bibfnamefont {P.}~\bibnamefont {Schwille}},\ }\bibfield
  {title} {\bibinfo {title} {Spatial regulators for bacterial cell division
  self-organize into surface waves in vitro},\ }\href
  {https://doi.org/10.1126/science.1154413} {\bibfield  {journal} {\bibinfo
  {journal} {Science}\ }\textbf {\bibinfo {volume} {320}},\ \bibinfo {pages}
  {789} (\bibinfo {year} {2008})}\BibitemShut {NoStop}%
\bibitem [{\citenamefont {Cantrell}\ and\ \citenamefont
  {Cosner}(2003)}]{ecology}%
  \BibitemOpen
  \bibfield  {author} {\bibinfo {author} {\bibfnamefont {R.~S.}\ \bibnamefont
  {Cantrell}}\ and\ \bibinfo {author} {\bibfnamefont {C.}~\bibnamefont
  {Cosner}},\ }\href {https://doi.org/https://doi.org/10.1002/0470871296}
  {\emph {\bibinfo {title} {{Spatial Ecology via Reaction-Diffusion
  Equations}}}}\ (\bibinfo  {publisher} {John Wiley \& Sons, Ltd},\ \bibinfo
  {year} {2003})\BibitemShut {NoStop}%
\bibitem [{\citenamefont {Milde}\ \emph {et~al.}(2013)\citenamefont {Milde},
  \citenamefont {K\"ohler}, \citenamefont {Seidel}, \citenamefont {Eng},
  \citenamefont {Bauer}, \citenamefont {Chacon}, \citenamefont {Kindervater},
  \citenamefont {M\"uhlbauer}, \citenamefont {Pfleiderer}, \citenamefont
  {Buhrandt}, \citenamefont {Sch\"utte},\ and\ \citenamefont {Rosch}}]{Milde}%
  \BibitemOpen
  \bibfield  {author} {\bibinfo {author} {\bibfnamefont {P.}~\bibnamefont
  {Milde}}, \bibinfo {author} {\bibfnamefont {D.}~\bibnamefont {K\"ohler}},
  \bibinfo {author} {\bibfnamefont {J.}~\bibnamefont {Seidel}}, \bibinfo
  {author} {\bibfnamefont {L.~M.}\ \bibnamefont {Eng}}, \bibinfo {author}
  {\bibfnamefont {A.}~\bibnamefont {Bauer}}, \bibinfo {author} {\bibfnamefont
  {A.}~\bibnamefont {Chacon}}, \bibinfo {author} {\bibfnamefont
  {J.}~\bibnamefont {Kindervater}}, \bibinfo {author} {\bibfnamefont
  {S.}~\bibnamefont {M\"uhlbauer}}, \bibinfo {author} {\bibfnamefont
  {C.}~\bibnamefont {Pfleiderer}}, \bibinfo {author} {\bibfnamefont
  {S.}~\bibnamefont {Buhrandt}}, \bibinfo {author} {\bibfnamefont
  {C.}~\bibnamefont {Sch\"utte}},\ and\ \bibinfo {author} {\bibfnamefont
  {A.}~\bibnamefont {Rosch}},\ }\bibfield  {title} {\bibinfo {title} {Unwinding
  of a skyrmion lattice by magnetic monopoles},\ }\href
  {https://doi.org/10.1126/science.1234657} {\bibfield  {journal} {\bibinfo
  {journal} {Science}\ }\textbf {\bibinfo {volume} {340}},\ \bibinfo {pages}
  {1076} (\bibinfo {year} {2013})}\BibitemShut {NoStop}%
\bibitem [{\citenamefont {Park}\ \emph {et~al.}(2014)\citenamefont {Park},
  \citenamefont {Yu}, \citenamefont {Aizawa}, \citenamefont {Tanigaki},
  \citenamefont {Akashi}, \citenamefont {Takahashi}, \citenamefont {Matsuda},
  \citenamefont {Kanazawa}, \citenamefont {Onose}, \citenamefont {Shindo} \emph
  {et~al.}}]{Tokura2014three}%
  \BibitemOpen
  \bibfield  {author} {\bibinfo {author} {\bibfnamefont {H.~S.}\ \bibnamefont
  {Park}}, \bibinfo {author} {\bibfnamefont {X.}~\bibnamefont {Yu}}, \bibinfo
  {author} {\bibfnamefont {S.}~\bibnamefont {Aizawa}}, \bibinfo {author}
  {\bibfnamefont {T.}~\bibnamefont {Tanigaki}}, \bibinfo {author}
  {\bibfnamefont {T.}~\bibnamefont {Akashi}}, \bibinfo {author} {\bibfnamefont
  {Y.}~\bibnamefont {Takahashi}}, \bibinfo {author} {\bibfnamefont
  {T.}~\bibnamefont {Matsuda}}, \bibinfo {author} {\bibfnamefont
  {N.}~\bibnamefont {Kanazawa}}, \bibinfo {author} {\bibfnamefont
  {Y.}~\bibnamefont {Onose}}, \bibinfo {author} {\bibfnamefont
  {D.}~\bibnamefont {Shindo}}, \emph {et~al.},\ }\bibfield  {title} {\bibinfo
  {title} {Observation of the magnetic flux and three-dimensional structure of
  skyrmion lattices by electron holography},\ }\href
  {https://doi.org/10.1038/nnano.2014.52} {\bibfield  {journal} {\bibinfo
  {journal} {Nature Nanotech.}\ }\textbf {\bibinfo {volume} {9}},\ \bibinfo
  {pages} {337} (\bibinfo {year} {2014})}\BibitemShut {NoStop}%
\bibitem [{\citenamefont {Nagase}\ \emph {et~al.}(2019)\citenamefont {Nagase},
  \citenamefont {Komatsu}, \citenamefont {So}, \citenamefont {Ishida},
  \citenamefont {Yoshida}, \citenamefont {Kawaguchi}, \citenamefont {Tanaka},
  \citenamefont {Saitoh}, \citenamefont {Ikarashi}, \citenamefont {Kuwahara},\
  and\ \citenamefont {Nagao}}]{nagase2019smectic}%
  \BibitemOpen
  \bibfield  {author} {\bibinfo {author} {\bibfnamefont {T.}~\bibnamefont
  {Nagase}}, \bibinfo {author} {\bibfnamefont {M.}~\bibnamefont {Komatsu}},
  \bibinfo {author} {\bibfnamefont {Y.~G.}\ \bibnamefont {So}}, \bibinfo
  {author} {\bibfnamefont {T.}~\bibnamefont {Ishida}}, \bibinfo {author}
  {\bibfnamefont {H.}~\bibnamefont {Yoshida}}, \bibinfo {author} {\bibfnamefont
  {Y.}~\bibnamefont {Kawaguchi}}, \bibinfo {author} {\bibfnamefont
  {Y.}~\bibnamefont {Tanaka}}, \bibinfo {author} {\bibfnamefont
  {K.}~\bibnamefont {Saitoh}}, \bibinfo {author} {\bibfnamefont
  {N.}~\bibnamefont {Ikarashi}}, \bibinfo {author} {\bibfnamefont
  {M.}~\bibnamefont {Kuwahara}},\ and\ \bibinfo {author} {\bibfnamefont
  {M.}~\bibnamefont {Nagao}},\ }\bibfield  {title} {\bibinfo {title} {Smectic
  liquid-crystalline structure of skyrmions in chiral magnet
  {${\mathrm{Co}}_{8.5}{\mathrm{Zn}}_{7.5}{\mathrm{Mn}}_{4}(110)$} thin film},\
  }\href {https://doi.org/10.1103/PhysRevLett.123.137203} {\bibfield  {journal}
  {\bibinfo  {journal} {Phys. Rev. Lett.}\ }\textbf {\bibinfo {volume} {123}},\
  \bibinfo {pages} {137203} (\bibinfo {year} {2019})}\BibitemShut {NoStop}%
\bibitem [{\citenamefont {Seki}\ \emph {et~al.}(2020)\citenamefont {Seki},
  \citenamefont {Garst}, \citenamefont {Waizner}, \citenamefont {Takagi},
  \citenamefont {Khanh}, \citenamefont {Okamura}, \citenamefont {Kondou},
  \citenamefont {Kagawa}, \citenamefont {Otani},\ and\ \citenamefont
  {Tokura}}]{Tokura2020-propagation}%
  \BibitemOpen
  \bibfield  {author} {\bibinfo {author} {\bibfnamefont {S.}~\bibnamefont
  {Seki}}, \bibinfo {author} {\bibfnamefont {M.}~\bibnamefont {Garst}},
  \bibinfo {author} {\bibfnamefont {J.}~\bibnamefont {Waizner}}, \bibinfo
  {author} {\bibfnamefont {R.}~\bibnamefont {Takagi}}, \bibinfo {author}
  {\bibfnamefont {N.}~\bibnamefont {Khanh}}, \bibinfo {author} {\bibfnamefont
  {Y.}~\bibnamefont {Okamura}}, \bibinfo {author} {\bibfnamefont
  {K.}~\bibnamefont {Kondou}}, \bibinfo {author} {\bibfnamefont
  {F.}~\bibnamefont {Kagawa}}, \bibinfo {author} {\bibfnamefont
  {Y.}~\bibnamefont {Otani}},\ and\ \bibinfo {author} {\bibfnamefont
  {Y.}~\bibnamefont {Tokura}},\ }\bibfield  {title} {\bibinfo {title}
  {Propagation dynamics of spin excitations along skyrmion strings},\ }\href
  {https://doi.org/10.1038/s41467-019-14095-0} {\bibfield  {journal} {\bibinfo
  {journal} {Nature Commun.}\ }\textbf {\bibinfo {volume} {11}},\ \bibinfo
  {pages} {256} (\bibinfo {year} {2020})}\BibitemShut {NoStop}%
\end{thebibliography}

%

\end{document}